\documentclass[twocolumn,appendixfloats]{emulateapj}
\usepackage{amsmath}
\usepackage{graphicx}
\usepackage{rotating}
\usepackage{longtable}

\usepackage{natbib}
\def\micron{{$\, {\mu}$m}}

\def\arcsec{{\mbox{$^{\prime\prime}$}}}
\def\lir{{L$_{IR}$}}
\def\mirslope{{F$_{\nu}$[30$\mu$m]/F$_{\nu}$[15$\mu$m]}}
\def\eqw{{EQW$_{6.2\mu m}$}}
\def\sil{{$s_{9.7\mu m}$}}

\shorttitle{MIR Properties of Nearby LIRGs}
\shortauthors{Stierwalt et al.}

\begin{document}
\title{Mid-Infrared Properties of Nearby Luminous Infrared Galaxies I:
  Spitzer IRS Spectra for the GOALS Sample}
\author{S. Stierwalt\altaffilmark{1,2}, L. Armus\altaffilmark{1}, J.A. Surace\altaffilmark{1}, H. Inami\altaffilmark{1,3}, A.O. Petric\altaffilmark{1,4}, T. Diaz-Santos\altaffilmark{1}, S. Haan\altaffilmark{1,5}, V. Charmandaris\altaffilmark{6,7}, J. Howell\altaffilmark{1}, D.C. Kim\altaffilmark{8}, J. Marshall\altaffilmark{1},  J.M. Mazzarella\altaffilmark{9}, H.W.W. Spoon\altaffilmark{10}, S. Veilleux\altaffilmark{11}, A. Evans\altaffilmark{2,8}, D. B. Sanders\altaffilmark{12}, P. Appleton\altaffilmark{13}, G. Bothun\altaffilmark{14}, C.R. Bridge\altaffilmark{4}, B. Chan\altaffilmark{9}, D. Frayer\altaffilmark{15}, K. Iwasawa\altaffilmark{16}, L.J. Kewley\altaffilmark{12}, S. Lord\altaffilmark{9}, B.F. Madore\altaffilmark{17}, J.E. Melbourne\altaffilmark{4}, E.J. Murphy\altaffilmark{17}, J.A. Rich\altaffilmark{12}, B. Schulz\altaffilmark{13}, E. Sturm\altaffilmark{18}, V. U\altaffilmark{12}, T. Vavilkin\altaffilmark{19}, K. Xu\altaffilmark{9}}

\altaffiltext{1}{Spitzer Science Center, California Institute of Technology, 1200 E. California Blvd., Pasadena, CA 91125. {\textit{e-mail:}} sabrinas@virginia.edu}
\altaffiltext{2}{Department of Astronomy, University of Virginia, P.O. Box 400325, Charlottesville, VA 22904.}
\altaffiltext{3}{National Optical Astronomy Observatory, 950 N. Cherry Ave, Tucson, AZ 85719.}
\altaffiltext{4}{Department of Astronomy, California Institute of Technology, 1200 E. California Blvd., Pasadena, CA 91125.}
\altaffiltext{5}{CSIRO Astronomy \& Space Science, Marsfield NSW 2122, Australia.}
\altaffiltext{6}{Department of Physics and ITCP, University of Crete, GR-71003 Heraklion, Greece.}
\altaffiltext{7}{IESL/Foundation for Research and Technology - Hellas,  GR-71110, Heraklion, Greece and Chercheur Associ\'e, Observatoire de Paris, F-75014, Paris, France.}
\altaffiltext{8}{National Radio Astronomy Observatory, 520 Edgemont Road, Charlottesville, VA 22903.}
\altaffiltext{9}{Infrared Processing \& Analysis Center, MS 100-22, California Institute of Technology, Pasadena, CA 91125.}
\altaffiltext{10}{Department of Astronomy, Cornell University, Ithaca, NY, 14853.}
\altaffiltext{11}{Astronomy Department, University of Maryland, College Park, MD 20742.}
\altaffiltext{12}{Institute for Astronomy, University of Hawaii, 2680 Woodlawn Drive, Honolulu, HI 96825.}
\altaffiltext{13}{NASA Herschel Science Center, 770 S. Wilson Ave., Pasadena, CA 91125.}
\altaffiltext{14}{Physics Department, University of Oregon, Eugene, OR 97402.}
\altaffiltext{15}{National Radio Astronomy Observatory, P.O. Box 2, Green Bank, WV 24944.}
\altaffiltext{16}{INAF-Observatorio Astronomico di Bologna, Via Ranzani 1, Bologna, Italy.}
\altaffiltext{17}{The Observatories, Carnegie Institute of Washington, 813 Santa Barbara Street, Pasadena, CA 91101.}
\altaffiltext{18}{MPE, Postfach 1312, 85741 Garching Germany.}
\altaffiltext{19}{Department of Physics and Astronomy, SUNY Stony Brook, Stony Brook, NY, 11794.}



\begin{abstract}
The Great Observatories All-Sky LIRG Survey (GOALS) is a comprehensive, multiwavelength study of luminous infrared galaxies (LIRGs) in the local universe. Here we present low resolution Spitzer IRS spectra covering 5-38\micron~and provide a basic analysis of the mid-IR spectral properties observed for nearby LIRGs. In a companion paper, we discuss detailed fits to the spectra and compare the LIRGs to other classes of galaxies. The GOALS sample of 244 nuclei in 180 luminous ($10^{11} \le L_{IR}/L_{\odot} < 10^{12}$) and 22 ultraluminous ($L_{IR}/L_{\odot} \ge 10^{12}$) IR galaxies represents a complete subset of the IRAS Revised Bright Galaxy Sample and covers a range of merger stages, morphologies and spectral types. The majority ($>$60\%) of the GOALS LIRGs have high 6.2\micron~PAH equivalent widths (EQW$_{6.2\mu m} >$ 0.4\micron) and low levels of silicate absorption ($s_{9.7\mu m} > $ -1.0). There is a general trend among the U/LIRGs for both silicate depth and mid-infrared (MIR) slope to increase with increasing \lir. U/LIRGs in the late to final stages of a merger also have, on average, steeper MIR slopes and higher levels of dust obscuration. Together, these trends suggest that as gas \& dust is funneled towards the center of a coalescing merger, the nuclei become more compact and more obscured. As a result, the dust temperature increases leading also to a steeper MIR slope. The sources that depart from these correlations have very low PAH equivalent width (\eqw\ $<$ 0.1\micron) consistent with their emission being dominated by an AGN in the MIR. These extremely low PAH equivalent width sources separate into two distinct types: relatively unobscured sources with a very hot dust component (and thus very shallow MIR slopes) and heavily dust obscured nuclei with a steep temperature gradient. The most heavily dust obscured sources are also the most compact in their MIR emission, suggesting that the obscuring (cool) dust is associated with the outer regions of the starburst and not simply a measure of the dust along the line of sight through a large, dusty disk. A marked decline is seen for the fraction of high EQW (star formation dominated) sources as the merger progresses. The decline is accompanied by an increase in the fraction of composite sources while the fraction of sources where an AGN dominates the MIR emission remains low. When compared to the MIR spectra of submillimeter galaxies (SMGs) at z$\sim$2, both the average GOALS LIRG and ULIRG spectra are more absorbed at 9.7\micron~and the average GOALS LIRG has more PAH emission. However, when the AGN contributions to both the local GOALS LIRGs and the high-z SMGs are removed, the average local starbursting LIRG closely resembles the starburst-dominated SMGs.
\end{abstract}


\section{Introduction}
A principal achievement of the Infrared Astronomical Satellite (IRAS) was
the discovery of a large population of 
galaxies whose bolometric luminosities were dominated by emission in
the infrared. At the highest luminosities, local ultraluminous infrared
galaxies (ULIRGs; L$_{IR} \ge 10^{12}$L$_{\odot}$) have been heavily
studied \citep{armusULIRGs, sandersULIRGs, murphyULIRGs, spoonULIRGs,
  vandanaULIRGs, rigULIRGs, genzelULIRGs}, and a clear formation picture
has been pieced together to explain their extreme emission in the
infrared: more than 90\% of local ULIRGs are the products of major mergers
between molecular gas-rich galaxies. The large amounts of gas that are
funneled into the centers of these mergers lead 
to intense star formation, the feeding of a central AGN, extremely compact reservoirs of molecular gas, and
infrared luminosities on the order of ten times their optical
luminosities. 

While ULIRGs constitute only 3\% of the IRAS Revised
Bright Galaxy Sample \citep[RBGS;][]{rbgs}, at just slightly lower luminosities, luminous infrared galaxies
(LIRGs; $10^{11}$M$_{\odot} \le $ L$_{IR} < 10^{12}$M$_{\odot}$) make up almost 1/3 of
the IR sources and have formation histories that are far
less straightforward. In the local universe, there is evidence that
galaxy-galaxy interactions and mergers drive the large IR luminosities
in some LIRGs \citep{sandersLIRGs} and many high-z submillimeter
galaxies (SMGs) show hints of disturbed optical and radio morphologies
\citep{blain, dasyra}. However, at least 20\% and as many as 40\% of local
LIRGs may have no history of major tidal
interactions (Howell et al., ${\it{in~prep}}$). LIRGs are also
represented across the full range of merger stages, unlike ULIRGs
which are almost always at the very end stages of coalescing. 

Although LIRGs are relatively rare in the local
universe, their comoving number
density increases by more than 100 times from the current epoch to z$\sim$1,
\citep{lefloch, magnelli} until LIRGs dominate the total IR energy density at
redshifts of z$\sim$1-2 when star formation in the universe was at its
peak \citep{caputi}. Piecing together the formation mechanisms
and subsequent evolution of these LIRG systems is thus vital to
understanding the processes governing star formation and black hole accretion, the main
sources of emitting power in the IR.

The Great Observatories All-sky LIRG Survey \citep[GOALS;][]{GOALS} represents a
complete subset of the RBGS comprising 180 LIRGs and 22 ULIRGs and aims to provide a multiwavelength understanding
of the formation and evolution of local LIRGs as a class of galaxy. As part
of the Spitzer Legacy survey, a complete set of IR imaging (Infrared
Array Camera (IRAC) at 3.6, 4.5, 5, and 8\micron, and Multiband
Imaging Photometer (MIPS) at 24, 70, and
160\micron) and IR spectroscopy at both high and low resolution
(Infrared Spectrograph (IRS) from 5-38\micron) is available for the entire
sample. In addition, imaging in the near-IR/optical \citep[Hubble Space Telescope NICMOS and
ACS;][Kim et al., ${\it{in~prep}}$]{haanHST}, the UV \citep[Galaxy Evolution Explorer near- and far-UV;][]{GALEX}, and the
X-Ray \citep[Chandra;][]{CHANDRAGOALS} are available for large subsets of the sample. 

In this paper, we present the mid-infrared (MIR) spectra for 244 galaxy nuclei in
the 202 nearby GOALS U/LIRG systems taken with the low resolution module on
the Spitzer Infrared Spectrograph \citep[IRS;][]{IRS}. The MIR
properties derived directly from such a large, complete sample of LIRG
spectra will allow us to place these intermediate-luminosity systems into the context of both
the extensive previous local ULIRG studies as well as those for lower
luminosity, star-forming or starbursting systems \citep{brandlEW,
  jdsings, SSGSS, 5muses}.

Full spectral decompositions, including fits to the gas and dust
features as well as the SEDs covered by the IRS data, for the entire sample along with the
comparison of MIR galaxy properties to those at other wavelengths will be presented
in Stierwalt et al. (2013b). The analysis presented here focuses on
properties derived directly from the MIR spectra. 
In Section \ref{obs}, we present the low resolution MIR spectra
observed with the Spitzer IRS Short-Low and Long-Low modules and describe our data
reduction methods.  In Section \ref{results} we give the distributions
of the MIR properties and investigate correlations with \lir\ and
compactness. In Section \ref{mergersection}, we follow each MIR property
through the merging process, and we place
our results into a high redshift context through comparisons to MIR spectra
of submillimeter galaxies at z$\sim$2 in Section \ref{smgsec}.
Finally, our summary and conclusions are presented in Section \ref{conc}. 

\section{Observations \& Data Reduction}\label{obs}

\subsection{The Sample}
The GOALS sample consists of 244 galaxy nuclei in 180 luminous and 22
ultraluminous nearby IR
galaxies. New spectra were obtained using the staring
mode for the IRS Short-Low (SL: 5.5-14.5\micron) and
Long-Low (LL: 14-38\micron) modules for 157 galaxies
(PID 30323; PI L. Armus). Integration times were determined from nuclear flux densities measured
from IRAC and MIPS images and range from
45-120 seconds in SL and 30-120 seconds in
LL. Secondary nuclei were targeted when the MIPS 24\micron~flux ratio of primary to secondary
was $\leq$ 5. Archival spectroscopic observations were used for the remaining 45
systems and borrowed most heavily from staring program PIDs 105, 3247, \& 20549 and mapping program PIDs 73, 3269, \& 30577.

All 202 systems are nearby but cover a range of distances (15 Mpc $<$
D $<$ 400 Mpc) and so the resulting projected IRS slit size varies
from source to source. At the median galaxy distance of 100 Mpc, the
nuclear spectrum covers the central 1.8 kpc in SL and the central 5.2
kpc in LL. 

\subsection{Data Reduction}
Staring mode spectroscopic data were reduced using the S17 and S18.7 IRS pipelines from the
Spitzer Science
Center\footnote{http://ssc.spitzer.caltech.edu/irs/features.html}. For
most sources, off-source
nods were used to perform background sky subtraction. In the cases of
more extended objects, dedicated background pointings were used to
determine the sky surface brightness. One dimensional spectra were
extracted using the standard extraction aperture and point source
calibration modes in
SPICE\footnote{http://ssc.spitzer.caltech.edu/postbcd/doc/spice.pdf} which
employs a tapered extraction aperture that averages roughly to a size
of 10$\farcs6 \times 36\farcs$6 in LL and $3\farcs7 \times 9\farcs5$ in SL. After masking bad pixels,
multiple nods were averaged to produce the final spectrum. 

 Of the archival data, 27 spectra were taken in staring mode and were
 reduced as described above. For the remaining 18 systems, spectra
 were extracted from low resolution mapping mode data using CUBISM
 \citep{cubism}. Two-dimensional BCDs were assembled, obvious bad
 pixels were removed and nuclear spectra were extracted. In two
 cases (CGCG011-076 and IC1623B), smaller apertures were necessary to avoid
 other sources in the Long-Low maps, but for most sources 2$\times$5 pixel extraction apertures
 centered on the galaxy's nucleus were used to resemble as closely as
 possible the results that would have been achieved with staring mode
 observations. However, since the tapered aperture used by SPICE
 cannot be completely reproduced by the square apertures in CUBISM, a
 further mapping-to-staring-mode correction was applied to all spectra
 derived from low resolution archival maps. The correction, a
 multiplicative factor that is a function of wavelength, was derived
 from NGC6240, a star-forming merger remnant typical of the GOALS
 sample for which both staring and mapping data were obtained. The
 correction function varies from 1.3 to 2.7 over SL wavelengths and from 1.7 to 2.3 over LL wavelengths.  

\begin{figure*}
\hspace{-2.0cm}
\begin{sideways}
\begin{minipage}[t]{8.5in}
\centering 
\hspace{-1.0cm}
\includegraphics[height=1.85in,width=1.85in,clip,viewport=10 10 235 235]{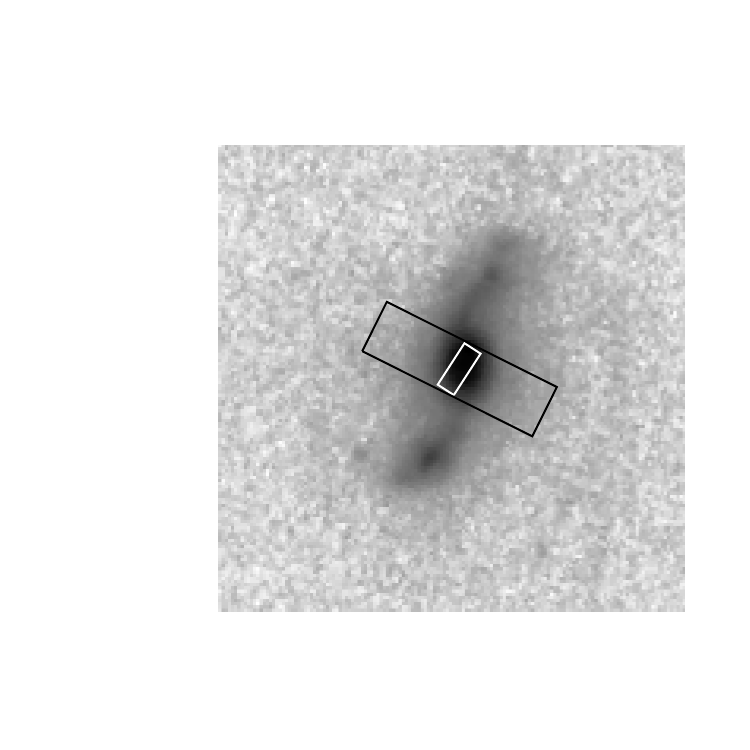}
\hspace{-1.8cm}
\includegraphics[width=3.3in,height=1.5in,clip]{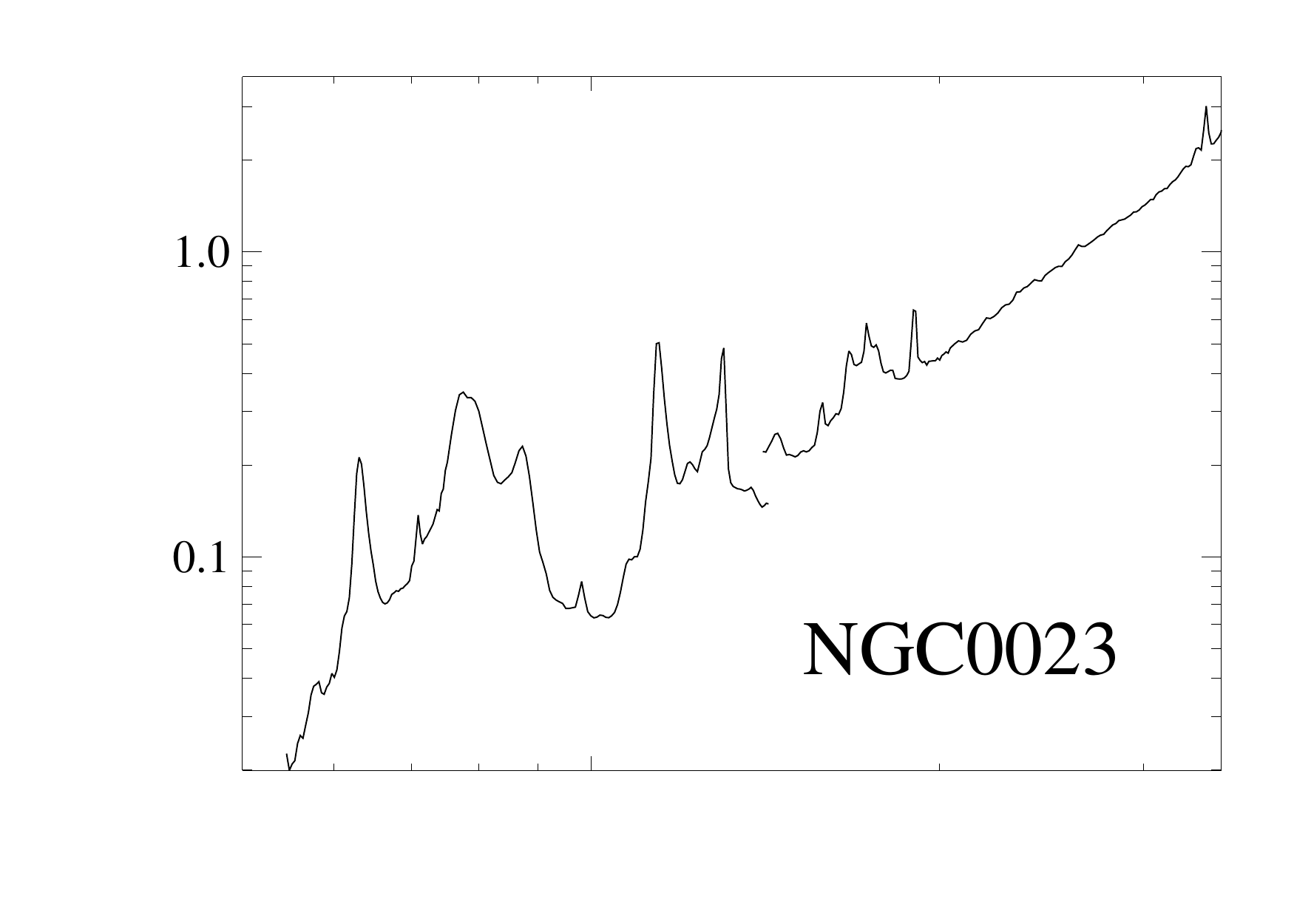}
\hspace{-1.0cm}
\includegraphics[height=1.85in,width=1.85in,clip,viewport=10 10 235 235]{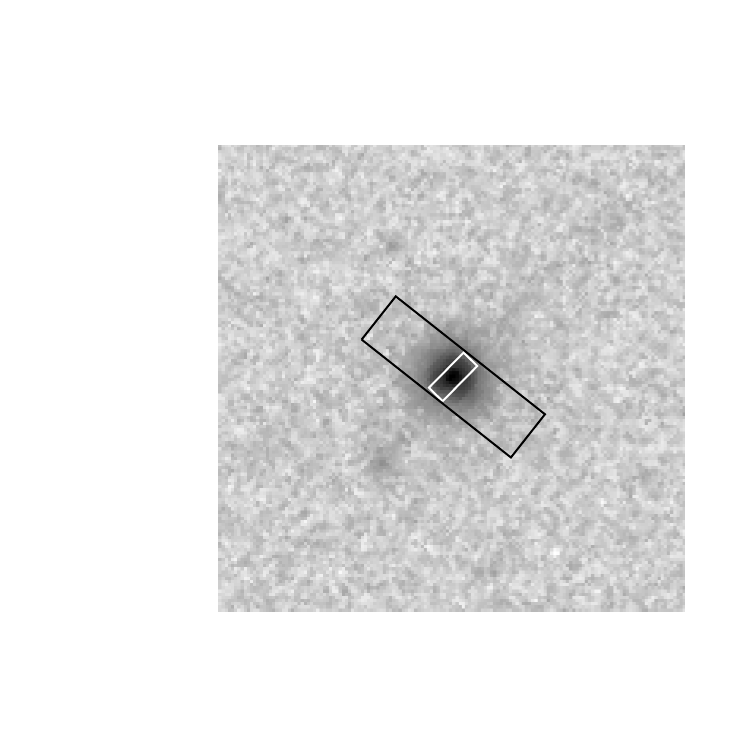}
\hspace{-1.8cm}
\includegraphics[width=3.3in,height=1.5in,clip]{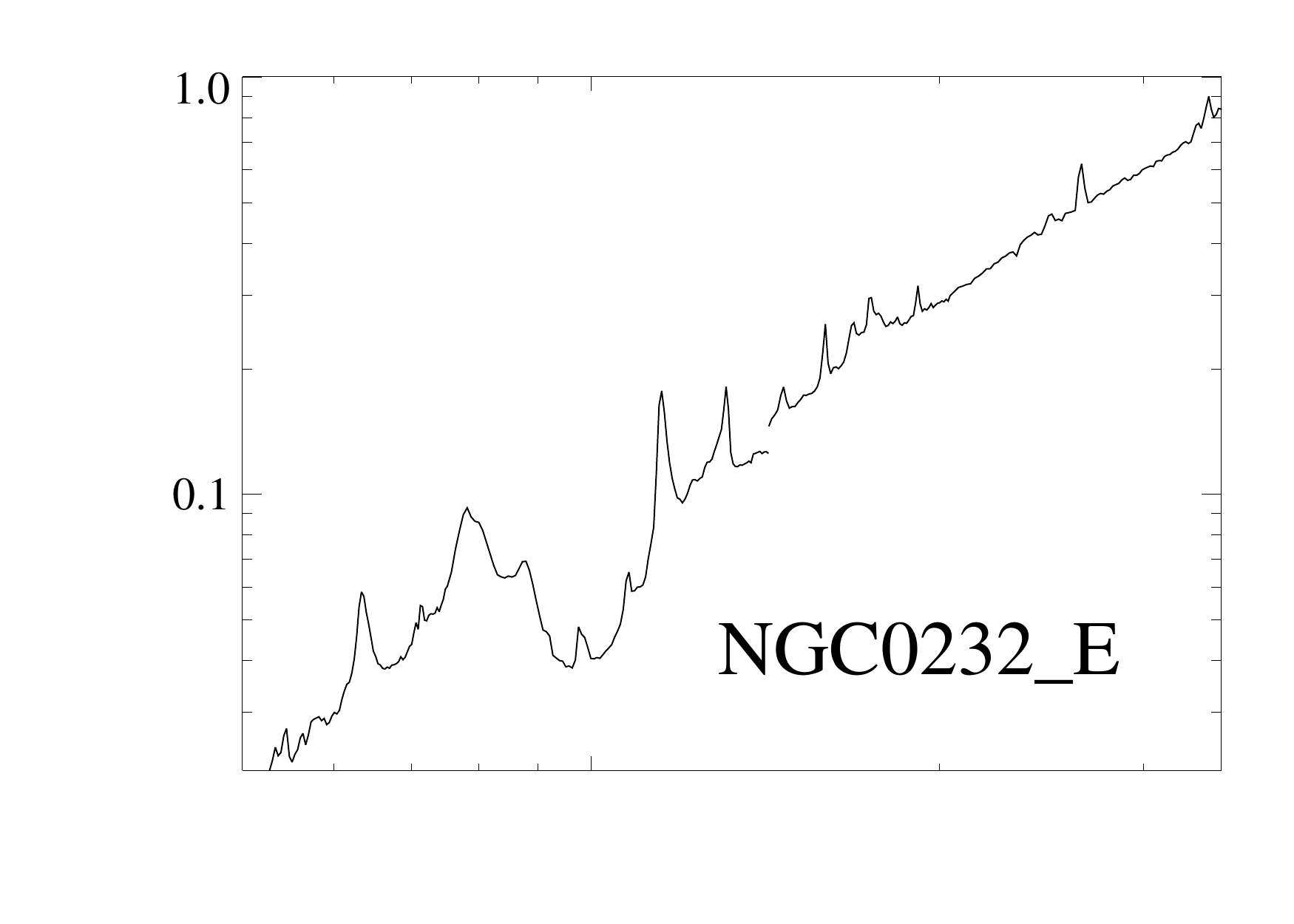}
\end{minipage}
\hfill 
\end{sideways}
\hspace{-1.95cm}
\begin{sideways}
\begin{minipage}[t]{8.5in}
\centering 
\hspace{-1.0cm}
\includegraphics[height=1.85in,width=1.85in,clip,viewport=10 10 235 235]{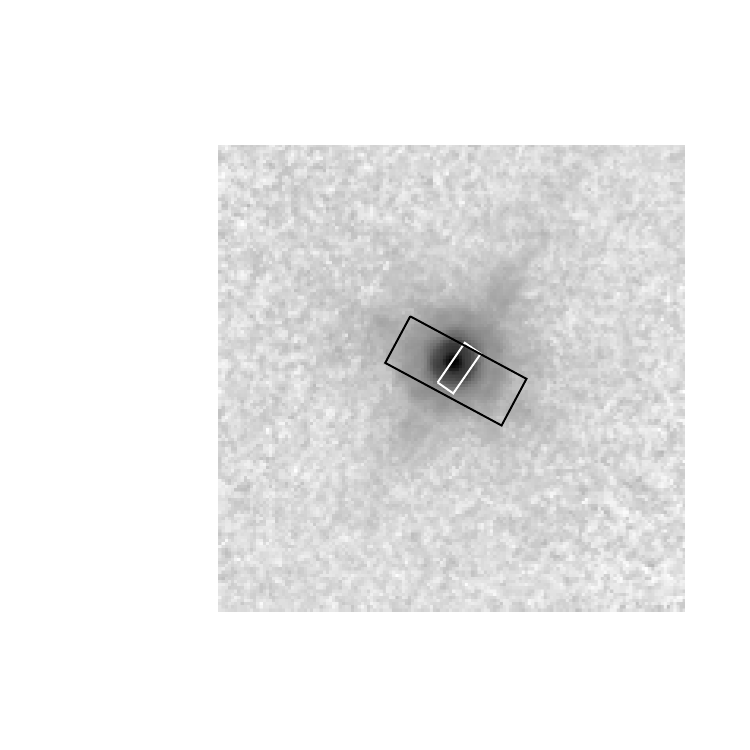}
\hspace{-1.8cm}
\includegraphics[width=3.3in,height=1.5in,clip]{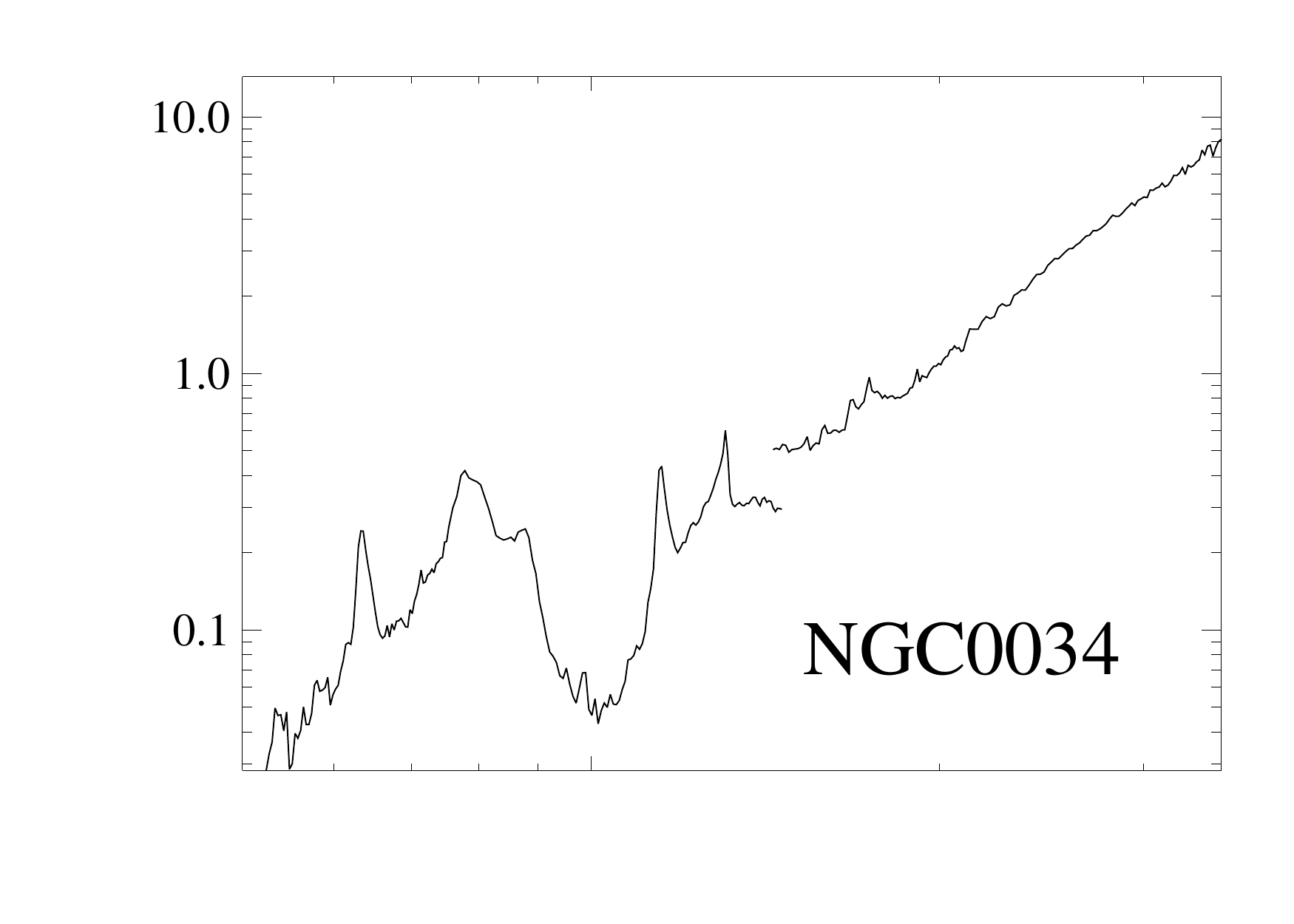}
\hspace{-1.0cm}
\includegraphics[height=1.85in,width=1.85in,clip,viewport=10 10 235 235]{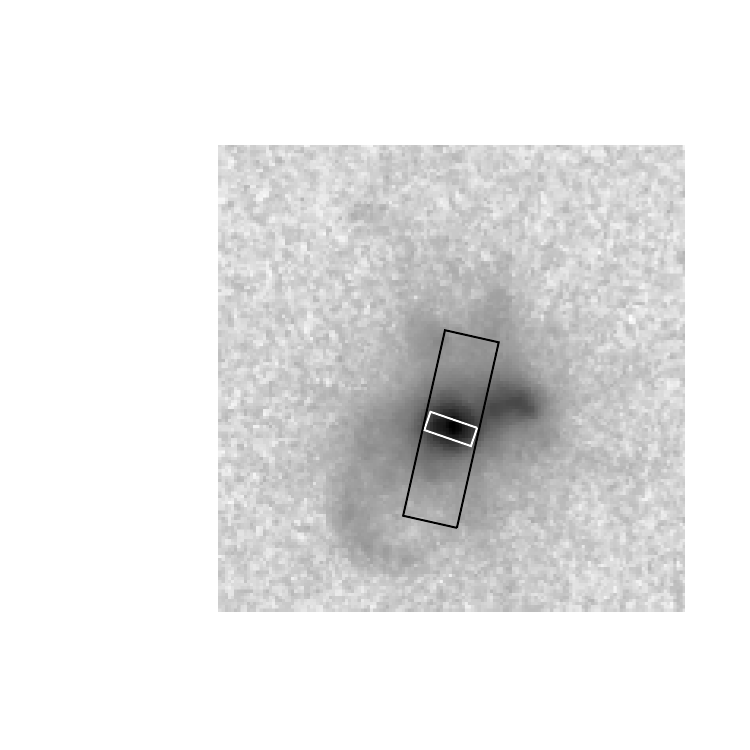}
\hspace{-1.8cm}
\includegraphics[width=3.3in,height=1.5in,clip]{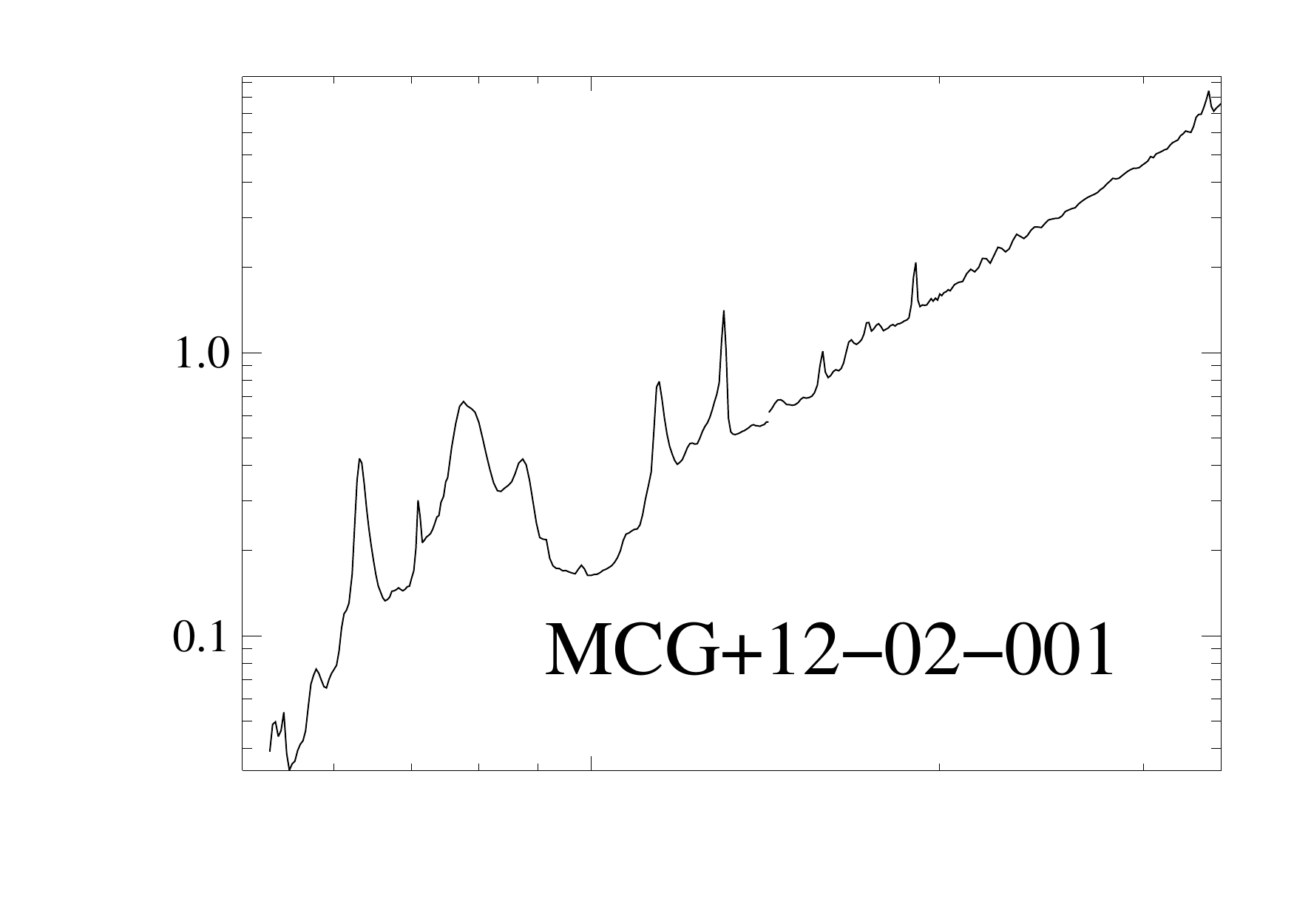}
\end{minipage}
\hfill 
\end{sideways}
\hspace{-1.95cm}
\begin{sideways}
\begin{minipage}[t]{8.5in}
\centering 
\hspace{-1.0cm}
\includegraphics[height=1.85in,width=1.85in,clip,viewport=10 10 235 235]{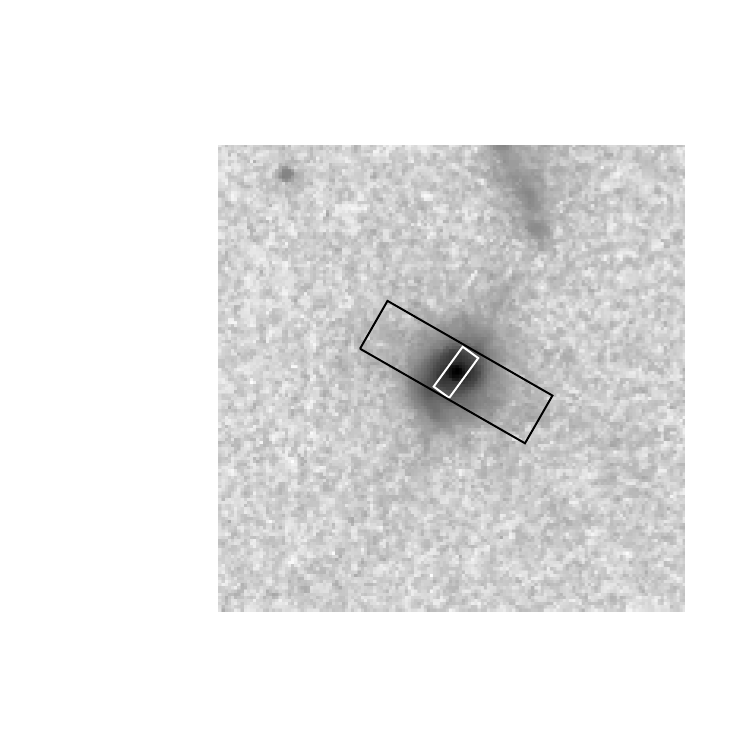}
\hspace{-1.8cm}
\includegraphics[width=3.3in,height=1.5in,clip]{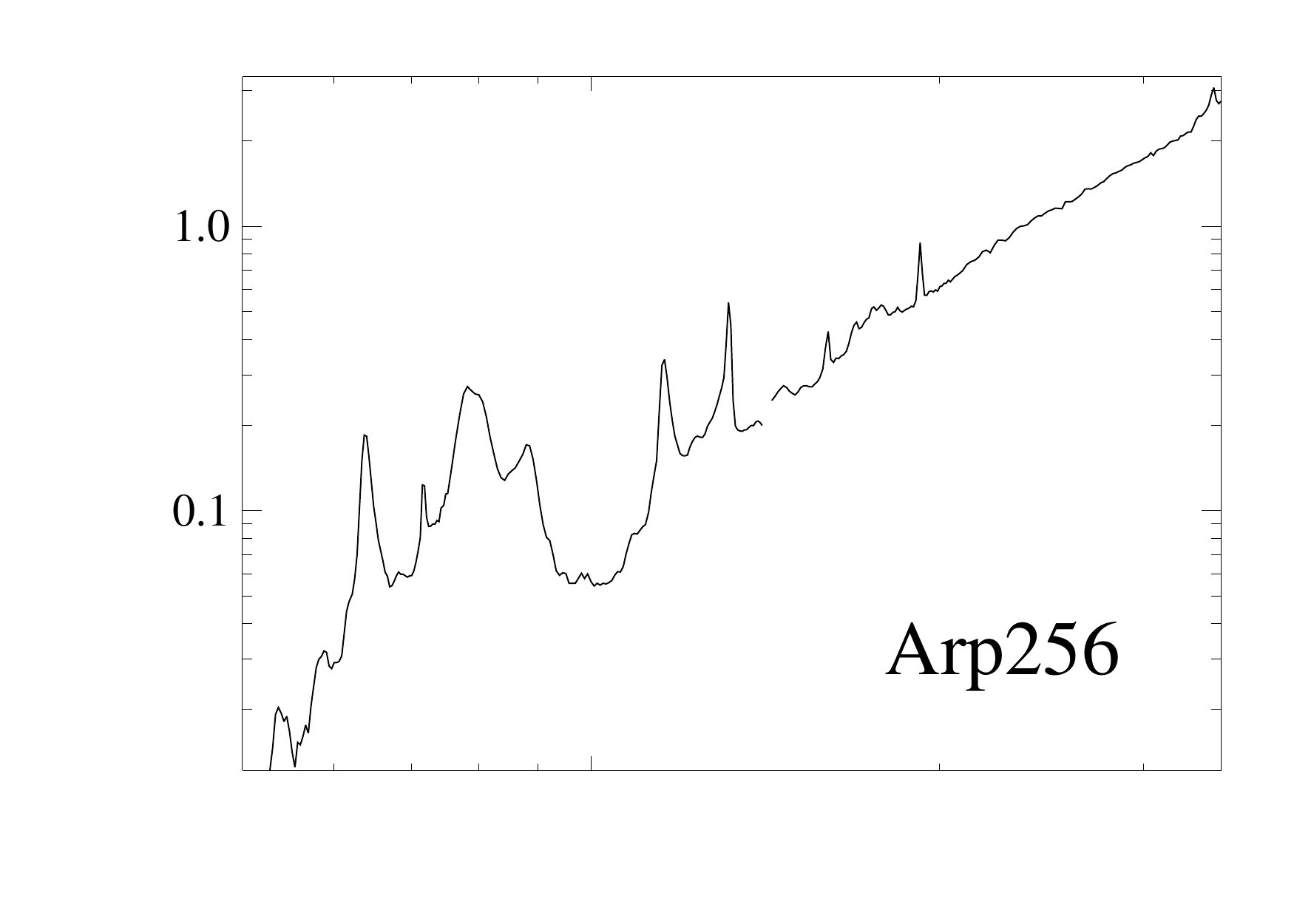}
\hspace{-1.0cm}
\includegraphics[height=1.85in,width=1.85in,clip,viewport=10 10 235 235]{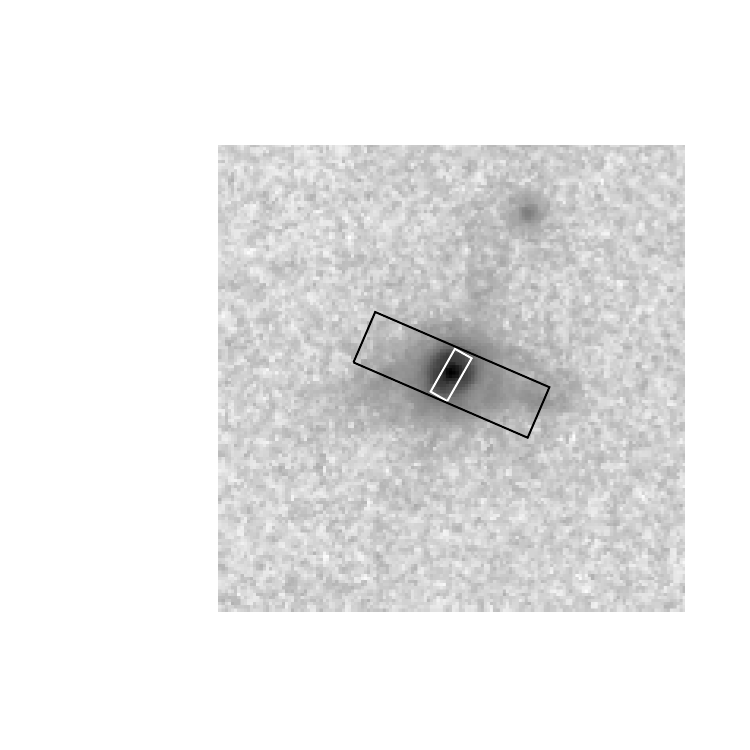}
\hspace{-1.8cm}
\includegraphics[width=3.3in,height=1.5in,clip]{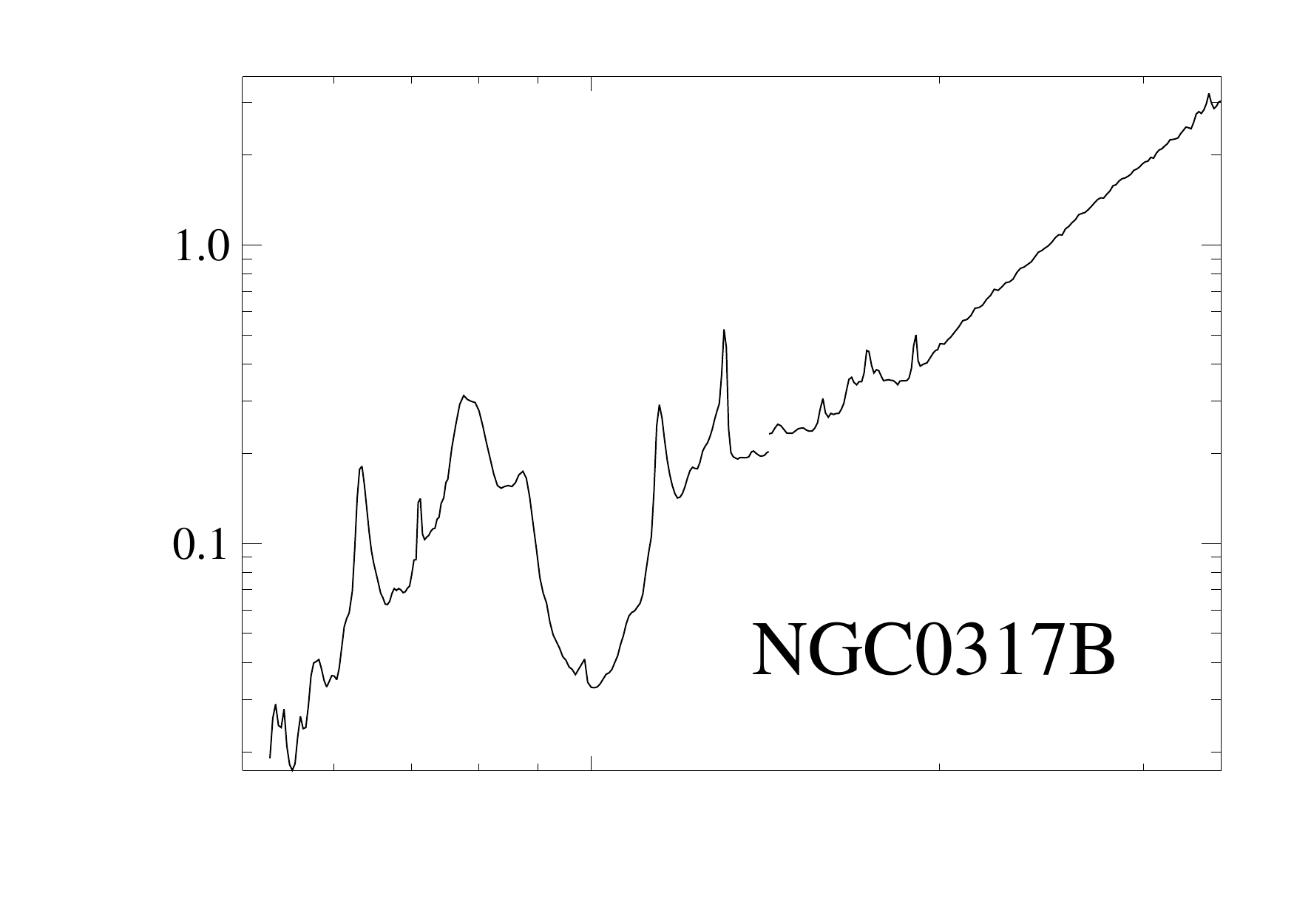}
\end{minipage}
\hfill 
\end{sideways}
\hspace{-1.95cm}
\begin{sideways}
\begin{minipage}[t]{8.5in}
\centering 
\hspace{-1.0cm}
\includegraphics[height=1.85in,width=1.85in,clip,viewport=10 10 235 235]{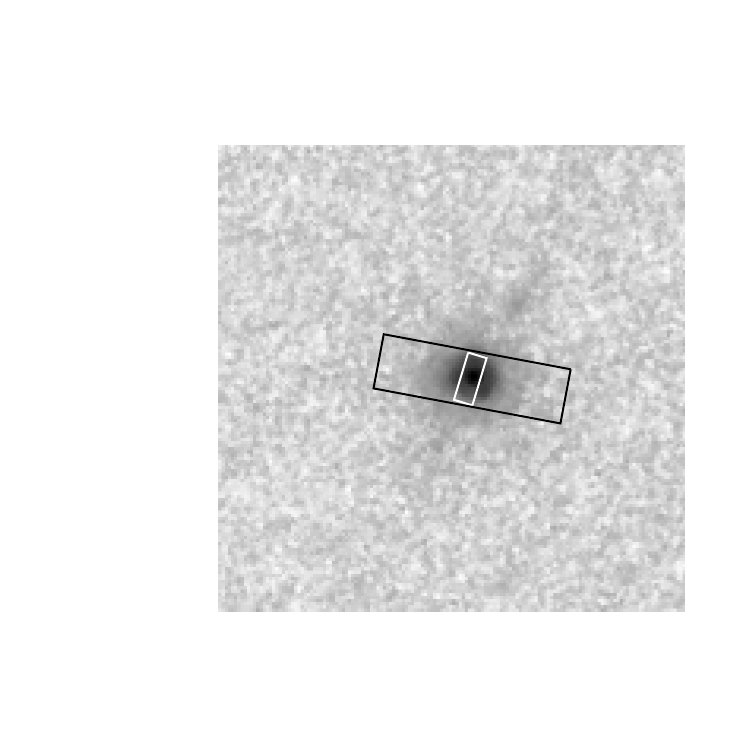}
\hspace{-1.8cm}
\includegraphics[width=3.3in,height=1.5in,clip]{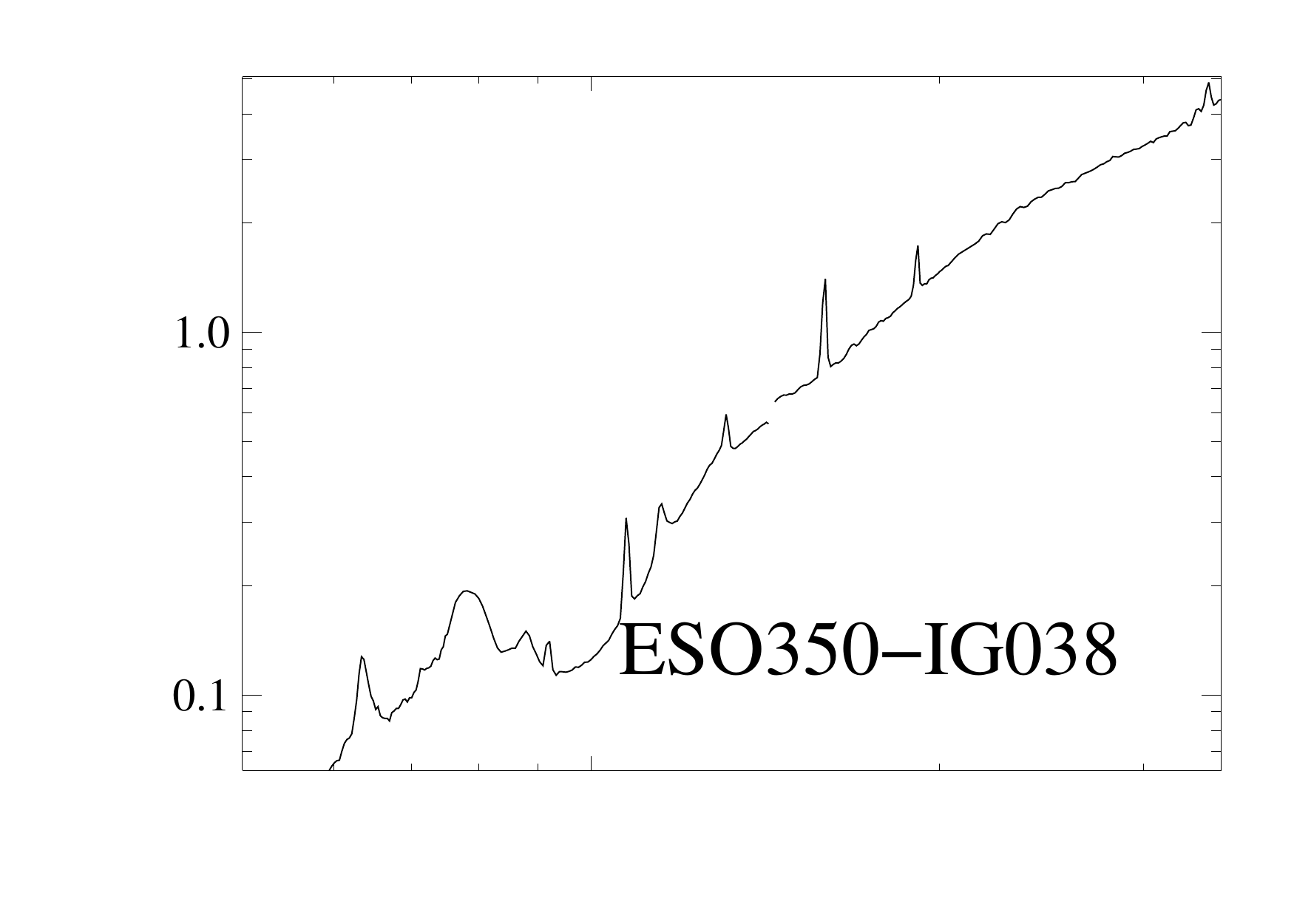}
\hspace{-1.0cm}
\includegraphics[height=1.85in,width=1.85in,clip,viewport=10 10 235 235]{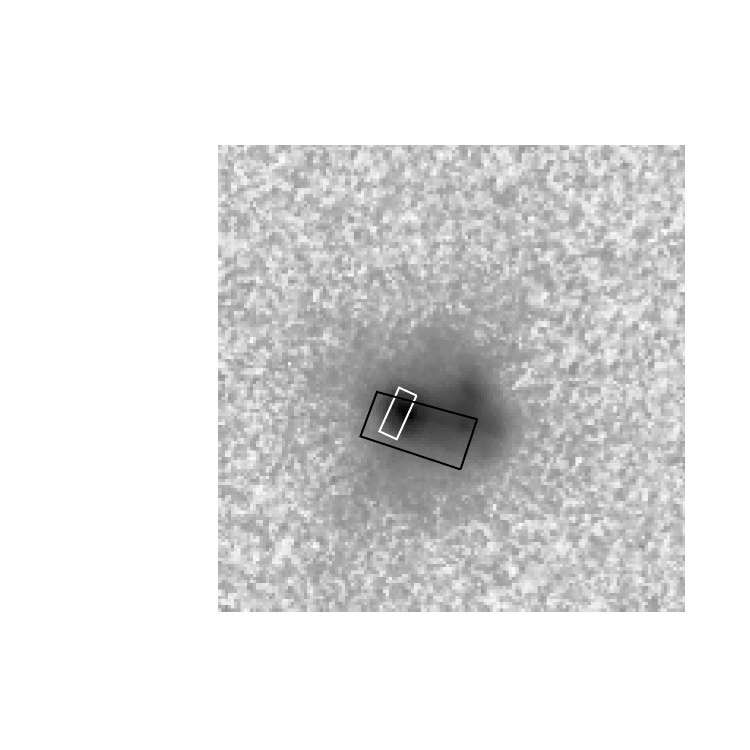}
\hspace{-1.8cm}
\includegraphics[width=3.3in,height=1.5in,clip]{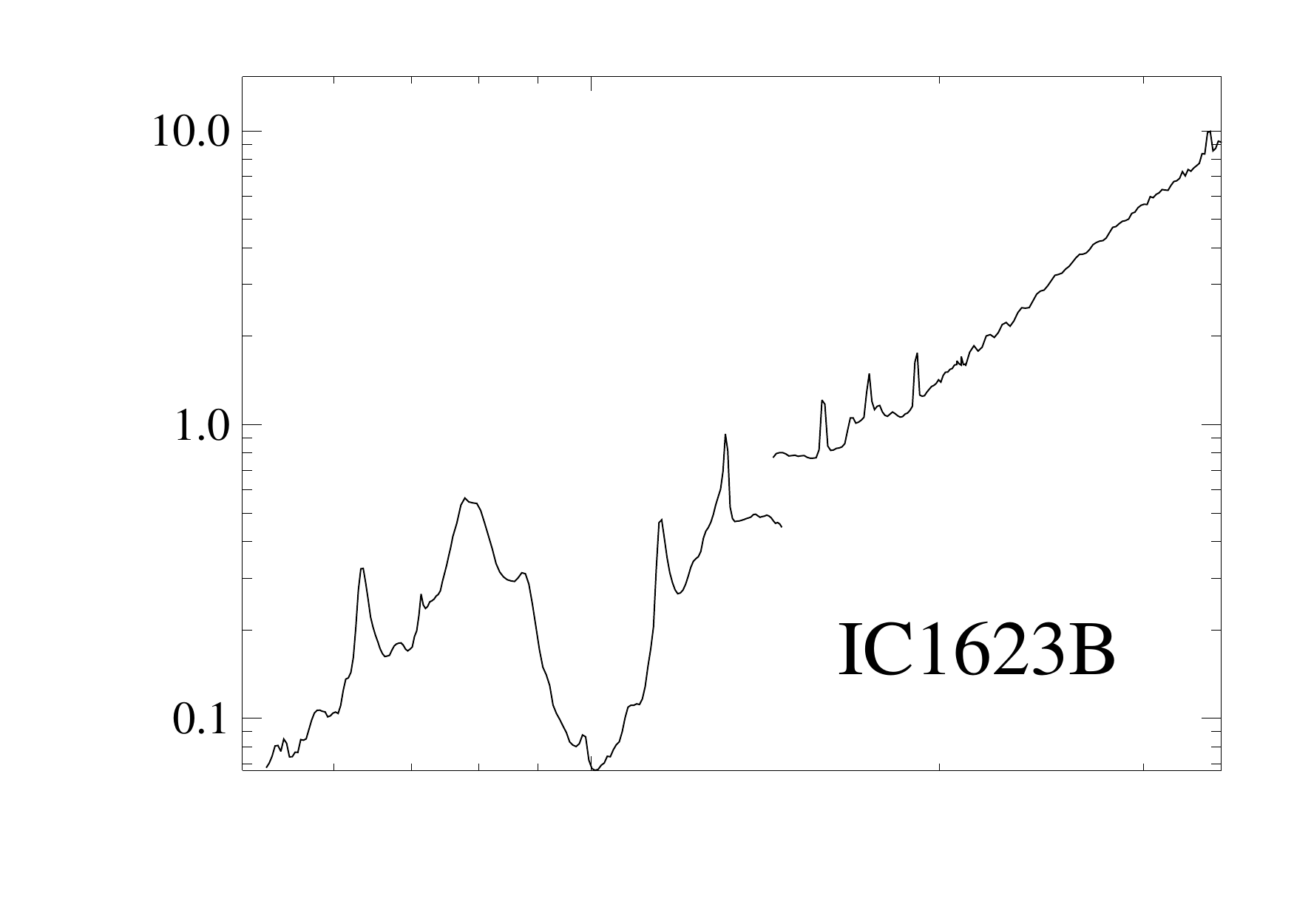}
\end{minipage}
\hfill 
\end{sideways}
\hspace{-1.95cm}
\begin{sideways}
\begin{minipage}[t]{8.5in}
\centering 
\hspace{-1.0cm}
\includegraphics[height=1.85in,width=1.85in,clip,viewport=10 10 235 235]{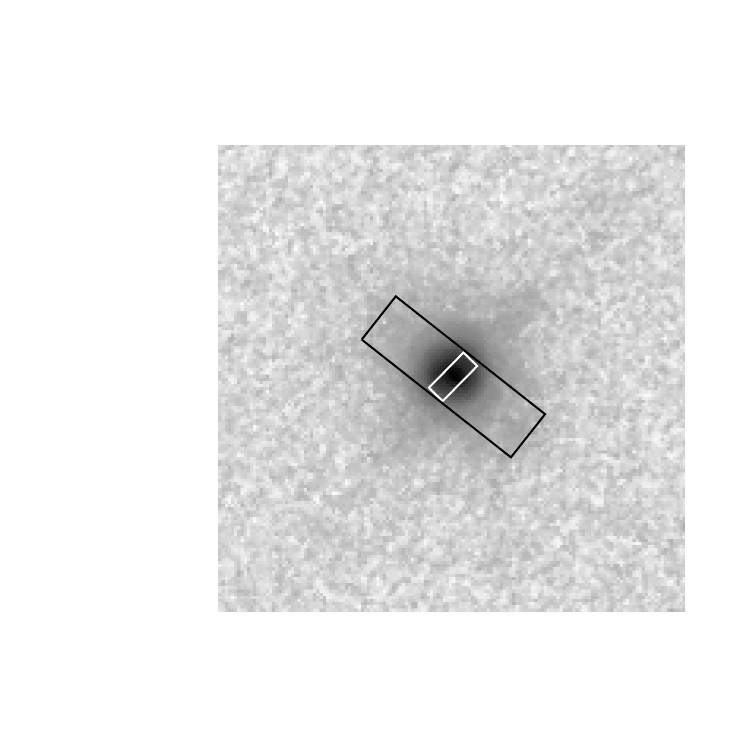}
\hspace{-1.8cm}
\includegraphics[width=3.3in,height=1.5in,clip]{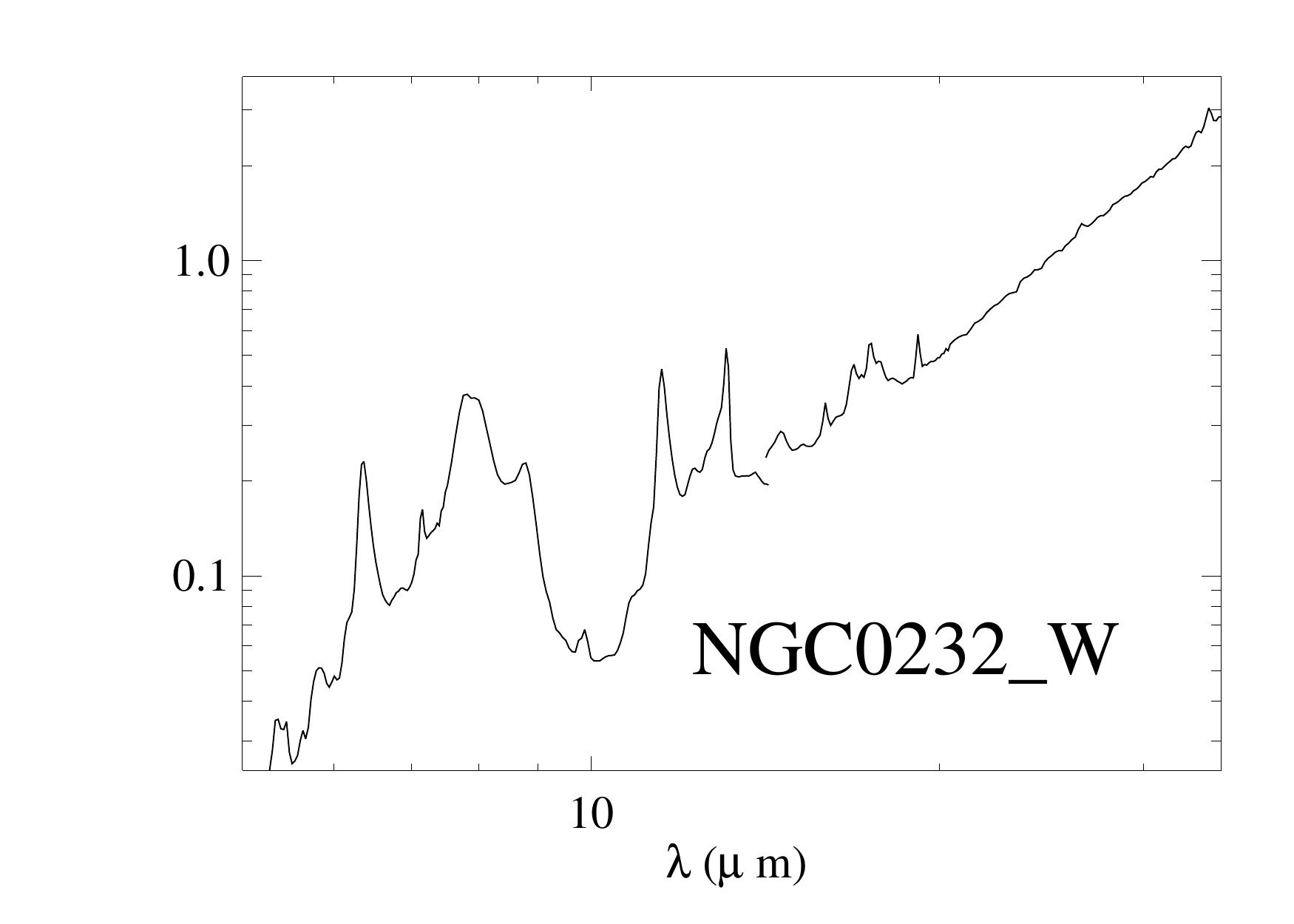}
\hspace{-1.0cm}
\includegraphics[height=1.85in,width=1.85in,clip,viewport=10 10 235 235]{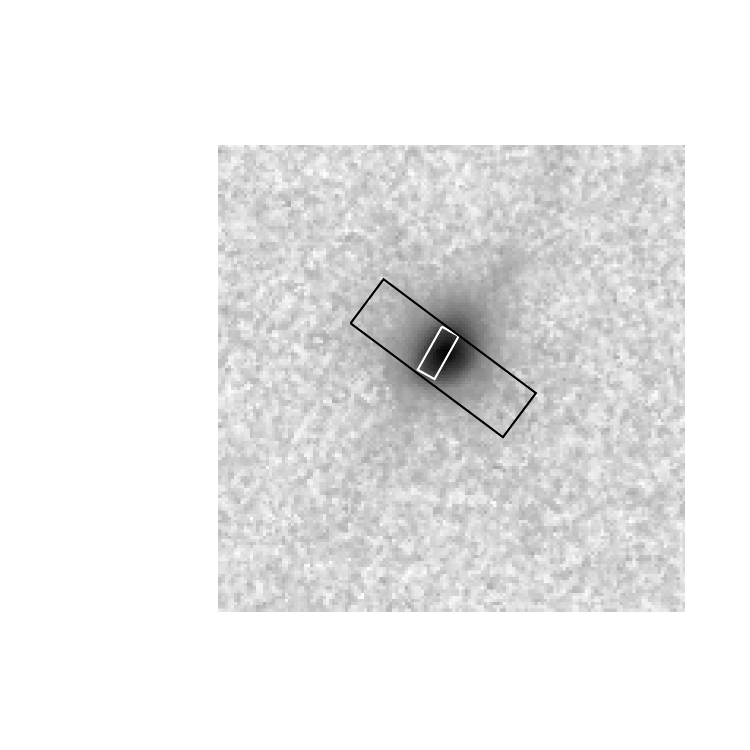}
\hspace{-1.8cm}
\includegraphics[width=3.3in,height=1.5in,clip]{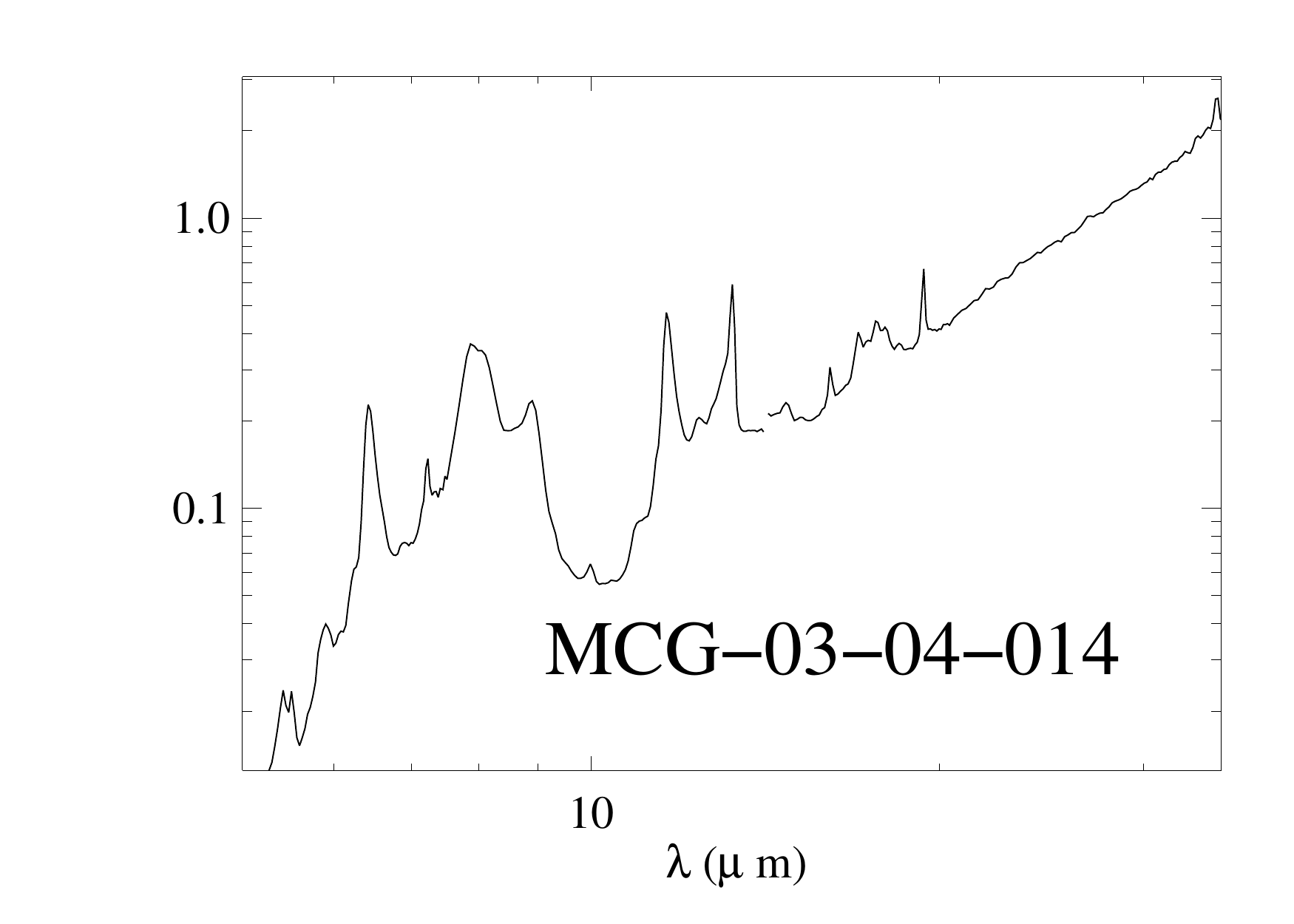}
\end{minipage}
\hfill 
\end{sideways}
\begin{sideways}
\begin{minipage}[t]{8.5in}
\centering 
\caption{IRAC 8\micron~(Channel 4) images with Spitzer IRS Short-Low (white)
  and Long-Low (black) extraction aperture projections
  and low resolution spectra for six GOALS galaxy nuclei. Each
  image is 45\arcsec~by 45\arcsec. Spectral flux densities are given
  in [Jy]. Similar images for the remainder of the sample are
  available as online
material and can also be found at {\it{http://goals.ipac.caltech.edu}}.\label{minifig}}
\end{minipage}
\end{sideways}
\end{figure*}

The IRAC 8\micron~(Channel 4) images for six example GOALS galaxy
nuclei are shown in Figure \ref{minifig} with the SL and LL extraction
aperture projections overlaid (in the case of staring mode data) or
with the CUBISM extraction windows overlaid (in the case of mapping mode data). The low resolution IRS spectrum for each source is also
presented along with each MIR image. Spectra for the remainder of the
galaxy nuclei, ordered by right ascension, are available as online
material and can also be found at {\it{http://goals.ipac.caltech.edu/}}. 
For five galaxies (IIIZw035, IRASF03359+1523, MCG+08-18-013,
IRASF17132+5313, and MCG-01-60-022), the archival SL staring mode observations
were not centered on the galaxy nucleus, so the SL slit
overlays are not shown and the extracted spectra were not used in our
analysis. Complete IRS observations were not available for an
additional 6 galaxies (no LL spectra: NGC2388, 
NGC4922, and VV705; no SL spectra: IRASF08339+6517; no
IRS data: ESO550-IG025 and IC4518). One galaxy \citep[NGC1068;][]{howell1068} saturates the spectrograph and so is also not shown. 

\subsection{Scale Factors}
For each spectrum a
break occurs between the SL and LL modules near 14\micron~due to the
larger LL slit, which covers nine times the area covered by the SL
slit. The scale factors required to match the SL flux to the LL
flux are not applied to the spectra in Figures \ref{minifig} and A1
but are calculated from the overlap in the SL1 and LL2 modules and
presented in Table \ref{bigtable}. Scale factors are not given for any source missing either SL or LL
data. For a small minority of cases (CGCG448-020, ESO077-IG014,
ESO173-G015, ESO255-IG007, ESO343-IG013, ESO440-IG058 (northern nuclei
only), IRAS03582+6012,
IRASF06076-2139, NGC5653, NGC6090, NGC3690 (western nuclei only), and NGC5256), the scale factor
is not given because the placement of the LL slit
covered multiple nuclei while the smaller SL slit covered only
one. 

The median scale factors are 1.22 and 1.70 for the staring and mapping mode data
respectively. The larger median scale factor for mapping data most likely
reflects a selection bias toward mapping more extended sources. Twelve
scale factors are $<$ 1 (i.e. more flux is recovered from SL than from
LL), but for all twelve, the scale factors are also $>$0.9 and thus
represent normal statistical scatter for sources with scale factors
near unity. No clear correlation is observed between the scale factors
and galaxy distance, but at distances $>$ 300 Mpc, a cutoff that includes 6 sources, the scale
factors are all $<$ 1.2. Similarly, at distances closer than 30 Mpc,
there are three GOALS sources that all have scale factors $>$ 1.6.

The scale factors are applied as a uniform multiplicative factor
across the entirety of the SL spectra and thus boost equally the PAH
fluxes, the continuum, and the absorption features. Since
calculations of the equivalent width of the 6.2\micron~PAH and the depth
of the silicate feature at 9.7\micron~(EQW$_{6.2\mu m}$ and $s_{9.7\mu
  m}$; see next section) both use measurements of feature flux relative to the continuum,
neither are affected by the scaling of the SL spectrum at these low redshifts. The MIR slopes
(F$_{\nu}$[30$\mu$m]/F$_{\nu}$[15$\mu$m]) are also unaffected as they only
rely on data from the (unscaled) LL portion of the spectrum.

\subsection{$s_{9.7\mu m}$, MIR Slope, \& EQW$_{6.2\mu m}$}
Silicate depths at 9.7\micron~($s_{9.7\mu m}$) were measured directly from
the MIR spectra via: $s_{\lambda} = log(f_{\lambda}/C_{\lambda})$ where
$f_{\lambda}$ is the measured flux at the central wavelength of the
absorption feature and $C_{\lambda}$ is where the level of the
continuum flux would be in the absence of the absorption feature,
based on an extrapolation to the surrounding continuum. Thus, a
positive value, $s_{\lambda} > 0$, suggests emission at that
wavelength and the deeper the absorption, the lower the s$_{9.7\mu
  m}$ value.

The fluxes
F$_{\nu}$ at 15\micron~and at 30\micron~were determined from the average
of eight data points surrounding each wavelength and were then used to
calculate the MIR slope. The wavelength regions used fell within
$\sim$14.7-15.4\micron~for F$_{\nu}$[15\micron] and
$\sim$29.5-30.8\micron~for F$_{\nu}$[30\micron].

Equivalent widths for the 6.2\micron~PAH feature (EQW$_{6.2\mu m}$) were measured
for each spectrum using the method outlined in
\cite{brandlEW}. 
Briefly, a spline fit was used to estimate the
continuum surrounding the 6.2\micron~PAH feature, and the continuum fit was
subtracted from the spectrum. In most cases, anchor points in determining the continuum were set at
5.15\micron~$< \lambda < $5.31\micron, 5.8\micron~$< \lambda < $
5.9\micron, 6.5\micron~$< \lambda < $ 6.8\micron, and 7.1\micron~$<
\lambda < $ 7.2\micron, but each spectrum was visually
inspected to make sure no features or bad points occurred in these
ranges. The PAH flux was then measured using direct integration. 
The 6.2\micron~feature was selected for the EQW calculation because, of the
five brightest PAH features, it is the least affected by silicate
absorption at 9.7\micron~and 18.5\micron, and it is not blended with other PAH features. 
However, in some cases the 6.2\micron~PAH feature partially overlaps with the
absorption feature due to water ice at 6.0\micron. For those sources
found by the spectral fitting to have $\tau_{ice} > $0 (see Stierwalt
et al. 2013b), the ice absorption was assumed
to affect the underlying continuum but not the PAH emission, and the EQW was
calculated accordingly. Four galaxies have only upper limits placed on
their EQW$_{6.2\mu m}$: IRAS05223+1908, MCG-03-34-064, NGC4418, and
IRAS08572+3015.

\subsection{Merger Stages}\label{mergsec}
Merger stages for the entire sample were determined via visual
inspection of the IRAC 3.6\micron~(Channel 1) images. Each galaxy was
assigned one of the following five designations: `N' for nonmergers (no
sign of merger activity or massive neighbors), `a' for pre-mergers
(galaxy pairs prior to a first encounter), `b' for early-stage mergers
(post-first encounter with galaxy disks still symmetric and in tact
but with signs of tidal tails), `c' for mid-stage mergers (showing amorphous
disks, tidal tails, and other signs of merger activity), or `d' for
late-stage mergers (two nuclei in a common envelope). Given the
resolution of the IRAC images ($\sim$2\arcsec), late stage mergers can be
easily mistaken for nonmergers in the 3.6-\micron\ images. To
alleviate this problem, any galaxies classified as nonmergers or early
stage mergers in
the IRAC images with available higher resolution imaging in the literature that clearly showed signs of a late stage major merger were changed accordingly. We also use the literature to identify spectroscopic pairs which resulted in reclassifying some nonmergers as pre-mergers.

For a subset of 78 GOALS galaxies (all with log(\lir/L$_{\odot}$) $>$11.4), we have additional merger
classifications based on available HST B, I,
and H-band images. The higher resolution of this imaging enables a
more detailed classification system with more finely tuned merger
stage designations (stages 0 through 6). These merger stages were
already described and presented in
\cite{haanHST}, but we reproduce and discuss them here to aid with
cross-referencing the two classification schemes.

\section{Mid-Infrared Properties of Nearby LIRGs}\label{results}

\subsection{LIRG vs ULIRG Distributions}
\begin{figure}[h!]
\begin{center}
\includegraphics[height=2.3in,width=3in]{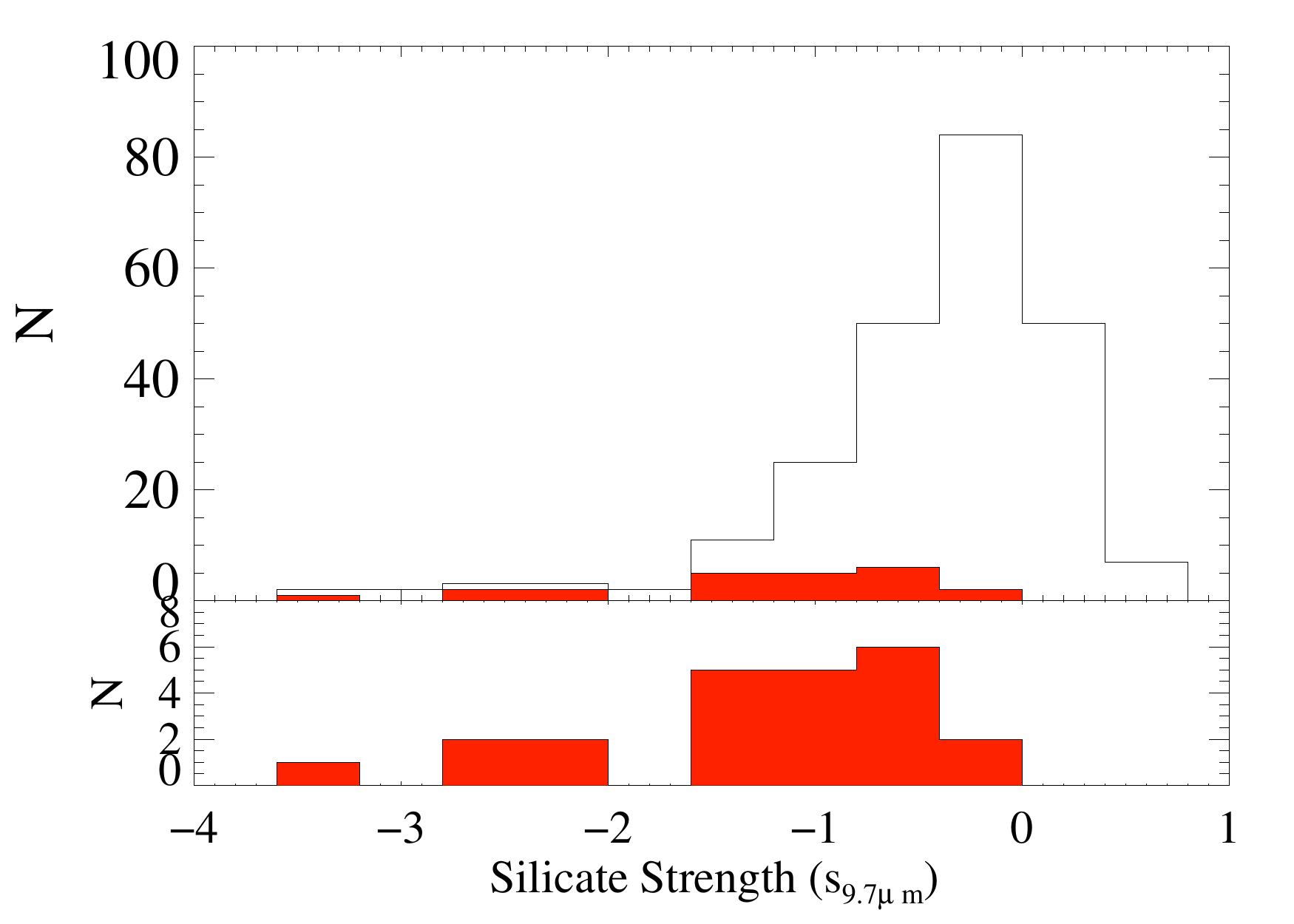}
\includegraphics[height=2.3in,width=3in]{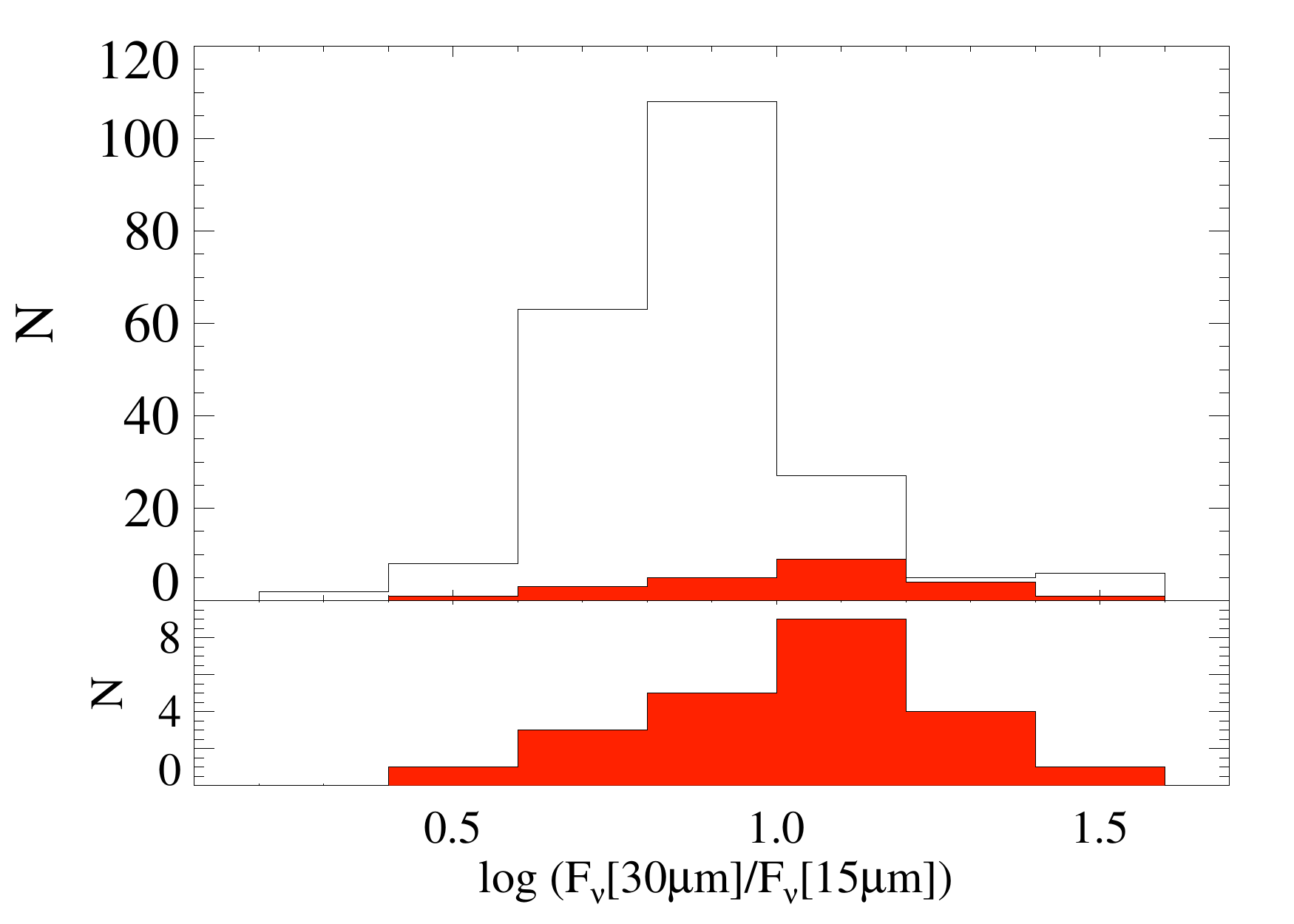}
\includegraphics[height=2.3in,width=3in]{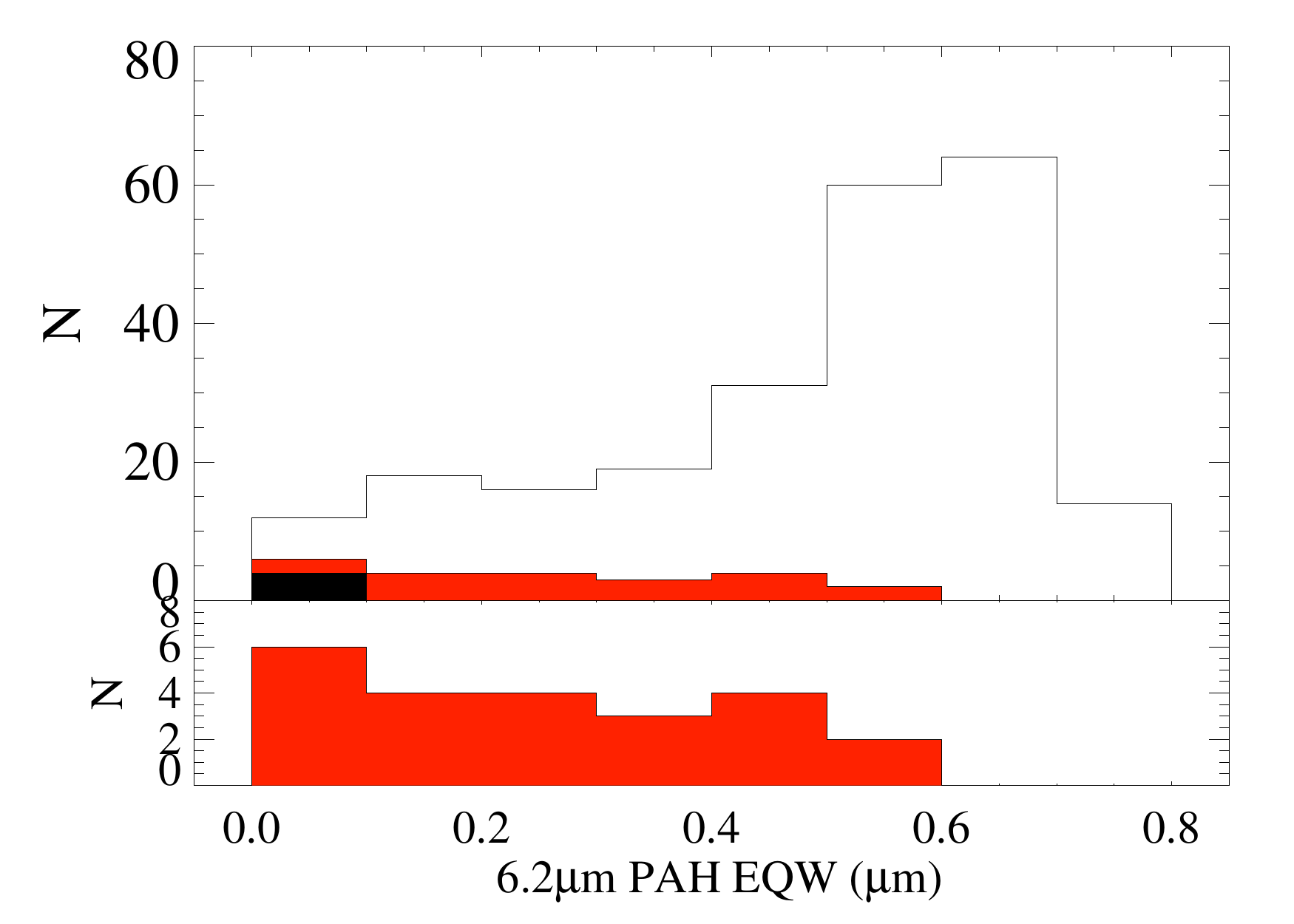}
\caption{Distributions of MIR spectral parameters (upper panels). Top: silicate 
  strength at 9.7\micron, Middle: logarithm of MIR slope, and Bottom: equivalent width of the
  6.2\micron~PAH feature. On average, GOALS ULIRGs
  (filled red histograms) have deeper silicate
  absorption depths, steeper MIR slopes, andlower equivalent widths than the GOALS sample as a
  whole (white histograms). The lower panel on each plot shows the
  same GOALS ULIRG distributions with a smaller y-scale. The filled solid black portion of the lowest bin of the EQW$_{6.2\mu m}$ histogram represents the four sources for which only upper
  limits are measured. 
\label{histplots}}
\end{center}
\end{figure}

Silicate depths, MIR slopes, PAH equivalent widths, and all associated uncertainties for the
GOALS sample, in addition to the SL-to-LL scale factors and merger stages, are presented in Table
\ref{bigtable}, and the distributions of \eqw,\sil, and MIR slope are shown in Figure
\ref{histplots}. The EQW$_{6.2\mu m}$ and $s_{9.7\mu m}$ parameters are not
given for the five sources with off-centered SL spectra, and MIR slopes
are not presented for the four sources without available either SL or
LL spectra or for the 12 sources for which multiple nuclei are observed
within the LL slit. 

As shown in Figure \ref{histplots}, the majority of LIRGs (63\%) are dominated by PAH emission (EQW$_{6.2\mu m} > $0.4\micron),
show little to no silicate absorption ($s_{9.7\mu m} > $ -1), and have MIR slopes of 4 $<$
F$_{\nu}$[30\micron]/F$_{\nu}$[15\micron] $<$ 10. Only six LIRGs have
deep silicate absorption with $s_{9.7\mu m} <$~-1.75 (NGC4418,
IRAS03582+6012\_E, ESO203-IG001, IRASF10038-3338, IRASF12224-0624, and ESO60-IG016). The remainder of the LIRGs show
 weak to no silicate absorption with a significant fraction (23\%) of LIRGs showing
silicates in emission at 9.7\micron, including 11\% with \sil\ $>$ 0.15. A few of the LIRGs with $s_{9.7\mu m} >$~0 are likely AGN-dominated (\eqw\ $<$ 0.27\micron) and thus any absorption at 9.7\micron\ is filled in by an excess of hot dust. However, most are lower luminosity galaxies with 90\% having log(L$_{IR}$/L$_{\odot}$)
$<$ 11.25. These LIRGs are likely analagous to the unobscured starburst NGC 7714, a galaxy whose IR emission is fueled almost entirely by star formation \citep{cafe}. The silicate
strengths in the LIRGs have a median of $s_{9.7\mu m}$ =
-0.25 $\pm$ 0.58
 and range
from the heavily obscured NGC4418 at  $s_{9.7\mu m}$ = -3.51 $\pm$
0.09 to
NGC5395, the southern component of the LIRG system Arp84, which shows
silicates in emission ($s_{9.7\mu m}$ = 0.52 $\pm$ 0.07). Five LIRGs are
continuum dominated and show at most only weak PAH or line features (EQW$_{6.2\mu m} < $~0.04\micron~and $s_{9.7\mu m}
> $~-0.2; MCG-03-34-064, IRAS05223+1908, NGC1275, NGC7674, and
AM0702-601\_N).

While the majority of LIRGs favor the high end of the distribution in both EQW$_{6.2\mu m}$ and $s_{9.7\mu m}$, they are found clustered in an
intermediate range of MIR slopes with a median of
F$_{\nu}$[30\micron]/F$_{\nu}$[15\micron] = 7.11$\pm$4.74. The MIR slopes
measured for the LIRGs range from
F$_{\nu}$[30\micron]/F$_{\nu}$[15\micron] = 2.00 $\pm$ 0.01 in
IRAS05223+1908 which shows a near power-law spectrum in the MIR to
F$_{\nu}$[30\micron]/F$_{\nu}$[15\micron] = 35.40 $\pm$ 1.38 in IRAS10173+0828.

For those LIRGs
with measurable 6.2\micron~PAH EQWs, the values range from
EQW$_{6.2\mu m}$ = 0.005\micron~$\pm$ 0.003\micron~for the northeastern component of
the LIRG pair IRAS03582+6012 to EQW$_{6.2\mu m}$ = 0.78\micron~$\pm$
0.01\micron~for the most southeastern of the three galaxies
composing the LIRG system IRAS17578-0400. 
The distribution for all of the GOALS LIRGs has a median of
EQW$_{6.2\mu m}$ = 0.55\micron~$\pm$0.18\micron. The same median value was
found for a sample of lower luminosity starbursting galaxies
\citep{brandlEW}.
Tight limits are placed on
the EQW for the three LIRGs and one ULIRG without a 6.2\micron~PAH
detection: IRAS05223+1908 at $<$0.043\micron, MCG-03-34-064 at
$<$0.044\micron, NGC4418 at $<$0.066\micron, and IRAS08572+3915 at $<$0.081\micron.

The GOALS ULIRGs, represented by the solid red histograms in Figure \ref{histplots}, show a clear offset from the LIRGs in their distributions for all three fundamental
properties. The ULIRGs
have a higher median flux density ratio
(F$_{\nu}$[30$\mu$m]/F$_{\nu}$[15$\mu$m] = 12.54$\pm$5.41), a lower
median PAH equivalent width (EQW$_{6.2\mu m}$ =
0.30\micron$\pm$0.17\micron), and deeper
median silicate
absorption ($s_{9.7\mu m}$ = -1.05 $\pm$ 0.85). The GOALS ULIRGs span nearly the
full range of MIR slopes covered by LIRGs but are not found with
EQW$_{6.2\mu m} >$0.52\micron~or with $s_{9.7\mu m} > $-0.15. 
Comparing the derived values for the 22 ULIRGs in GOALS with the
larger samples from \cite{spoon} (104 ULIRGs) and \cite{quest} (QUEST;
50 ULIRGs), we find that the
larger numbers of ULIRGs in these samples result in a larger spread in
MIR properties (i.e. 6.2\micron\ PAH EQWs up to 0.8\micron\ and
silicate depths up to 0.2; \cite{spoon}). However, the median values are
consistent with ULIRGs having lower EQW$_{6.2\mu m}$, deeper
$s_{9.7\mu m}$, and steeper MIR slope than LIRGs: median EQW$_{6.2\mu m}$ = 0.15\micron\ \& $s_{9.7\mu m}$ = -1.47
\citep{spoon} and median F$_{\nu}$[30$\mu$m]/F$_{\nu}$[15$\mu$m] = 11.6 \citep{quest}.

The results of a Kolmogorov-Smirnov (KS) test give probabilities of $<$0.01\% that the chance deviations between the distributions of \eqw, \sil, and MIR slope for GOALS LIRGs vs ULIRGs are expected to be larger assuming they are derived from the same parent sample. In other
words, the two samples are significantly different. These probabilities decrease by several
orders of magnitude when the QUEST and \cite{spoon} ULIRGs are included. When the GOALS
ULIRGs are compared to the larger ULIRG samples, the KS test
suggests the chance deviations in their
distributions in MIR slope, \eqw, and \sil\ are expected to be larger with
probabilities of 80\%, 40\%, and 30\%, i.e. it is likely
the GOALS ULIRGs and the \cite{spoon} \& \cite{quest} samples are derived from the same parent sample. 

\subsection{Correlations with L$_{IR}$}%

\begin{figure}[h!]
\begin{center}
\includegraphics[height=2.5in,width=3.2in]{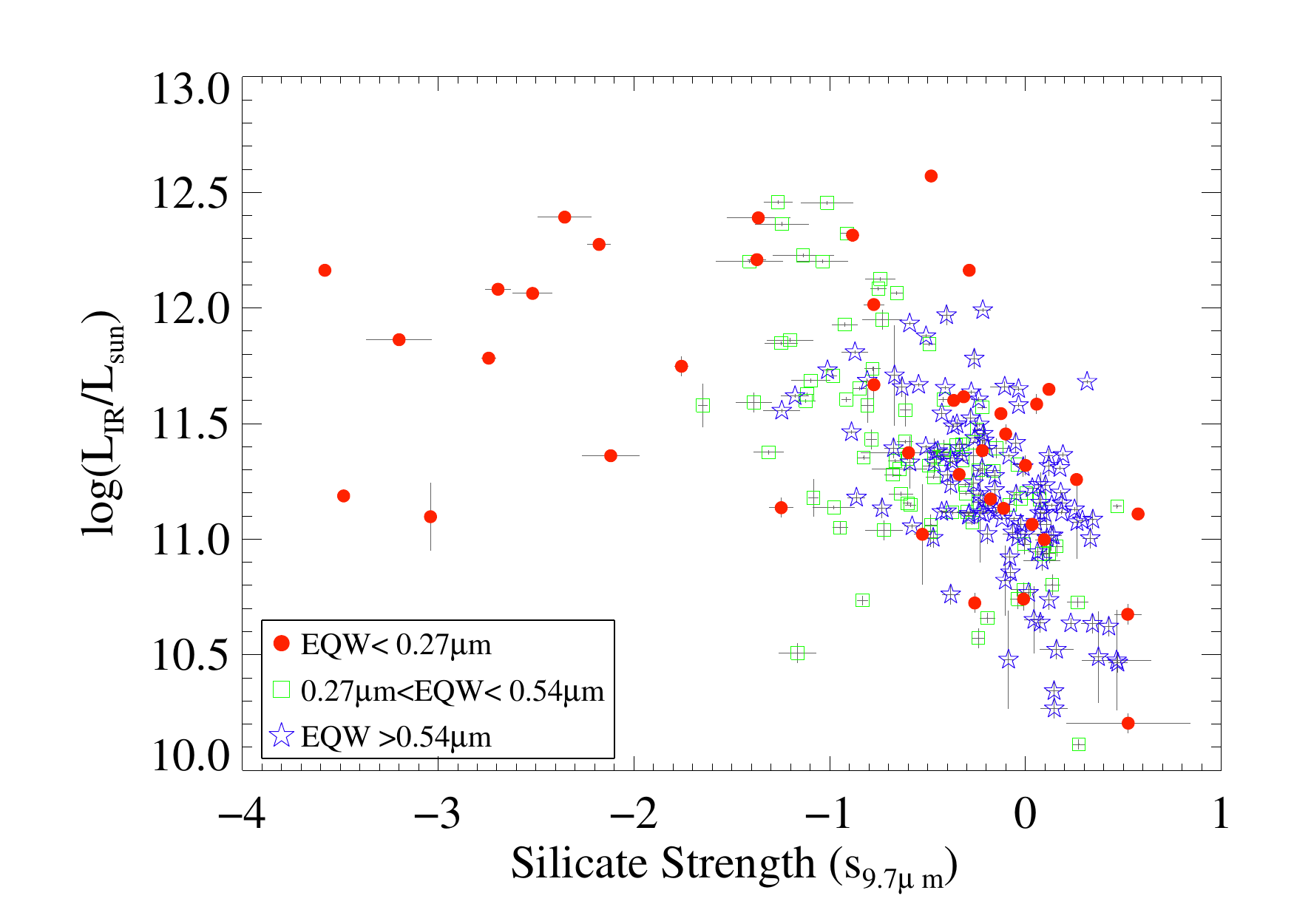}
\includegraphics[height=2.5in,width=3.2in]{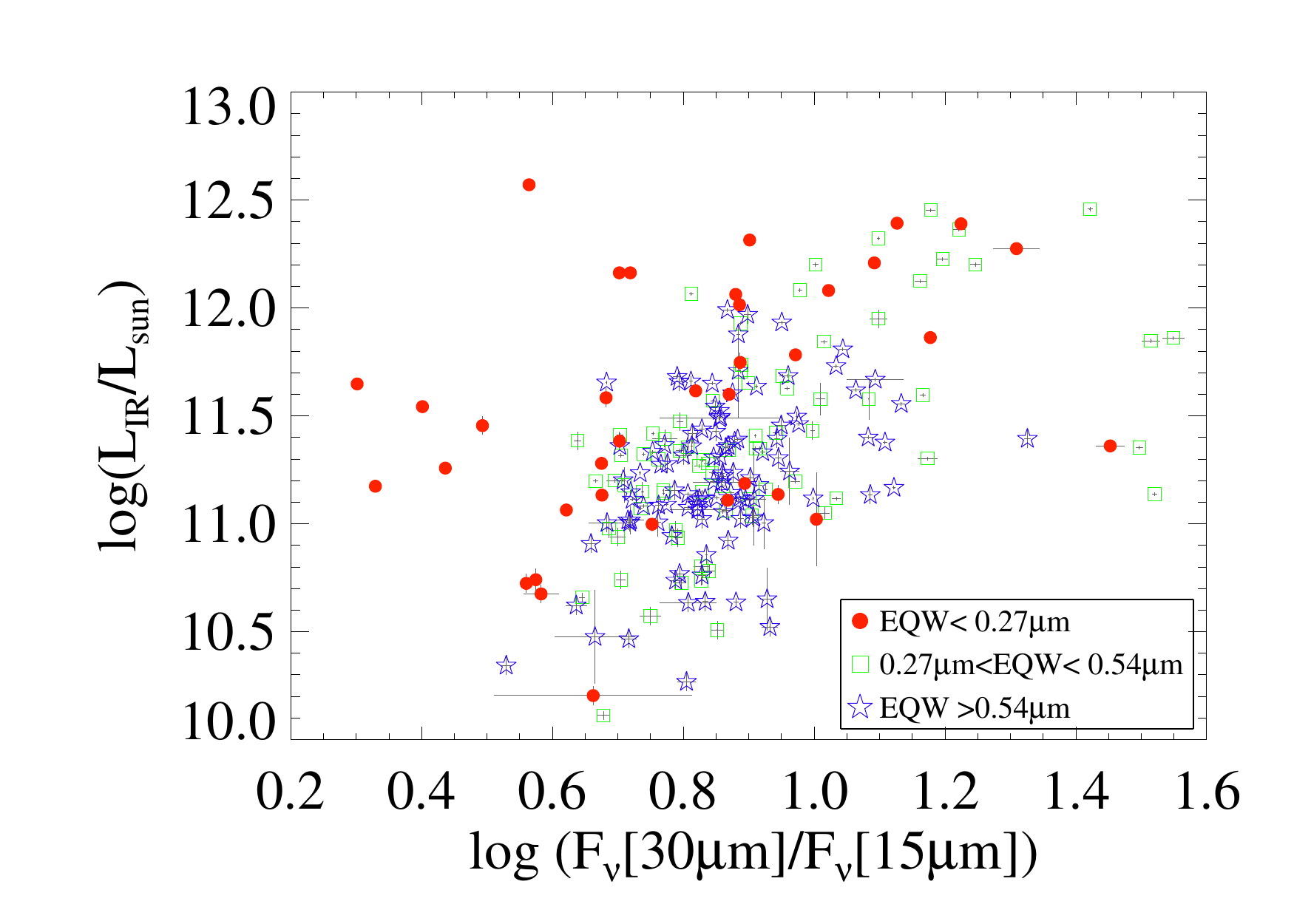}
\includegraphics[height=2.5in,width=3.2in]{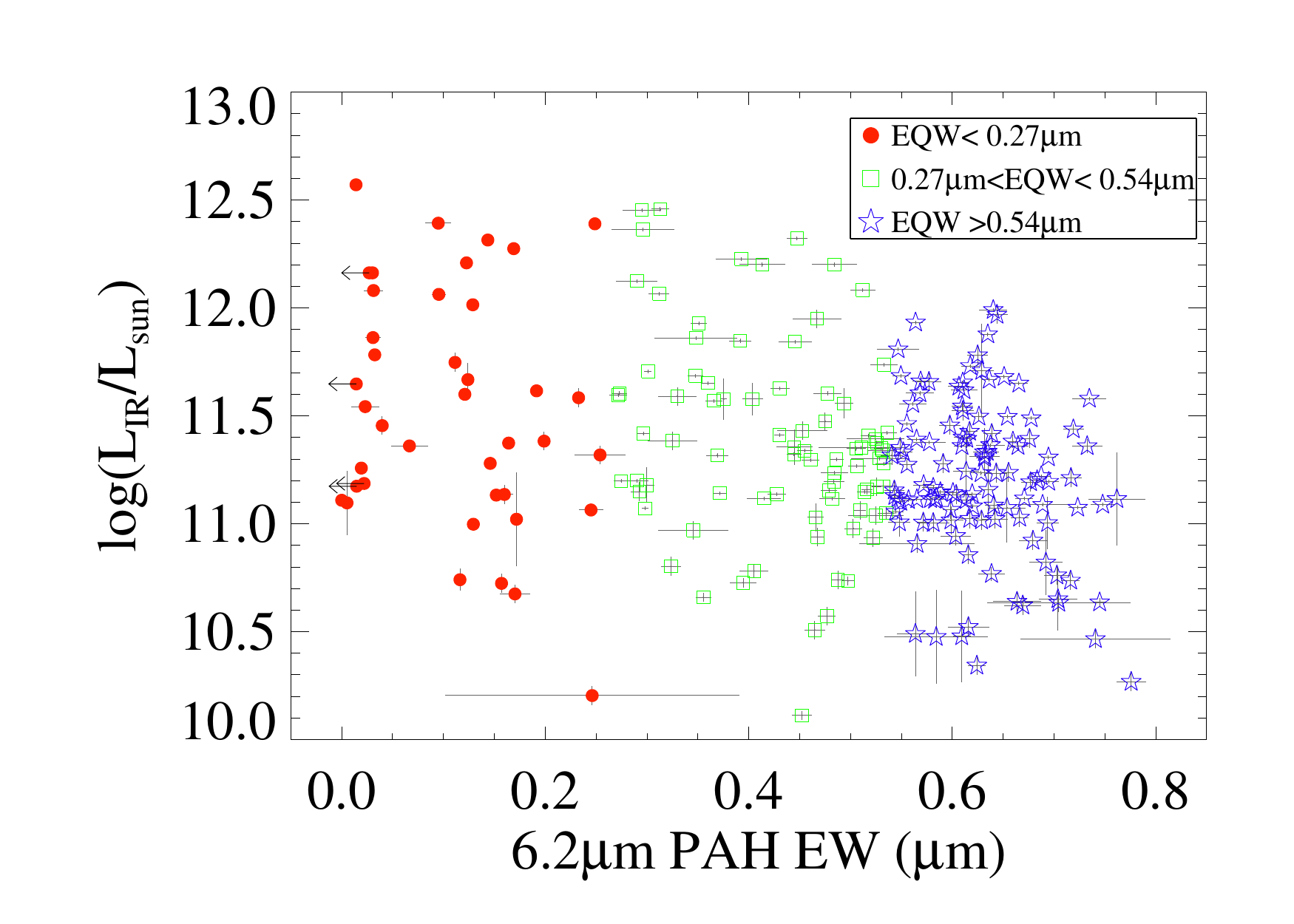}
\caption{Distribution of MIR spectral parameters with L$_{IR}$ color-coded by EQW$_{6.2\mu m}$. Top: silicate 
  strength at 9.7\micron, Middle: logarithm of MIR slope, and Bottom: equivalent width of the
  6.2\micron~PAH feature. There is
  a loose trend among LIRGs for increasing silicate
  depth and MIR slope with increasing L$_{IR}$. However, LIRGs span nearly the full range of EQW$_{6.2\mu m}$ at
any given luminosity. 
\label{LIRplots}}
\end{center}
\end{figure}

Figure \ref{LIRplots} shows the distributions of $s_{9.7\mu m}$, MIR slope, and 
EQW$_{6.2\mu m}$ as a function of IR luminosity,
L$_{IR}$. The IR luminosities for all 202 U/LIRG systems were presented in
\cite{GOALS} and derived using the definitions of \cite{sandersLIRGs}\footnote{$L_{IR}/L_{\odot} = 4\pi(D_L[m])^2 (F_{IR}[W
m^{-2}])/3.826\times10^{26}[W m^{-2}]$ and $F_{IR} = 1.8\times
10^{-14}(13.48f_{12\mu m} + 5.16f_{25\mu m} + 2.58f_{60\mu m} +
  f_{100\mu m}[W m^{-2}])$}. In cases of multiple nuclei, the total L$_{IR}$ for the system is
divided according to the ratio of the fluxes at 70\micron\ for each
nuclei. In a small number of cases, 70\micron\ images are not available
and so 24\micron\ flux ratios are used instead.

There is a general trend among the U/LIRGs for both silicate
depth and MIR slope to increase with increasing \lir. The sources that
depart from these correlations at deep levels of silicate obscuration
(top panel) or shallow MIR slopes (middle panel) have, in both
cases, very low PAH equivalent width (\eqw\ $<$ 0.27\micron) and are thus
likely dominated by emission from an AGN. Increasingly luminous systems become increasingly dust obscured until
 a turnover occurs at \sil$\sim$ -1.5, above which the buried AGN
 candidates show no further correlation between \sil\ and \lir. As \lir\
 decreases, the MIR slopes flatten until \mirslope$\lesssim$ 0.5,
 below which the relatively unobscured AGN have high \lir\ given their
 slopes.
ULIRGs have an average \eqw\ that is lower than that for LIRGs, but sources with a large range of luminosities are found at each equivalent width (lower panel) so there is not a tight correlation between \lir\ and \eqw.

\subsection{Disentangling $s_{9.7\mu m}$, MIR Slope, \& EQW$_{6.2\mu m}$}

\begin{figure*}[h!]
\begin{center}
\includegraphics[height=3.5in,width=7in,viewport=0 0 500 225,clip]{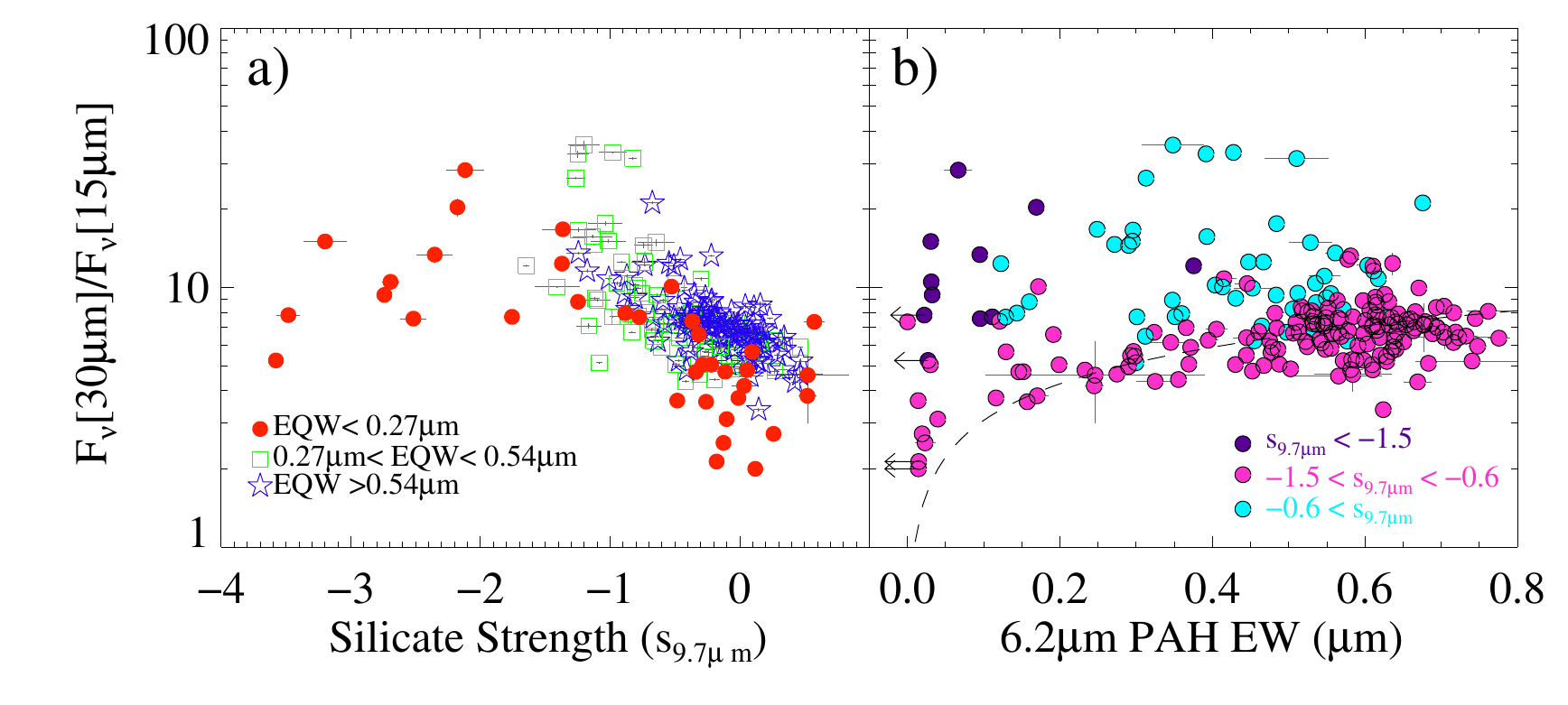}
\caption{Distribution of MIR slope
  (F$_{\nu}$[30\micron]/F$_{\nu}$[15\micron]) versus a) silicate
  absorption at 9.7\micron~color-coded by EQW$_{6.2\mu m}$ and b)
  EQW$_{6.2\mu m}$ color-coded by $s_{9.7\mu m}$. GOALS sources with
  EQW$_{6.2\mu m} > $0.27\micron~(green squares + blue stars in
  panel a, right side of panel b) show a rough correlation between
  increasing silicate depth and increasing MIR slope (a) and follow the
  correlation between EQW$_{6.2\mu m}$ and MIR slope observed in the
  24-\micron~selected 5MUSES sample (dashed line, b; \cite{5muses}). At low EQW
  (EQW$_{6.2\mu m} < $0.27\micron; red circles in panel a, left side
  of panel b), relatively unobscured AGN-dominated sources all have MIR slopes
  below $\sim$4. However, at the deepest levels of silicate
  absorption (left side of panel a,
  purple circles in panel b), the MIR slope is no longer a clear
  indicator of temperature and so the heavily obscured sources do not
  follow the trend in panel a, and the location of the 15-\micron~continuum between the 9.7\micron~and
  18.5\micron~absorption features lead to elevated MIR slopes in panel
  b. Although far less
numerous (only 18\% of GOALS nuclei have EQW$_{6.2\mu m} <$ 0.27\micron), the lowest equivalent width sources cover a wider range of
\lir, MIR slope, and $s_{9.7\mu m}$ than those sources of higher EQW$_{6.2\mu m}$ that
make up the majority of the sample.
\label{dotplots}}
\end{center}
\end{figure*}

To further disentangle the relationship between the 3 main MIR parameters, we
examine the $s_{9.7\mu m}$ and EQW$_{6.2\mu m}$ versus MIR slope
parameter spaces in Figure \ref{dotplots}. The distribution of \sil\
with MIR slope is color-coded by \eqw\ (panel $a$) while the
distribution of \eqw\ with MIR slope is color-coded by \sil\ (panel $b$). 

In the majority of LIRGs, star formation dominates the intense MIR
emission and the similar conditions in the photodissociation regions
resulting in this (PAH-dominated) emission lead to similar MIR
properties among the bulk of the GOALS sample. In Figure
\ref{dotplots}a, the majority of GOALS galaxies, those with \eqw\ $>$
0.27\micron\ (green squares and blue stars), show a rough correlation  between increasing MIR slope and increasing silicate depth (lower
$s_{9.7\mu m}$). The starburst galaxies of \cite{brandlEW} which span
mostly luminosities below 10$^{11}$L$_{\odot}$ exhibit a similar relationship between $\tau_{9.7\mu m}$ and
F$_{\nu}$[30$\mu$m]/F$_{\nu}$[15$\mu$m] with the same slope.
The trend in Figure \ref{dotplots}a suggests that the average dust temperature rises as a consequence of the nuclei becoming more obscured and
compact. As the dust temperature increases, the rising portion of the blackbody emission spectrum shifts to shorter wavelengths, and warmer sources have
increasingly more flux at 30\micron~as seen for the GOALS U/LIRGs with $s_{9.7\mu m} > $-1.5.

Most of the sources with low PAH equivalent width, however, do not follow these simple trends in MIR properties. In Figure \ref{dotplots}a, these low-EQW sources (red
circles) are
split roughly into two populations: those that are relatively unobscured with shallow
MIR slopes and those heavily obscured sources ($s_{9.7\mu
  m}$$<$-1.5) with steep MIR slopes. A similar split is observed in Figure \ref{dotplots}b: for \eqw\ $<$0.27\micron, the heavily
obscured sources (purple circles) are found at steeper MIR slopes
while the relatively unobscured sources (magenta circles) are found at
the shallowest flux density ratios.  

An increasingly significant hot dust component from an AGN leads
both to a decrease in EQW$_{6.2\mu m}$ and to a flatter MIR slope. 
For the GOALS sources with the strongest, relatively unobscured AGN
(\sil$\gtrsim$-0.6; \eqw$\lesssim$0.05\micron), an upper limit to the
MIR slope can be set at F$_{\nu}$[30$\mu$m]/F$_{\nu}$[15$\mu$m]$<$4
from both panels in Figure \ref{dotplots}. These galaxies (including
IRAS05223+1908 and UGC08058) are represented by the red circles in the
lower right corner of panel $a$ and the magenta circles in the lower
left corner of panel $b$. No other sources 
are found with flatter MIR slopes. 
This limit agrees with that
observed for the starbursts of \cite{brandlEW} 
and for the QUEST ULIRGs of \cite{quest}. 
Relatively unobscured AGN can thus be identified based on their low MIR flux density ratio alone. 


In the most heavily obscured, low EQW galaxies, however, the MIR continuum slopes
are steeper due to the buried, hot source. These galaxies (red
circles in the left half of Figure \ref{dotplots}a and purple circles
in Figure \ref{dotplots}b) have steep MIR slopes for the same reason
sources with \sil$\sim$-1.5 in Figure \ref{dotplots}a have steep MIR
slopes: most of the warm dust emission is hidden behind a large amount of cooler dust.
A comparison to the 5 mJy Unbiased Spitzer Extragalactic Survey \citep[5MUSES;][]{5muses} highlights the difference between the low and high \sil\ sources at low \eqw. The 5MUSES sample is 24-\micron\ selected (indicating the presence of hot dust) but lacks the heavily obscured sources found in GOALS. The distributions for the two samples in Figure \ref{dotplots}b are roughly the same (5MUSES is represented by the dashed line although there is significant scatter about this line; see \cite{5muses}) - both show the lowest \eqw\ sources have the shallowest MIR slopes - except 5MUSES lacks the obscured low \eqw\ galaxies (purple circles in Figure \ref{dotplots}b). 

\begin{figure*}[htp!]
\begin{center}
\includegraphics[height=3.5in,width=4.5in]{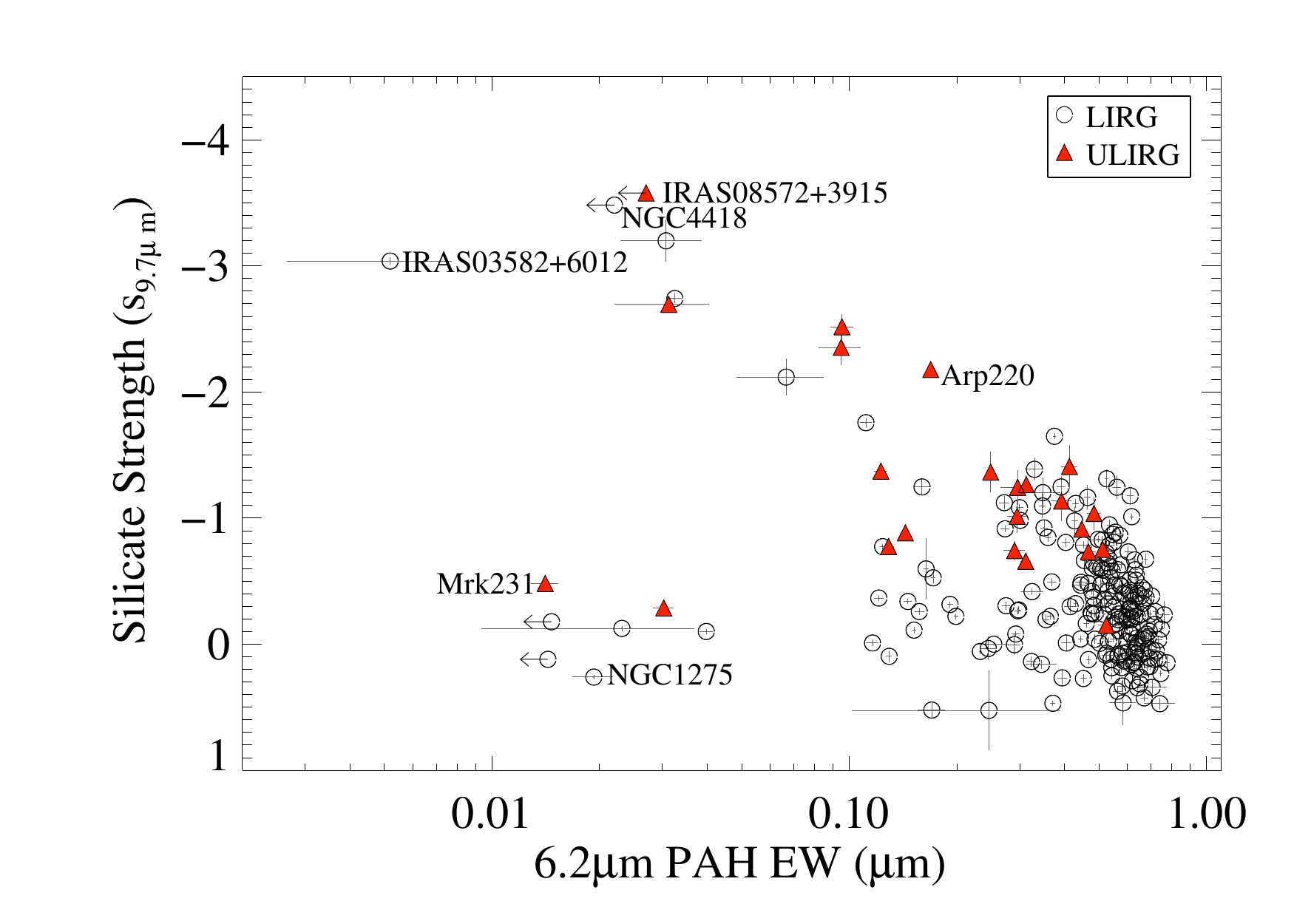}
\caption{Equivalent width of the 6.2\micron~PAH versus silicate
  strength for the LIRGs (open circles) and ULIRGs (red triangles) of the GOALS
  sample. The majority ($>$60\%) of the LIRGs are found at EQW$_{6.2\mu m}
  >$ 0.4$\mu$m and $s_{9.7\mu m}$ $>$ -1.0, while nearly all ULIRGs have EQW$_{6.2\mu m}
  <$ 0.5$\mu$m and $s_{9.7\mu m}$ $<$ -0.5. Sources at low EQW are divided
  into two distinct branches (silicate-dominated versus continuum-dominated).
\label{spoonplot}}
\end{center}
\end{figure*}

The apparent strength of the 9.7\micron~silicate feature (i.e. the depth of the absorption feature that does not account for any silicate emission, $s_{9.7\mu m}$) is shown
versus EQW$_{6.2\mu m}$ in Figure \ref{spoonplot} for GOALS LIRGs
(open circles)
and ULIRGs (red triangles). 
No galaxies are observed with both high equivalent widths $and$ large
levels of silicate absorption. However, at low equivalent widths, two
distinct branches, similar to those seen by \cite{spoon}, emerge that clearly
distinguish the lower equivalent width (EQW$_{6.2\mu m} <$ 0.1\micron)
sources with minimal to no silicate absorption ($s_{9.7\mu m}$ $>$ -0.5) from
those dominated by silicate absorption (heavily obscured sources; $s_{9.7\mu m}$ $<$-1.75). Sources with
intermediate levels of silicate absorption are not found at low equivalent
widths.

As shown in Figure \ref{spoonplot}, the highly absorbed sources are
not limited to ULIRGs. At values of $s_{9.7\mu m}$ $<$ -1.75, the
GOALS sample includes five ULIRGs (labeled in Figure \ref{spoonplot})
as well as the dense, compact nascent starburst LIRG NGC4418
\citep{spoon4418, nascent, evans4418} and five additional LIRGs that span a large range of LIRG luminosities: IRASF12224-0624 (log(L$_{IR}$/L$_{\odot}$) = 11.36), IRAS03582+6012 (log(L$_{IR}$/L$_{\odot}$) = 11.42), IRASF10038-3338 (log(L$_{IR}$/L$_{\odot}$) = 11.78), ESO60-IG016 (log(L$_{IR}$/L$_{\odot}$) = 11.82), and ESO203-IG001 (log(L$_{IR}$/L$_{\odot}$) = 11.86). 

\subsection{Compactness}\label{obscured}

\begin{figure}[htp!]
\includegraphics[height=2.9in,width=4in,viewport=20 0 590 345,clip]{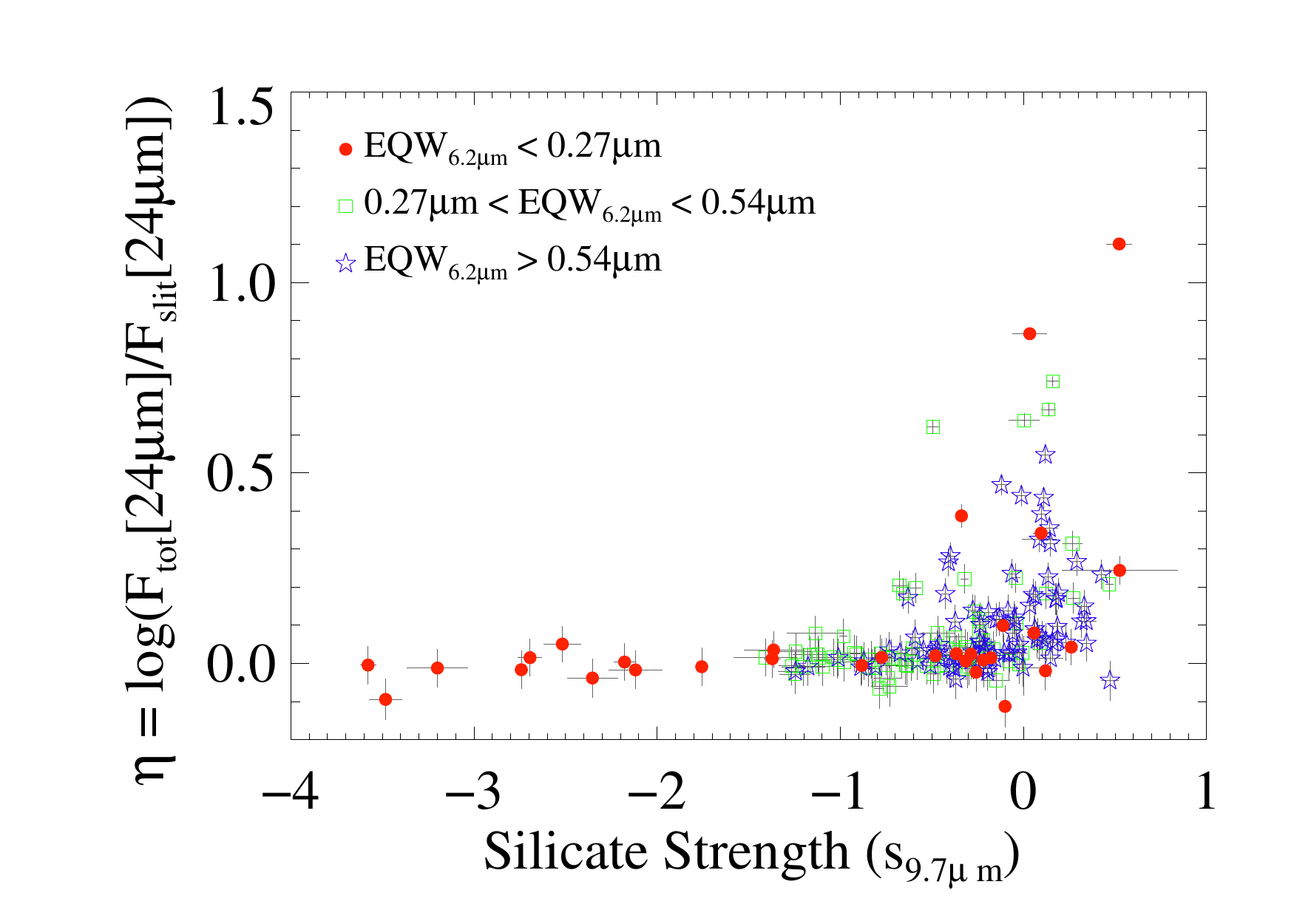}
\caption{Silicate strength versus $\eta$, a measure of the total-to-slit
  flux ratio at
  24\micron. GOALS
  LIRGs and ULIRGs are color coded by 6.2\micron~PAH equivalent width
  with low equivalent width (AGN-dominated) sources (EQW$_{6.2\mu m} <$ 0.27\micron)
  represented by red circles and high equivalent width
  (starburst-dominated) sources (EQW$_{6.2\mu
  m} >$ 0.54\micron) represented by blue stars. Intermediate EQW
(composite) sources
are shown by orange (0.27\micron~$<$ EQW$_{6.2\mu m} <$ 0.41\micron)
and green (0.41\micron~$<$ EQW$_{6.2\mu m} <$ 0.54\micron)
squares. Heavily obscures sources have no extended component to their
24\micron~emission ($\eta\sim$0).
\label{etafig}}
\end{figure}

The most heavily obscured nuclei among the GOALS galaxies are also the most compact: they all show little-to-no MIR emission extending outside of the IRS slit.
In Figure \ref{etafig}, the silicate strength is plotted
against $\eta$, a parameter that represents the fraction of the emission at 24\micron~intercepted by the IRS slit:
\begin{equation}
\eta = log (F_{tot}^{MIPS}[24\mu m] / F_{slit}^{IRS}[24\mu m])\label{eqeta}
\end{equation}
\noindent where $F_{tot}^{MIPS}[24\mu m]$ is the total flux of a
source as measured from its MIPS 24\micron~image (Mazzarella et al.,
$in~prep$) and
$F_{slit}^{IRS}[24\mu m]$ is the flux within the IRS slit derived by
convolving the MIPS-24\micron~filter with the low resolution IRS
spectrum. The most obscured sources ($s_{9.7\mu m}$ $<$ -1.75) all have
$\eta\sim$ 0 (i.e. all of the flux measured by the larger MIPS field of
view at 24\micron~is also recovered within the much smaller IRS slit).

Although distance effects could act to disguise an extended component
in comparisons of total versus intra-slit fluxes if all of the
obscured sources were the most distant, the median distance for the
heavily obscured, low $\eta$ nuclei is only 60\% larger than the
median distance for the less obscured sources (190 Mpc vs 115
Mpc) suggesting distance alone cannot be driving the difference in
$\eta$. Even more importantly, the heavily obscured, low $\eta$
tail of the distribution in Figure \ref{etafig} includes the two closest, obscured LIRGs, NGC4418 at 36.5
Mpc and $s_{9.7\mu m} =$-3.51 $\pm$ 0.09 and NGC3690 at 50.7 Mpc and
$s_{9.7\mu m} =$-1.65 $\pm$ 0.02. Both of these LIRGs would have been easily
resolved had they shown any extended MIR emission.
Additionally, the existence of galaxies with high $\eta$ and low \sil\ with distances
well above 100
Mpc indicate that extended sources can still be resolved even at larger distances.

The fraction of resolved emission $within$
the IRS slit is also much lower for the obscured sources. The fraction
of extended emission (FEE$_{13.2\mu m}$) is defined by
\cite{taniopI} as the fraction of emission within the IRS slit
originating outside of the unresolved component (i.e. a point source
at that distance). For the most obscured GOALS nuclei, the average
$\langle$FEE$_{13.2\mu m}\rangle =$ 0.07 compared to
$\langle$FEE$_{13.2\mu m}\rangle =$ 0.39 for the remaining (weakly
obscured or unobscured) LIRGs. As discussed in detail in
\cite{taniopI}, such a dramatic difference in FEE
between obscured and unobscured nuclei cannot be the result of
distance effects alone.

Together the low $\eta$, the low FEE$_{13.2\mu m}$, and their
inclusion of nearby sources suggest the nuclei
in these heavily obscured sources dominate the 24-\micron~emission
from their parent galaxies, and so the most heavily obscured LIRGs and
ULIRGs also have the most compact MIR continuum emission. Given their low
EQW$_{6.2\mu m}$, if extreme levels of obscuration are not simply
masking the PAH emission, the higher densities in these nuclei may create an
environment where PAH dust grains are not present or the conditions
are not appropriate for exciting them (i.e. lacking in
photodissociation regions). Alternatively, the low EQW$_{6.2\mu m}$
may indicate an increase in the continuum flux at 6\micron~due to a
rise in dust temperature. 
None of the low $\eta$, high \sil\ nuclei are observed to be [NeV]
emitters \citep{petric}, most likely because their large optical depths
obscure any line emission that would be present at 14.3\micron.

\section{Tracing MIR Properties through Merger Stage}\label{mergersection}
\begin{figure}[htp]
\begin{center}
\includegraphics[height=2.1in,width=3in]{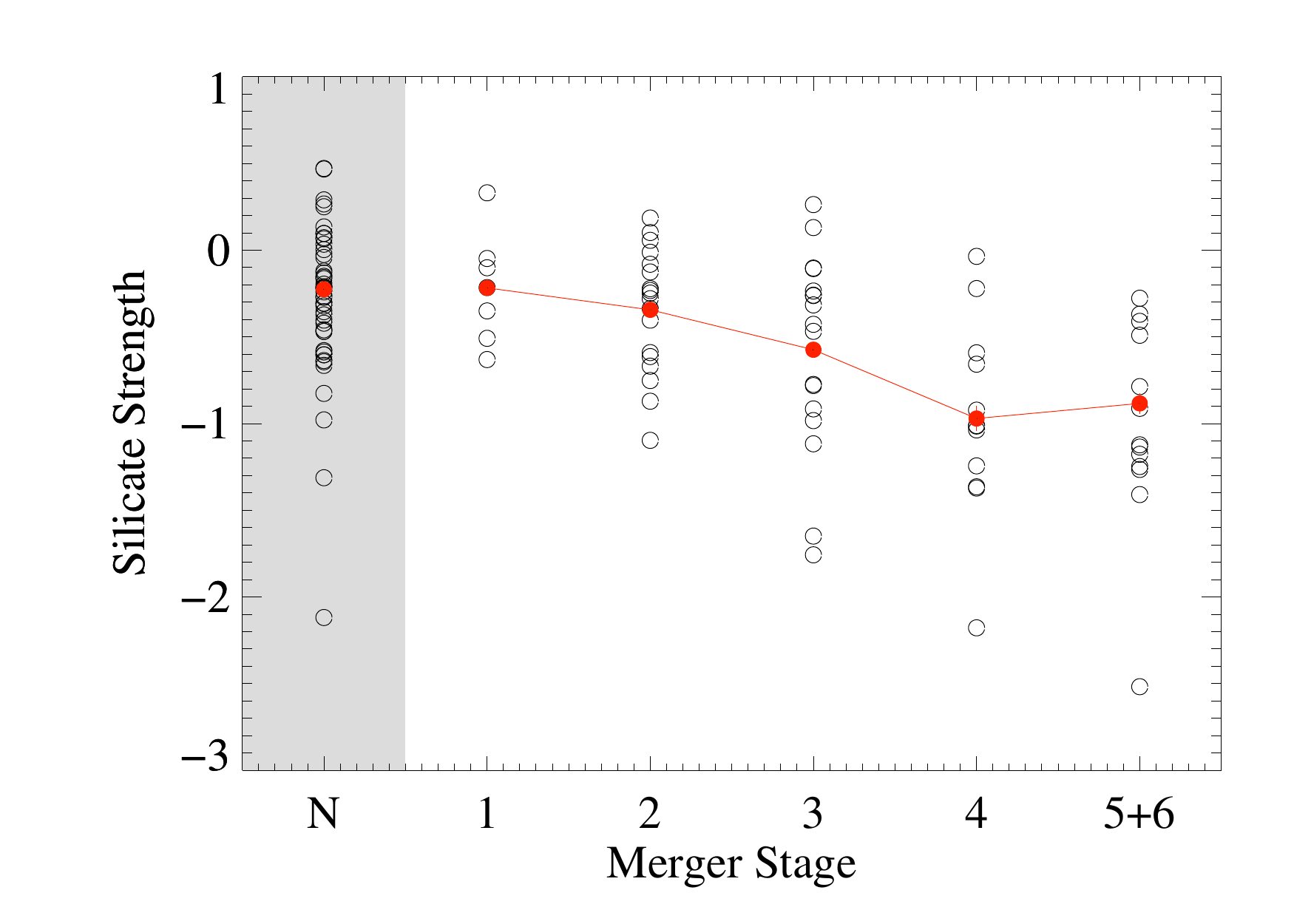}
\includegraphics[height=2.1in,width=3in]{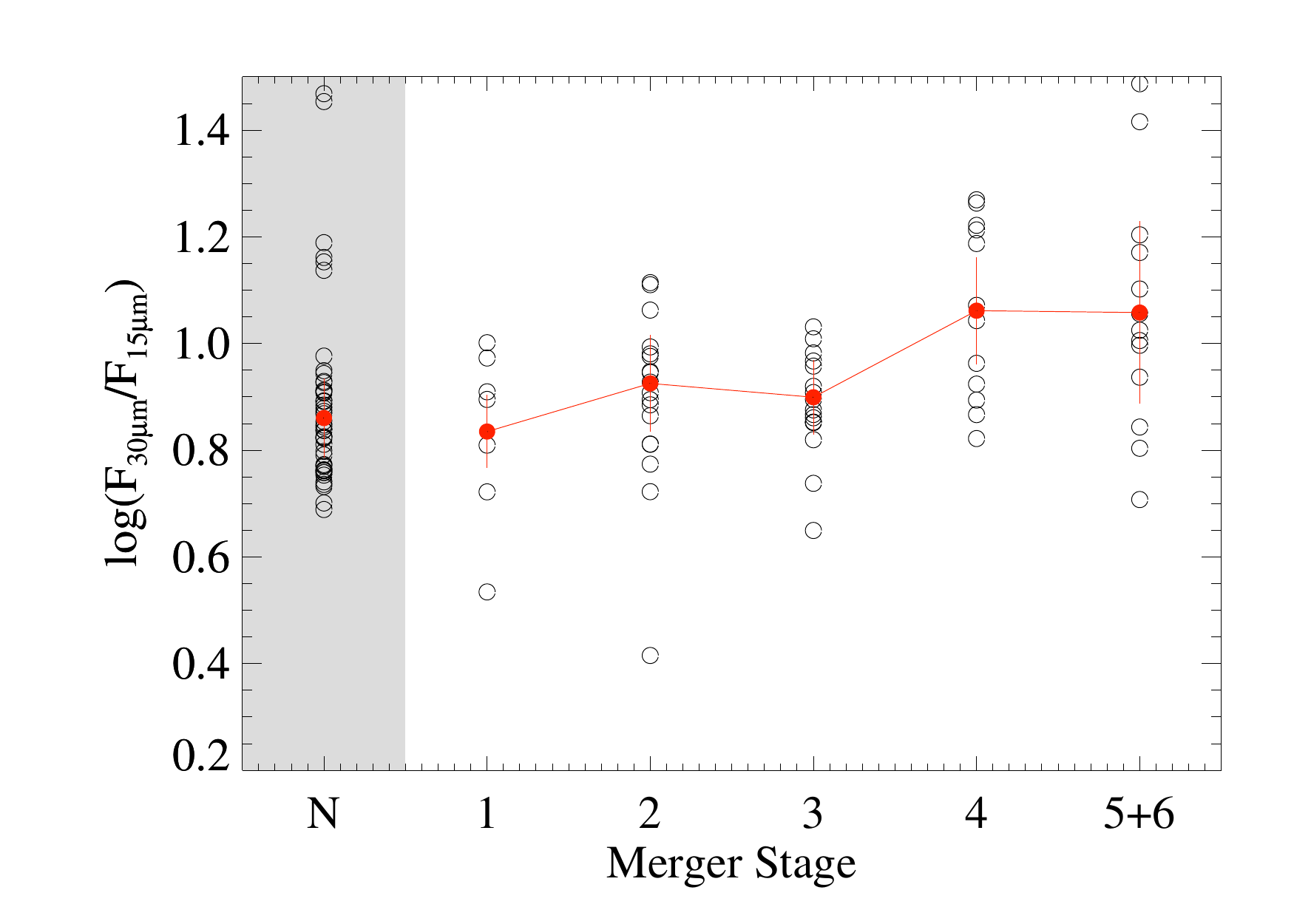}
\includegraphics[height=2.1in,width=3in]{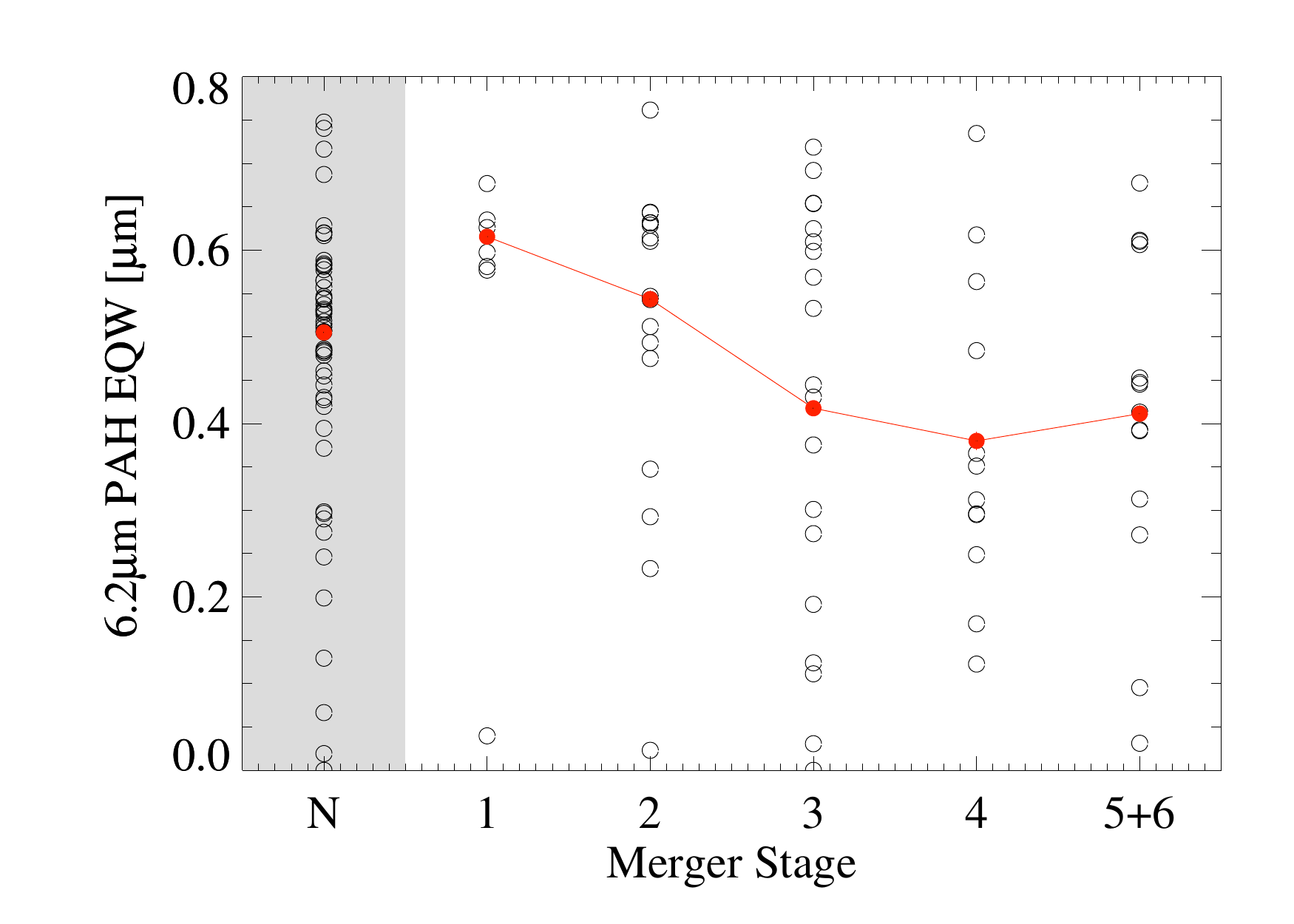}
\caption{MIR properties of GOALS galaxies traced through merger stage. Top: silicate strength ($s_{9.7\mu m}$), Middle: MIR slope (log(F$_{\nu}$[30\micron]/F$_{\nu}$[15\micron]), and Bottom: \eqw. Mergers (Stages $1$-$6$) are represented by the 78 GOALS galaxies for which high resolution HST imaging is available \citep[][; see Column (11) in Table \ref{bigtable}]{haanHST}. Nonmergers (Stage $N$) are classified using IRAC 3.6\micron~images and the literature (see Section \ref{mergsec} for details). Mean values for each merger stage clipped at
  3$\sigma$ are shown in red with their associated standard deviations.
\label{mstages}}
\end{center}
\end{figure}

In Figure \ref{mstages}, silicate strength, MIR slope, and PAH
equivalent width are traced through
merger stage for GOALS galaxies. To look for subtle differences in
MIR properties throughout the merging process, we focus on only those
sources that have HST classifications (column 11 in Table
\ref{bigtable}). Since this subset contains only six galaxies with no
indication of a merger (stage $0$), we include all of the nonmergers (Stage $N$)
from the IRAC-based classifications (column 10 in Table \ref{bigtable}) to derive a more secure median for each spectral property. Although the HST data samples only LIRGs with log(\lir/L$_{\odot}$)$>$11.4, the dense sampling of merger stages made possible by the deep, high spatial resolution optical and NIR images provides a much finer look at the spectral changes along the merger sequence. Mean values for each merger stage clipped at
3$\sigma$ are shown in red with their associated standard deviations. 

As the mergers progress and gas \& dust is funneled towards the center, galaxies become on average more obscured with
steeper MIR slopes. Silicate depths of \sil$\lesssim$ -1 are only
reached at merger stages of $3$ and later. No LIRG systems in merger
stage $1$ have \mirslope$>$ 1, while the average MIR slope is $>$1 for
the later stages $4$-$6$. These two results agree with several studies
finding higher \lir\ at later merger stages since, as shown in Figure \ref{LIRplots},
increasing MIR slope and silicate depth are also linked to higher
\lir\ in LIRGs. As merging galaxies coalesce, the
nuclei become more compact and more obscured, and, as a result, the
dust temperature increases leading to a steeper MIR slope as discussed
in Section \ref{results}. 



There is some indication that lower PAH equivalent widths are favored at later merger stages
but this is mostly dominated by the fact that only
starburst-dominated galaxies (\eqw$>$0.54\micron) are observed in
stage $1$. (The one exception is southern component of the LIRG system
AM0702-601.) For all other merger stages, the full range of \eqw\ is
observed.

\begin{figure}[htp]
\begin{center}
\includegraphics[height=2in,width=3in]{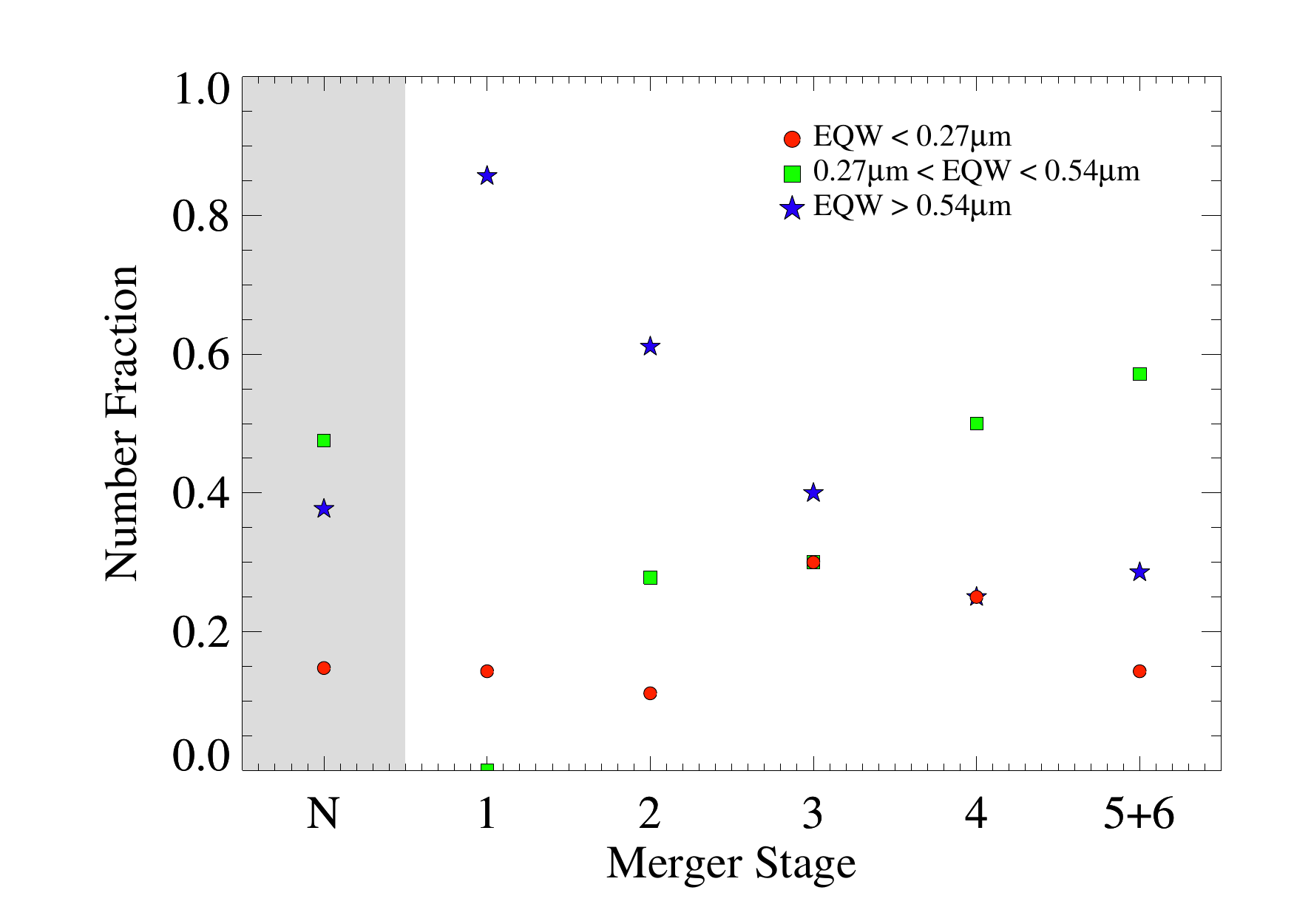}
\caption{Overall AGN fraction (as determined by EQW of the 6.2$\mu$m PAH) traced through merger stage for the GOALS sample. Merger stages are classified as described in Figure \ref{mstages}. A marked decline is seen for the fraction of high EQW
  (star formation dominated; blue stars) sources as the merger progresses. This
  decline is accompanied by an increased contribution not from the
  strongest AGN (red circles) which remain low but from the composite sources
  (i.e. weaker AGN that are not yet entirely dominant over star
  formation in the MIR; green squares). 
\label{mstagesEW}}
\end{center}
\end{figure}

 A clearer link between PAH equivalent width and merger stage is observed when galaxies are binned by
their \eqw (and thus the likely AGN contribution to the MIR). In
Figure \ref{mstagesEW} the LIRGs are divided into three EQW$_{6.2\mu m}$ bins indicating AGN dominated sources (red circles), composite
sources (green squares), and starbursts (blue stars). Starbursts
clearly play a dominant role at early merger stages as was also shown
by \cite{petric} and \cite{haanHST}, but the decline in the starburst
contribution is not balanced by an increase in AGN-dominated
sources. The contribution from LIRGs with an AGN dominating in the MIR stays at a roughly constant
fraction throughout the merger process, but $composite$ sources
(i.e. the weaker AGN that are not yet entirely dominant over star formation in the MIR) show a
marked increase at later merger stages. This may indicate that the timescales for the AGN to begin to dominate the MIR emission are longer than the merger timescale (a few hundred million years). In both Figures \ref{mstages} and \ref{mstagesEW}, the nonmerging
LIRGs cover nearly the full range of every MIR property
investigated. 

\section{Comparisons to Submillimeter Galaxies}\label{smgsec}
\begin{figure}[htp!]
\begin{center}
\includegraphics[height=2.5in,width=3.2in]{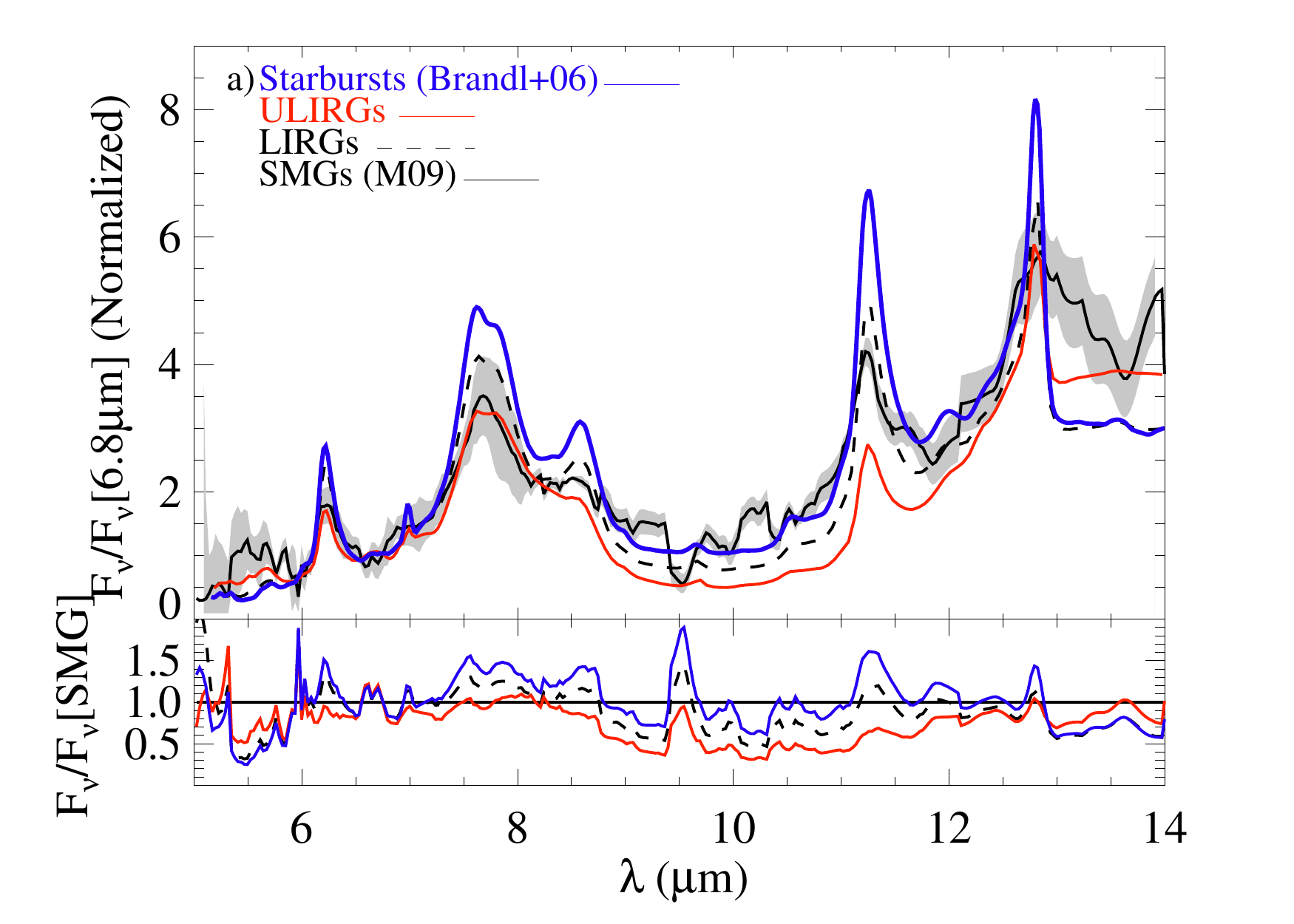}
\includegraphics[height=2.5in,width=3.2in]{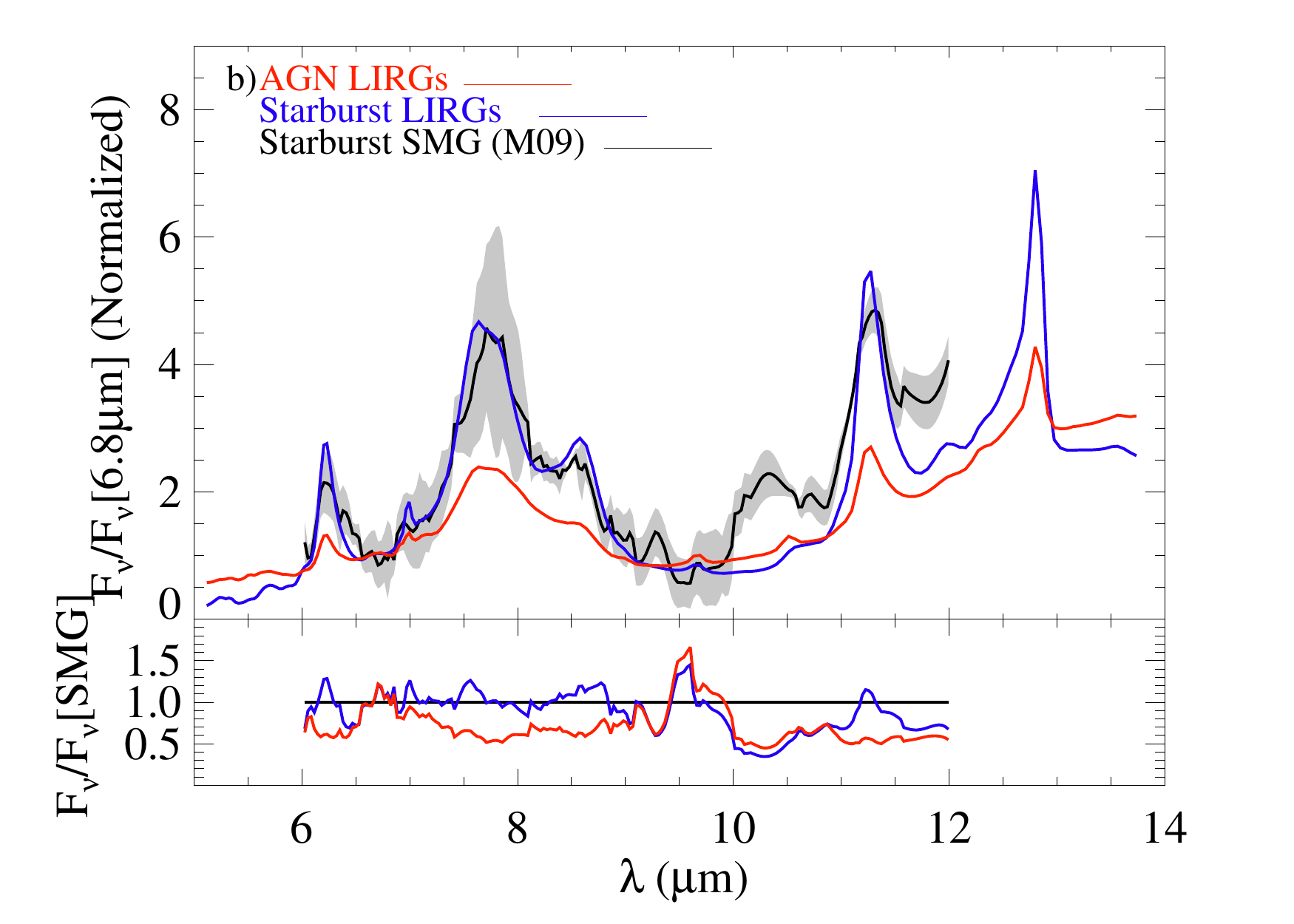}
\caption{Comparison of GOALS average LIRG and ULIRG spectra to average submillimeter
  galaxy spectra from \cite{SMGkarin}: a) the composite SMG spectrum (black) is
  less obscured than the average ULIRG (red) and the average LIRG
  (dashed) but has weaker PAH emission than local starbursts
  (blue). b) after removing the AGN-dominated systems from both the
  average SMG and the average LIRG, the average starburst SMG spectrum
  (black) is well-represented by the starburst LIRGs
  (EQW$_{6.2\mu m} > $0.54$\mu$m; blue) but not the AGN-dominated
  LIRGs (EQW$_{6.2\mu m} < $0.27$\mu$m; red) with the exception of a
  feature at $\sim$10.5\micron. All average spectra are
  normalized at $\lambda = 6.8$\micron~and the shaded gray area
  represents the 1-$\sigma$ standard deviation to the averaged SMG
  spectrum. Residuals are shown in the bottom panels.\label{smg}}
\end{center}
\end{figure}

The dust-enshrouded, strongly starbursting nature of LIRGs makes them
obvious candidates for possible local analogs to the dusty submillimeter
galaxies (SMGs) that make a significant contribution to the
global star formation rate density at higher redshifts. 
In Figure \ref{smg}, we compare different subsets of the GOALS MIR
spectra with the average SMG spectra from
\cite{SMGkarin} (hereafter M09) derived from a sample of 24 SMGs at
redshifts of 0.65 $<$ z
$<$3.2. All average spectra for both the LIRGs and the SMGs are
normalized at 6.8\micron. In agreement with the conclusions of
\cite{vandanaULIRGs} and M09, the average local ULIRG spectrum (red line
in Figure \ref{smg}a) is more absorbed than the average SMG spectrum
(black line) but has weaker PAH emission. Although GOALS LIRGs are less obscured than ULIRGs
on average, the average LIRG spectrum (dashed line) is still more
absorbed than the average SMG while also showing stronger PAH emission
at 6.2\micron, 7.7\micron, and 11.3\micron. Even when the nuclear emission
of galaxies is removed, the spectrum of the extended component of LIRGs
does not resemble that of the total SMG composite \citep{tanio2}. The average local
starburst \citep[blue line;][]{brandlEW} shows a similar level of
silicate absorption but much stronger PAH emission compared to the
average SMG. 

The fraction of MIR emission attributed to AGN overall for the GOALS
LIRGs is only 12\% \citep{petric}, and M09 observed a contribution of
$<$32\% from AGN to the total bolometric luminosity in SMGs. However,
as seen in Figures \ref{minifig} and A1, those sources
dominated by AGN have MIR spectra that are vastly different from those with
strong PAH emission. The low equivalent width sources (i.e. those with MIR
emission that is most likely AGN-dominated) also show a larger scatter
in their MIR properties, as discussed in Section \ref{results}. To reduce possible confusion caused by this AGN
contribution, we also compare the average SMG spectrum for only those
SMGs without AGN indicators in the MIR (i.e starburst SMGs with
EQW$_{7.7\mu m} > $1\micron~\& $\alpha_{MIR} < $0.5; M09) to average
LIRG spectra with and without an AGN contribution in
Figure \ref{smg}b. 

The average AGN-dominated LIRG spectrum (LIRGs with
EQW$_{6.2\mu m} < $0.27\micron; red line) clearly does not resemble
the average starburst SMG (solid black line). However the average
starburst-dominated LIRG spectra (EQW$_{6.2\mu m} > $0.54\micron; blue
line) is a better match to the
average starburst SMG. All three average spectra agree within 15\%
below 10\micron~(see the residuals in the lower panel). 

Between 10-11\micron, all three average LIRG spectra in Figure \ref{smg}b
agree closely, but no subset of the GOALS LIRGs or ULIRGs
reproduces the emission feature observed in the average starburst SMG spectrum near
10.5\micron. Although the feature is also not detected in the average
(low resolution) local
starburst spectrum (see
Figure \ref{smg}a), both the [SIV] emission line at 10.51\micron~and a
PAH feature at 10.60\micron~are clearly seen in the average of the high
resolution IRS spectra of the same starburst sample \citep[see Figure
4 of][]{jero}. 
The feature detected in the low resolution SMG
spectra is likely a blend of these two features~\citep{sturmM82, jero,
  jdsings}, but may be dominated by the PAH feature emission since it remains
faint and unresolved at low resolution.
\cite{jero} also detect a third feature at
10.75\micron~that they associate with PAH emission due to its close
correlation with the 11.3\micron~PAH which may contribute to the
emission in the SMGs. 

\section{Summary \& Conclusions}\label{conc}
We presented low resolution IRS spectra for 244 galaxy nuclei in the
GOALS sample of 180 LIRGs and 22 ULIRGs. The GOALS galaxies cover a
range of spectral types, silicate strengths, and merger stages, and
represent a complete subset of the IRAS Revised Bright Galaxy Sample. We
investigated the MIR properties directly measured from the spectra
and discovered the following:

\noindent 1) Local LIRGs cover a large range of MIR properties and any single LIRG cannot represent the class as a whole. LIRGs span 0.005\micron~$<$~EQW$_{6.2\mu m} < $0.78\micron~(with nondetections of the 6.2\micron~PAH reaching EQW$_{6.2\mu m} < $0.043\micron), -3.51 $< s_{9.7\mu m} <$ 0.052 (with 23\% of LIRGs showing silicate emission), and 2.00 $<$ F$_{\nu}$[30\micron]/F$_{\nu}$[15\micron] $<$ 35.40. However, the majority (63\%) of LIRGs have EQW$_{6.2\mu m} >$ 0.4, $s_{9.7\mu m}$ $>$ -1.0, and MIR slopes in the range of 4 $<$ F$_{\nu}$[30\micron]/F$_{\nu}$[15\micron] $<$ 10. 

\noindent 2) The GOALS ULIRGs span a narrower range of MIR properties than those covered by the LIRGs. When compared to LIRGs, the ULIRGs (L$_{IR} > 10^{12}$L$_{\odot}$) have a steeper median slope (F$_{\nu}$[30$\mu$m]/F$_{\nu}$[15$\mu$m] = 12.54 for the ULIRGs compared to F$_{\nu}$[30\micron]/F$_{\nu}$[15\micron] = 7.11 for the LIRGs), a lower mean equivalent width (EQW$_{6.2\mu m}$ = 0.30\micron~versus EQW$_{6.2\mu m}$ = 0.55\micron), and deeper average silicate absorption ($s_{9.7\mu m}$ = -1.05 versus $s_{9.7\mu m}$ = -0.25). 

\noindent 3) There is a general trend among the U/LIRGs for both silicate depth and MIR slope to increase with increasing \lir. As \lir\ increases, the temperature may rise as a consequence of the nuclei becoming more obscured and compact. As the dust temperature increases, the rising portion of the blackbody emission spectrum shifts to shorter wavelengths, and warmer sources have increasingly more flux at 30\micron, and thus steeper MIR slopes. The sources that depart from these correlations, in both cases, have very low PAH equivalent width (\eqw\ $<$ 0.1\micron) consistent with their MIR emission being dominated by an AGN. 

\noindent 4) Although less numerous (only 18\% of the sample), LIRGs with the largest contributions from AGN (those with EQW$_{6.2\mu m} <$ 0.27\micron) cover a wider range of MIR slopes and silicate strengths than those sources of higher equivalent width that make up the majority of the sample. The sources with extremely low PAH equivalent widths (\eqw$<$0.1\micron) separate into two distinct types: relatively unobscured sources with a very hot dust component (and thus very shallow MIR slopes) and heavily dust obscured nuclei with a steep temperature gradient. For the AGN-dominated LIRGs with low apparent obscuration, an upper limit to the MIR slope can be set at F$_{\nu}$[30$\mu$m]/F$_{\nu}$[15$\mu$m]$\sim$4. The most obscured nuclei, however, have steeper MIR slopes due to most of their warm dust emission being hidden behind a large amount of cooler dust.


suggesting
\noindent 5) The LIRGs most likely harboring buried AGN (the obscured nuclei with $s_{9.7\mu m}$$<$-1.75) all have EQW$_{6.2\mu m} < $0.2\micron~and lack any extended component to their MIR emission at 24\micron. Extreme levels of dust obscuration may simply be blocking PAH emission, or the higher densities in these nuclei may create an environment where PAH dust grains are not present or the conditions are not appropriate for exciting them (i.e. lacking in photodissociation regions). Their compact nature suggests that their obscuring (cool) dust is associated with the outer regions of the starburst and not simply a measure of the dust along the line of sight through a large, dusty disk.

\noindent 6) U/LIRGs in the late to final stages of a merger have, on average, steeper MIR slopes and higher levels of dust obscuration. As merging galaxies coalesce and gas \& dust is funneled towards the center, the nuclei become more compact and more obscured. As a result, the dust temperature increases leading also to a steeper MIR slope. A marked decline is seen for the fraction of high EQW (star formation dominated) sources as the merger progresses. The decline is accompanied by an increase in the fraction of composite sources while the fraction of sources where an AGN dominates the MIR emission remains low. 

\noindent 7) Despite their dusty and starbursty nature, the average nearby LIRG spectrum does not resemble the average composite (starburst $+$ AGN) MIR spectrum from submillimeter galaxies at z$\sim$2. Both the average LIRG and ULIRG spectra are more absorbed at 9.7\micron~and the average LIRG has more PAH emission. However, once the AGN contributions are removed from the average LIRG and from the average SMG spectra, the PAH emission and level of silicate absorption of the average spectrum for starburst-dominated SMGs \citep[i.e. those without AGN spectral signatures;][]{SMGkarin} are fit well by the average starburst-dominated local LIRG. 

The Spitzer Space Telescope is operated by the Jet Propulsion
Laboratory, California Institute of Technology, under NASA contract
1407. This research has made use of the NASA/IPAC Extragalactic
Database (NED) which is operated by the Jet Propulsion Laboratory,
California Institute of Technology, under contract with the National
Aeronautics and Space Administration. This research has made use of
the NASA/IPAC Infrared Science Archive, which is operated by the Jet
Propulsion Laboratory, California Institute of Technology, under
contract with the National Aeronautics Space Administration. We would
like to thank M. Cluver for many helpful discussions and
K. Men{\'e}ndez-Delmestre for sharing her average SMG spectra. \\

\bibliography{LIRGreferences}{}

\begin{thebibliography}{37}
\expandafter\ifx\csname natexlab\endcsname\relax\def\natexlab#1{#1}\fi

\bibitem[{{Armus} {et~al.}(2007){Armus}, {Charmandaris}, {Bernard-Salas},
  {Spoon}, {Marshall}, {Higdon}, {Desai}, {Teplitz}, {Hao}, {Devost}, {Brandl},
  {Wu}, {Sloan}, {Soifer}, {Houck}, \& {Herter}}]{armusULIRGs}
{Armus}, L., {Charmandaris}, V., {Bernard-Salas}, J., {et~al.} 2007, \apj, 656,
  148

\bibitem[{{Armus} {et~al.}(2009){Armus}, {Mazzarella}, {Evans}, {Surace},
  {Sanders}, {Iwasawa}, {Frayer}, {Howell}, {Chan}, {Petric}, {Vavilkin},
  {Kim}, {Haan}, {Inami}, {Murphy}, {Appleton}, {Barnes}, {Bothun}, {Bridge},
  {Charmandaris}, {Jensen}, {Kewley}, {Lord}, {Madore}, {Marshall},
  {Melbourne}, {Rich}, {Satyapal}, {Schulz}, {Spoon}, {Sturm}, {U}, {Veilleux},
  \& {Xu}}]{GOALS}
{Armus}, L., {Mazzarella}, J.~M., {Evans}, A.~S., {et~al.} 2009, \pasp, 121,
  559

\bibitem[{{Bernard-Salas} {et~al.}(2009){Bernard-Salas}, {Spoon},
  {Charmandaris}, {Lebouteiller}, {Farrah}, {Devost}, {Brandl}, {Wu}, {Armus},
  {Hao}, {Sloan}, {Weedman}, \& {Houck}}]{jero}
{Bernard-Salas}, J., {Spoon}, H.~W.~W., {Charmandaris}, V., {et~al.} 2009,
  \apjs, 184, 230

\bibitem[{{Blain} {et~al.}(2002){Blain}, {Smail}, {Ivison}, {Kneib}, \&
  {Frayer}}]{blain}
{Blain}, A.~W., {Smail}, I., {Ivison}, R.~J., {Kneib}, J.-P., \& {Frayer},
  D.~T. 2002, \physrep, 369, 111

\bibitem[{{Brandl} {et~al.}(2006){Brandl}, {Bernard-Salas}, {Spoon}, {Devost},
  {Sloan}, {Guilles}, {Wu}, {Houck}, {Weedman}, {Armus}, {Appleton}, {Soifer},
  {Charmandaris}, {Hao}, {Higdon}, {Marshall}, \& {Herter}}]{brandlEW}
{Brandl}, B.~R., {Bernard-Salas}, J., {Spoon}, H.~W.~W., {et~al.} 2006, \apj,
  653, 1129

\bibitem[{{Caputi} {et~al.}(2007){Caputi}, {Lagache}, {Yan}, {Dole},
  {Bavouzet}, {Le Floc'h}, {Choi}, {Helou}, \& {Reddy}}]{caputi}
{Caputi}, K.~I., {Lagache}, G., {Yan}, L., {et~al.} 2007, \apj, 660, 97

\bibitem[{{Dasyra} {et~al.}(2008){Dasyra}, {Yan}, {Helou}, {Surace}, {Sajina},
  \& {Colbert}}]{dasyra}
{Dasyra}, K.~M., {Yan}, L., {Helou}, G., {et~al.} 2008, \apj, 680, 232

\bibitem[{{Desai} {et~al.}(2007){Desai}, {Armus}, {Spoon}, {Charmandaris},
  {Bernard-Salas}, {Brandl}, {Farrah}, {Soifer}, {Teplitz}, {Ogle}, {Devost},
  {Higdon}, {Marshall}, \& {Houck}}]{vandanaULIRGs}
{Desai}, V., {Armus}, L., {Spoon}, H.~W.~W., {et~al.} 2007, \apj, 669, 810

\bibitem[{{D{\'{\i}}az-Santos} {et~al.}(2010){D{\'{\i}}az-Santos},
  {Charmandaris}, {Armus}, {Petric}, {Howell}, {Murphy}, {Mazzarella},
  {Veilleux}, {Bothun}, {Inami}, {Appleton}, {Evans}, {Haan}, {Marshall},
  {Sanders}, {Stierwalt}, \& {Surace}}]{taniopI}
{D{\'{\i}}az-Santos}, T., {Charmandaris}, V., {Armus}, L., {et~al.} 2010, \apj,
  723, 993

\bibitem[{{D{\'{\i}}az-Santos} {et~al.}(2011){D{\'{\i}}az-Santos},
  {Charmandaris}, {Armus}, {Stierwalt}, {Haan}, {Mazzarella}, {Howell},
  {Veilleux}, {Murphy}, {Petric}, {Appleton}, {Evans}, {Sanders}, \&
  {Surace}}]{tanio2}
---. 2011, \apj, 741, 32

\bibitem[{{Evans} {et~al.}(2006){Evans}, {Solomon}, {Tacconi}, {Vavilkin}, \&
  {Downes}}]{evans4418}
{Evans}, A.~S., {Solomon}, P.~M., {Tacconi}, L.~J., {Vavilkin}, T., \&
  {Downes}, D. 2006, \aj, 132, 2398

\bibitem[{{Genzel} {et~al.}(1998){Genzel}, {Lutz}, {Sturm}, {Egami}, {Kunze},
  {Moorwood}, {Rigopoulou}, {Spoon}, {Sternberg}, {Tacconi-Garman}, {Tacconi},
  \& {Thatte}}]{genzelULIRGs}
{Genzel}, R., {Lutz}, D., {Sturm}, E., {et~al.} 1998, \apj, 498, 579

\bibitem[{{Haan} {et~al.}(2011){Haan}, {Surace}, {Armus}, {Evans}, {Howell},
  {Mazzarella}, {Kim}, {Vavilkin}, {Inami}, {Sanders}, {Petric}, {Bridge},
  {Melbourne}, {Charmandaris}, {Diaz-Santos}, {Murphy}, {U}, {Stierwalt}, \&
  {Marshall}}]{haanHST}
{Haan}, S., {Surace}, J.~A., {Armus}, L., {et~al.} 2011, \aj, 141, 100

\bibitem[{{Houck} {et~al.}(2004){Houck}, {Roellig}, {van Cleve}, {Forrest},
  {Herter}, {Lawrence}, {Matthews}, {Reitsema}, {Soifer}, {Watson}, {Weedman},
  {Huisjen}, {Troeltzsch}, {Barry}, {Bernard-Salas}, {Blacken}, {Brandl},
  {Charmandaris}, {Devost}, {Gull}, {Hall}, {Henderson}, {Higdon}, {Pirger},
  {Schoenwald}, {Sloan}, {Uchida}, {Appleton}, {Armus}, {Burgdorf},
  {Fajardo-Acosta}, {Grillmair}, {Ingalls}, {Morris}, \& {Teplitz}}]{IRS}
{Houck}, J.~R., {Roellig}, T.~L., {van Cleve}, J., {et~al.} 2004, \apjs, 154,
  18

\bibitem[{{Howell} {et~al.}(2007){Howell}, {Mazzarella}, {Chan}, {Lord},
  {Surace}, {Frayer}, {Appleton}, {Armus}, {Evans}, {Bothun}, {Ishida}, {Kim},
  {Jensen}, {Madore}, {Sanders}, {Schulz}, {Vavilkin}, {Veilleux}, \&
  {Xu}}]{howell1068}
{Howell}, J.~H., {Mazzarella}, J.~M., {Chan}, B.~H.~P., {et~al.} 2007, \aj,
  134, 2086

\bibitem[{{Howell} {et~al.}(2010){Howell}, {Armus}, {Mazzarella}, {Evans},
  {Surace}, {Sanders}, {Petric}, {Appleton}, {Bothun}, {Bridge}, {Chan},
  {Charmandaris}, {Frayer}, {Haan}, {Inami}, {Kim}, {Lord}, {Madore},
  {Melbourne}, {Schulz}, {U}, {Vavilkin}, {Veilleux}, \& {Xu}}]{GALEX}
{Howell}, J.~H., {Armus}, L., {Mazzarella}, J.~M., {et~al.} 2010, \apj, 715,
  572

\bibitem[{{Iwasawa} {et~al.}(2011){Iwasawa}, {Sanders}, {Teng}, {U}, {Armus},
  {Evans}, {Howell}, {Komossa}, {Mazzarella}, {Petric}, {Surace}, {Vavilkin},
  {Veilleux}, \& {Trentham}}]{CHANDRAGOALS}
{Iwasawa}, K., {Sanders}, D.~B., {Teng}, S.~H., {et~al.} 2011, \aap, 529, A106

\bibitem[{{Le Floc'h} {et~al.}(2005){Le Floc'h}, {Papovich}, {Dole}, {Bell},
  {Lagache}, {Rieke}, {Egami}, {P{\'e}rez-Gonz{\'a}lez}, {Alonso-Herrero},
  {Rieke}, {Blaylock}, {Engelbracht}, {Gordon}, {Hines}, {Misselt}, {Morrison},
  \& {Mould}}]{lefloch}
{Le Floc'h}, E., {Papovich}, C., {Dole}, H., {et~al.} 2005, \apj, 632, 169

\bibitem[{{Magnelli} {et~al.}(2009){Magnelli}, {Elbaz}, {Chary}, {Dickinson},
  {Le Borgne}, {Frayer}, \& {Willmer}}]{magnelli}
{Magnelli}, B., {Elbaz}, D., {Chary}, R.~R., {et~al.} 2009, \aap, 496, 57

\bibitem[{{Marshall} {et~al.}(2007){Marshall}, {Herter}, {Armus},
  {Charmandaris}, {Spoon}, {Bernard-Salas}, \& {Houck}}]{cafe}
{Marshall}, J.~A., {Herter}, T.~L., {Armus}, L., {et~al.} 2007, \apj, 670, 129

\bibitem[{{Men{\'e}ndez-Delmestre} {et~al.}(2009){Men{\'e}ndez-Delmestre},
  {Blain}, {Smail}, {Alexander}, {Chapman}, {Armus}, {Frayer}, {Ivison}, \&
  {Teplitz}}]{SMGkarin}
{Men{\'e}ndez-Delmestre}, K., {Blain}, A.~W., {Smail}, I., {et~al.} 2009, \apj,
  699, 667

\bibitem[{{Murphy} {et~al.}(1996){Murphy}, {Armus}, {Matthews}, {Soifer},
  {Mazzarella}, {Shupe}, {Strauss}, \& {Neugebauer}}]{murphyULIRGs}
{Murphy}, Jr., T.~W., {Armus}, L., {Matthews}, K., {et~al.} 1996, \aj, 111,
  1025

\bibitem[{{O'Dowd} {et~al.}(2009){O'Dowd}, {Schiminovich}, {Johnson}, {Treyer},
  {Martin}, {Wyder}, {Charlot}, {Heckman}, {Martins}, {Seibert}, \& {van der
  Hulst}}]{SSGSS}
{O'Dowd}, M.~J., {Schiminovich}, D., {Johnson}, B.~D., {et~al.} 2009, \apj,
  705, 885

\bibitem[{{Petric} {et~al.}(2011){Petric}, {Armus}, {Howell}, {Chan},
  {Mazzarella}, {Evans}, {Surace}, {Sanders}, {Appleton}, {Charmandaris},
  {D{\'{\i}}az-Santos}, {Frayer}, {Haan}, {Inami}, {Iwasawa}, {Kim}, {Madore},
  {Marshall}, {Spoon}, {Stierwalt}, {Sturm}, {U}, {Vavilkin}, \&
  {Veilleux}}]{petric}
{Petric}, A.~O., {Armus}, L., {Howell}, J., {et~al.} 2011, \apj, 730, 28

\bibitem[{{Rigopoulou} {et~al.}(1999){Rigopoulou}, {Spoon}, {Genzel}, {Lutz},
  {Moorwood}, \& {Tran}}]{rigULIRGs}
{Rigopoulou}, D., {Spoon}, H.~W.~W., {Genzel}, R., {et~al.} 1999, \aj, 118,
  2625

\bibitem[{{Roussel} {et~al.}(2003){Roussel}, {Helou}, {Beck}, {Condon},
  {Bosma}, {Matthews}, \& {Jarrett}}]{nascent}
{Roussel}, H., {Helou}, G., {Beck}, R., {et~al.} 2003, \apj, 593, 733

\bibitem[{{Sanders} {et~al.}(2003){Sanders}, {Mazzarella}, {Kim}, {Surace}, \&
  {Soifer}}]{rbgs}
{Sanders}, D.~B., {Mazzarella}, J.~M., {Kim}, D.-C., {Surace}, J.~A., \&
  {Soifer}, B.~T. 2003, \aj, 126, 1607

\bibitem[{{Sanders} \& {Mirabel}(1996)}]{sandersLIRGs}
{Sanders}, D.~B., \& {Mirabel}, I.~F. 1996, \araa, 34, 749

\bibitem[{{Sanders} {et~al.}(1988){Sanders}, {Soifer}, {Elias}, {Madore},
  {Matthews}, {Neugebauer}, \& {Scoville}}]{sandersULIRGs}
{Sanders}, D.~B., {Soifer}, B.~T., {Elias}, J.~H., {et~al.} 1988, \apj, 325, 74

\bibitem[{{Smith} {et~al.}(2007{\natexlab{a}}){Smith}, {Armus}, {Dale},
  {Roussel}, {Sheth}, {Buckalew}, {Jarrett}, {Helou}, \& {Kennicutt}}]{cubism}
{Smith}, J.~D.~T., {Armus}, L., {Dale}, D.~A., {et~al.} 2007{\natexlab{a}},
  \pasp, 119, 1133

\bibitem[{{Smith} {et~al.}(2007{\natexlab{b}}){Smith}, {Draine}, {Dale},
  {Moustakas}, {Kennicutt}, {Helou}, {Armus}, {Roussel}, {Sheth}, {Bendo},
  {Buckalew}, {Calzetti}, {Engelbracht}, {Gordon}, {Hollenbach}, {Li},
  {Malhotra}, {Murphy}, \& {Walter}}]{jdsings}
{Smith}, J.~D.~T., {Draine}, B.~T., {Dale}, D.~A., {et~al.} 2007{\natexlab{b}},
  \apj, 656, 770

\bibitem[{{Spoon} {et~al.}(2001){Spoon}, {Keane}, {Tielens}, {Lutz}, \&
  {Moorwood}}]{spoon4418}
{Spoon}, H.~W.~W., {Keane}, J.~V., {Tielens}, A.~G.~G.~M., {Lutz}, D., \&
  {Moorwood}, A.~F.~M. 2001, \aap, 365, L353

\bibitem[{{Spoon} {et~al.}(2007){Spoon}, {Marshall}, {Houck}, {Elitzur}, {Hao},
  {Armus}, {Brandl}, \& {Charmandaris}}]{spoon}
{Spoon}, H.~W.~W., {Marshall}, J.~A., {Houck}, J.~R., {et~al.} 2007, \apjl,
  654, L49

\bibitem[{{Spoon} {et~al.}(2006){Spoon}, {Tielens}, {Armus}, {Sloan},
  {Sargent}, {Cami}, {Charmandaris}, {Houck}, \& {Soifer}}]{spoonULIRGs}
{Spoon}, H.~W.~W., {Tielens}, A.~G.~G.~M., {Armus}, L., {et~al.} 2006, \apj,
  638, 759

\bibitem[{{Stierwalt} {et~al.}(2013){Stierwalt}}]{paperII}{Stierwalt},
  S., {et~al.} 2013b, \apj, in prep

\bibitem[{{Sturm} {et~al.}(2000){Sturm}, {Lutz}, {Tran}, {Feuchtgruber},
  {Genzel}, {Kunze}, {Moorwood}, \& {Thornley}}]{sturmM82}
{Sturm}, E., {Lutz}, D., {Tran}, D., {et~al.} 2000, \aap, 358, 481

\bibitem[{{Veilleux} {et~al.}(2009){Veilleux}, {Rupke}, {Kim}, {Genzel},
  {Sturm}, {Lutz}, {Contursi}, {Schweitzer}, {Tacconi}, {Netzer}, {Sternberg},
  {Mihos}, {Baker}, {Mazzarella}, {Lord}, {Sanders}, {Stockton}, {Joseph}, \&
  {Barnes}}]{quest}
{Veilleux}, S., {Rupke}, D.~S.~N., {Kim}, D.-C., {et~al.} 2009, \apjs, 182, 628

\bibitem[{{Wu} {et~al.}(2010){Wu}, {Helou}, {Armus}, {Cormier}, {Shi}, {Dale},
  {Dasyra}, {Smith}, {Papovich}, {Draine}, {Rahman}, {Stierwalt}, {Fadda},
  {Lagache}, \& {Wright}}]{5muses}
{Wu}, Y., {Helou}, G., {Armus}, L., {et~al.} 2010, \apj, 723, 895

\end{thebibliography}
\newpage
\begin{sidewaystable}
\caption{Mid-IR Spectral Parameters for the GOALS Sample \label{bigtable}}
\resizebox{\linewidth}{!}{
\tabcolsep=3pt
\begin{tabular}{llclccrrlcc}
\hline
\hline
Source & ~~~~~~Short-Low RA/DEC & (PA) & ~~~~~~Long-Low RA/DEC & (PA) & 6.2$\mu m$ EQW($\sigma$) & s$_{9.7\mu m}$($\sigma$)~ & $F_{\nu}[30\mu m]/F_{\nu}[15\mu m]$($\sigma$) & Scale & Merger & MS\\
Name & ~~~~~~~~~~~~~~~~~~~[J2000] & [$^{\circ}$]~ & ~~~~~~~~~~~~~~~~~~~[J2000] & [$^{\circ}$]~ & [$\mu m$] & & & Factor & Stage & (HST)\\
\hline
NGC0023 & 00h09m53.4s +25d55m26.3s & (-32.8) & 00h09m53.4s +25d55m26.3s & (-116.6) & 0.58 (0.01) & 0.18 (0.04) &  6.81 (0.04) ~~~~~~~~~~~~~ &  1.49 & b & \\
NGC0034 & 00h11m06.4s -12d06m28.7s & ~~(35.3)$*$ & 00h11m06.5s -12d06m29.2s & ~~(117.7)$*$ & 0.45 (0.02) & -0.79 (0.03) &  9.94 (0.09) ~~~~~~~~~~~~~ &  1.70 & d & 5\\
Arp256 & 00h18m50.9s -10d22m36.6s & (-36.1) & 00h18m50.9s -10d22m36.6s & (-119.8) & 0.72 (0.01) & -0.26 (0.03) &  6.73 (0.07) ~~~~~~~~~~~~~ &  1.22 & b & 3\\
ESO350-IG038 & 00h36m52.5s -33d33m17.1s & (163.1) & 00h36m52.5s -33d33m17.0s & (79.4) & 0.15 (0.004)~~ & -0.34 (0.02) &  4.73 (0.01) ~~~~~~~~~~~~~ &  1.15 & c & \\
NGC0232\_W & 00h42m45.8s -23d33m40.7s & (-44.6) & 00h42m45.8s -23d33m40.7s & (-128.3) & 0.55 (0.01) & -0.38 (0.03) &  7.33 (0.03) ~~~~~~~~~~~~~ &  1.21 & b & \\
NGC0232\_E & 00h42m52.8s -23d32m27.5s & (-44.6) & 00h42m52.8s -23d32m27.5s & (-128.3) & 0.16 (0.005)~~ & -0.26 (0.03) &  3.63 (0.03) ~~~~~~~~~~~~~ &  1.16 & b & \\
MCG+12-02-001 & 00h54m03.9s +73d05m06.0s & (71.1) & 00h54m03.9s +73d05m05.9s & (-12.7) & 0.65 (0.01) & -0.24 (0.02) &  7.19 (0.04) ~~~~~~~~~~~~~ &  1.08 & c & 3\\
NGC0317B & 00h57m40.4s +43d47m32.1s & (-29.6) & 00h57m40.4s +43d47m32.1s & (-113.4) & 0.57 (0.01) & -0.86 (0.05) &  8.23 (0.09) ~~~~~~~~~~~~~ &  1.15 & a & \\
IC1623B & 01h07m47.6s -17d30m25.4s & ~~(25.4)$*$ & 01h07m47.3s -17d30m28.4s & ~~(19.1)$*$ & 0.30 (0.004)~~ & -0.98 (0.02) &  7.71 (0.01) ~~~~~~~~~~~~~ &  1.73 & c & 3\\
MCG-03-04-014 & 01h10m09.0s -16d51m09.6s & (150.7) & 01h10m08.9s -16d51m10.6s & (-126.8) & 0.67 (0.01) & -0.04 (0.02) &  6.98 (0.09) ~~~~~~~~~~~~~ &  1.16 & N & 0\\
ESO244-G012 & 01h18m08.3s -44d27m43.5s & (-79.0) & 01h18m08.3s -44d27m43.6s & (-162.7) & 0.66 (0.01) & -0.32 (0.13) &  7.39 (0.06) ~~~~~~~~~~~~~ &  1.19 & b & \\
CGCG436-030 & 01h20m02.6s +14d21m42.7s & (-27.1) & 01h20m02.6s +14d21m42.7s & (-110.9) & 0.35 (0.01) & -1.10 (0.10) &  8.94 (0.09) ~~~~~~~~~~~~~ &  1.10 & b & 2\\
ESO353-G020 & 01h34m51.2s -36d08m14.5s & (-58.4) & 01h34m51.2s -36d08m14.5s & (-142.1) & 0.54 (0.01) & -0.58 (0.04) &  7.26 (0.10) ~~~~~~~~~~~~~ &  1.42 & N & \\
RR032\_N & 01h36m23.4s -37d19m18.0s & (-59.0) & 01h36m23.4s -37d19m18.1s & (-142.7) & 0.52 (0.01) & 0.09 (0.03) &  6.19 (0.10) ~~~~~~~~~~~~~ &  1.28 & a & \\
RR032\_S & 01h36m24.1s -37d20m25.9s & (-58.9) & 01h36m24.1s -37d20m25.9s & (-142.7) & 0.70 (0.01) & -0.38 (0.04) &  6.74 (0.09) ~~~~~~~~~~~~~ &  1.14 & a & \\
IRASF01364-1042 & 01h38m52.9s -10d27m11.2s & (149.2) & 01h38m52.8s -10d27m12.0s & (-110.8) & 0.39 (0.01) & -1.27 (0.08) & 32.72 (1.04) ~~~~~~~~~~~~~ &  1.41 & d & 5\\
IIIZw035 & ~~~~~~~~~~~~~~~~~~~~~--- & & 01h44m30.5s +17d06m08.7s & (-109.1) & --- & ---~~~~~~ & 23.52 (0.24) ~~~~~~~~~~~~~ & ~~--- & a & 3\\
NGC0695 & 01h51m14.2s +22d34m56.4s & (-23.5) & 01h51m14.3s +22d34m55.4s & (-107.3) & 0.65 (0.01) & 0.31 (0.02) &  6.17 (0.09) ~~~~~~~~~~~~~ &  1.86 & N & \\
UGC01385 & 01h54m53.8s +36d55m04.0s & (-22.0) & 01h54m53.8s +36d55m04.1s & (-105.7) & 0.64 (0.01) & -0.00 (0.11) &  6.73 (0.08) ~~~~~~~~~~~~~ &  1.12 & a & \\
NGC0838\_W & 02h09m24.7s -10d08m09.6s & (162.7) & 02h09m24.7s -10d08m09.5s & (78.9) & 0.45 (0.01) & 0.27 (0.03) &  4.76 (0.10) ~~~~~~~~~~~~~ &  1.63 & a & \\
NGC0838\_E & 02h09m38.7s -10d08m47.5s & (162.6) & 02h09m38.7s -10d08m47.5s & (78.9) & 0.74 (0.01) & 0.23 (0.03) &  7.59 (0.06) ~~~~~~~~~~~~~ &  2.00 & a & \\
NGC0838\_S & 02h09m42.8s -10d11m02.3s & (162.6) & 02h09m42.8s -10d11m02.3s & (78.9) & 0.50 (0.01) & -0.83 (0.03) &  6.72 (0.04) ~~~~~~~~~~~~~ &  1.18 & a & \\
NGC0828 & 02h10m09.5s +39d11m24.6s & (-19.3) & 02h10m09.5s +39d11m24.7s & (-103.0) & 0.61 (0.01) & -0.09 (0.03) &  5.04 (0.05) ~~~~~~~~~~~~~ &  1.64 & d & \\
IC0214 & 02h14m05.5s +05d10m25.4s & (-26.2) & 02h14m05.4s +05d10m25.4s & (-110.0) & 0.64 (0.01) & -0.05 (0.02) &  6.51 (0.10) ~~~~~~~~~~~~~ &  1.34 & d & \\
NGC0877\_S & 02h17m53.2s +14d31m18.1s & (-24.5) & 02h17m53.2s +14d31m18.2s & (-108.3) & 0.46 (0.01) & -1.16 (0.10) &  7.11 (0.17) ~~~~~~~~~~~~~ &  1.36 & a & \\
NGC0877\_N & 02h17m59.7s +14d32m38.0s & (-24.5) & 02h17m59.6s +14d32m38.0s & (-108.3) & 0.35 (0.03) & 0.16 (0.03) &  6.14 (0.14) ~~~~~~~~~~~~~ &  2.10 & a & \\
MCG+05-06-036\_S & 02h23m19.0s +32d11m18.1s & (155.4) & 02h23m19.0s +32d11m18.2s & (71.7) & 0.54 (0.01) & 0.19 (0.03) &  6.12 (0.22) ~~~~~~~~~~~~~ &  1.46 & a & 2\\
MCG+05-06-036\_N & 02h23m22.0s +32d11m48.4s & (155.4) & 02h23m22.0s +32d11m48.5s & (71.7) & 0.48 (0.01) & -0.25 (0.02) &  6.23 (0.07) ~~~~~~~~~~~~~ &  1.29 & a & 2\\
UGC01845 & 02h24m08.0s +47d58m11.8s & (-31.2) & 02h24m08.0s +47d58m11.9s & (-115.0) & 0.58 (0.01) & -0.40 (0.03) &  7.71 (0.09) ~~~~~~~~~~~~~ &  1.18 & N & \\
NGC0958 & 02h30m42.9s -02d56m20.4s & (-27.6) & 02h30m42.9s -02d56m20.4s & (-111.3) & 0.29 (0.01) & 0.00 (0.08) &  4.96 (0.15) ~~~~~~~~~~~~~ &  1.77 & N & \\
\hline\end{tabular}}
\end{sidewaystable}
\begin{sidewaystable}
\resizebox{\linewidth}{!}{
\tabcolsep=3pt
\begin{tabular}{llclccrrlcc}
\hline
\hline
Source & ~~~~~~Short-Low RA/DEC & (PA) & ~~~~~~Long-Low RA/DEC & (PA) & 6.2$\mu m$ EQW($\sigma$) & s$_{9.7\mu m}$($\sigma$)~ & $F_{\nu}[30\mu m]/F_{\nu}[15\mu m]$($\sigma$) & Scale & Merger & MS\\
Name & ~~~~~~~~~~~~~~~~~~~[J2000] & [$^{\circ}$]~ & ~~~~~~~~~~~~~~~~~~~[J2000] & [$^{\circ}$]~ & [$\mu m$] & & & Factor & Stage & (HST)\\
\hline
NGC0992 & 02h37m25.5s +21d06m04.4s & (-21.8) & 02h37m25.5s +21d06m04.4s & (-105.5) & 0.72 (0.01) & 0.05 (0.04) &  6.41 (0.09) ~~~~~~~~~~~~~ &  1.28 & c & \\
UGC02238 & 02h46m17.5s +13d05m44.3s & (-24.7) & 02h46m17.5s +13d05m44.9s & (-108.2) & 0.63 (0.01) & -0.37 (0.04) &  5.66 (0.07) ~~~~~~~~~~~~~ &  1.14 & d & \\
IRASF02437+2122 & 02h46m39.1s +21d35m10.1s & (-22.3) & 02h46m39.1s +21d35m10.3s & (-104.0) & 0.16 (0.01) & -1.25 (0.06) &  8.81 (0.16) ~~~~~~~~~~~~~ &  1.06 & c & \\
UGC02369 & 02h54m01.8s +14d58m15.4s & (-22.9) & 02h54m01.8s +14d58m15.5s & (-106.7) & 0.57 (0.01) & -0.11 (0.07) &  6.48 (0.03) ~~~~~~~~~~~~~ &  1.30 & b & 3\\
UGC02608\_N & 03h15m01.5s +42d02m08.5s & (-21.8) & 03h15m01.5s +42d02m08.6s & (-105.6) & 0.20 (0.003)~~ & -0.22 (0.02) &  5.04 (0.02) ~~~~~~~~~~~~~ &  1.15 & N & \\
UGC02608\_S & 03h15m14.6s +41d58m49.9s & (-21.8) & 03h15m14.6s +41d58m49.9s & (-105.5) & 0.25 (0.14) & 0.52 (0.32) &  4.59 (1.60) ~~~~~~~~~~~~~ &  1.72 & N & \\
NGC1275 & 03h19m48.2s +41d30m41.9s & (157.6) & 03h19m48.2s +41d30m41.7s & (73.8) & 0.02 (0.003)~~ & 0.26 (0.02) &  2.73 (0.01) ~~~~~~~~~~~~~ &  1.01 & N & \\
IRASF03217+4022 & 03h25m05.4s +40d33m32.3s & (-22.2) & 03h25m05.4s +40d33m32.3s & (-105.9) & 0.55 (0.01) & -0.47 (0.03) &  8.34 (0.07) ~~~~~~~~~~~~~ &  1.17 & d & \\
NGC1365 & 03h33m36.4s -36d08m26.0s & (-51.0) & 03h33m36.4s -36d08m25.5s & (74.4) & 0.13 (0.002)~~ & 0.10 (0.05) &  5.65 (0.03) ~~~~~~~~~~~~~ &  1.67 & N & \\
IRASF03359+1523 & ~~~~~~~~~~~~~~~~~~~~~--- & & 03h38m47.1s +15d32m53.2s & (-102.3) & --- & ---~~~~~~ & 12.45 (0.16) ~~~~~~~~~~~~~ & ~~--- & d & \\
CGCG465-012\_N & 03h54m07.8s +15d59m24.1s & (-17.6) & 03h54m07.8s +15d59m24.2s & (-101.3) & 0.67 (0.02) & 0.43 (0.03) &  4.33 (0.16) ~~~~~~~~~~~~~ &  1.83 & a & \\
CGCG465-012\_S & 03h54m16.1s +15d55m43.2s & (-17.6) & 03h54m16.1s +15d55m43.3s & (-101.3) & 0.60 (0.01) & -0.02 (0.03) &  6.64 (0.66) ~~~~~~~~~~~~~ &  1.65 & c & \\
IRAS03582+6012\_W & 04h02m32.0s +60d20m38.3s & (-12.6) & 04h02m32.0s +60d20m38.3s & (-96.3) & 0.64 (0.01) & 0.08 (0.04) &  8.80 (0.08) ~~~~~~~~~~~~~ & ~~--- & c & \\
IRAS03582+6012\_E & 04h02m33.0s +60d20m41.8s & (-12.6) & 04h02m33.0s +60d20m41.8s & (-96.3) & 0.01 (0.003)~~ & -3.04 (0.03) &  6.39 (0.02) ~~~~~~~~~~~~~ & ~~--- & c & \\
UGC02982 & 04h12m22.6s +05d32m50.5s & (-15.8) & 04h12m22.6s +05d32m50.5s & (-99.6) & 0.68 (0.01) & 0.18 (0.03) &  5.11 (0.06) ~~~~~~~~~~~~~ &  2.02 & d & \\
ESO420-G013 & 04h13m49.7s -32d00m25.5s & (-49.4) & 04h13m49.7s -32d00m25.5s & (-133.2) & 0.30 (0.003)~~ & -0.27 (0.03) &  5.43 (0.04) ~~~~~~~~~~~~~ &  1.21 & N & \\
NGC1572 & 04h22m42.8s -40d36m03.6s & (-131.6) & 04h22m42.8s -40d36m03.6s & (144.7) & 0.46 (0.01) & -0.16 (0.03) &  6.76 (0.08) ~~~~~~~~~~~~~ &  1.20 & N & \\
IRAS04271+3849 & 04h30m33.1s +38d55m48.0s & (-10.8) & 04h30m33.1s +38d55m48.0s & (-94.5) & 0.61 (0.01) & -0.29 (0.04) &  7.85 (0.12) ~~~~~~~~~~~~~ &  1.25 & d & \\
NGC1614 & 04h33m59.8s -08d34m43.1s & (-26.7) & 04h33m59.8s -08d34m40.8s & (-110.4) & 0.61 (0.01) & -0.41 (0.02) &  4.81 (0.01) ~~~~~~~~~~~~~ &  1.40 & d & 5\\
UGC03094 & 04h35m33.9s +19d10m18.3s & (167.0) & 04h35m33.8s +19d10m17.5s & (-97.5) & 0.43 (0.005)~~ & -0.32 (0.02) &  5.05 (0.06) ~~~~~~~~~~~~~ &  1.16 & N & \\
ESO203-IG001 & 04h46m49.5s -48d33m30.1s & (-106.3) & 04h46m49.5s -48d33m30.1s & (170.0) & 0.03 (0.01) & -3.20 (0.17) & 15.05 (0.33) ~~~~~~~~~~~~~ &  0.98 & d & 3\\
MCG-05-12-006 & 04h52m05.0s -32d59m27.1s & (-141.0) & 04h52m05.0s -32d59m27.0s & (135.3) & 0.53 (0.01) & -0.02 (0.03) &  7.25 (0.07) ~~~~~~~~~~~~~ &  1.11 & N & \\
NGC1797 & 05h07m44.8s -08d01m08.7s & (-18.8) & 05h07m44.8s -08d01m08.7s & (-102.5) & 0.62 (0.01) & -0.20 (0.03) &  7.72 (0.08) ~~~~~~~~~~~~~ &  1.30 & a & \\
CGCG468-002\_W & 05h08m19.7s +17d21m47.5s & (173.1) & 05h08m19.7s +17d21m47.5s & (89.3) & 0.12 (0.005)~~ & -0.01 (0.03) &  3.75 (0.05) ~~~~~~~~~~~~~ &  1.04 & b & \\
CGCG468-002\_E & 05h08m21.2s +17d22m07.7s & (173.1) & 05h08m21.2s +17d22m07.7s & (89.3) & 0.54 (0.01) & -0.95 (0.04) & 10.38 (0.19) ~~~~~~~~~~~~~ &  1.03 & b & \\
IRAS05083+2441\_S & 05h11m25.9s +24d45m18.0s & (170.4) & 05h11m25.9s +24d45m18.0s & (86.6) & 0.72 (0.01) & -0.16 (0.04) &  7.99 (0.11) ~~~~~~~~~~~~~ &  1.17 & N & \\
VIIZw031 & 05h16m46.4s +79d40m12.7s & (155.0) & 05h16m46.3s +79d40m12.7s & (71.3) & 0.64 (0.01) & -0.22 (0.04) &  7.37 (0.11) ~~~~~~~~~~~~~ &  1.27 & N & 0\\
IRAS05129+5128 & 05h16m56.0s +51d31m57.3s & (2.3) & 05h16m56.0s +51d31m57.3s & (-81.5) & 0.54 (0.01) & -0.61 (0.04) &  8.75 (0.11) ~~~~~~~~~~~~~ &  1.27 & d & \\
IRASF05189-2524 & 05h21m01.3s -25d21m45.6s & (-2.5) & 05h21m01.4s -25d21m46.1s & (-86.2) & 0.03 (0.002)~~ & -0.29 (0.02) &  5.04 (0.01) ~~~~~~~~~~~~~ &  0.98 & d & \\
IRASF05187-1017 & 05h21m06.5s -10d14m46.8s & (-13.7) & 05h21m06.5s -10d14m46.2s & (-108.7) & 0.53 (0.03) & -0.64 (0.14) & 14.90 (0.51) ~~~~~~~~~~~~~ &  1.06 & N & \\
\hline
\end{tabular}}
\end{sidewaystable}
\begin{sidewaystable}
\resizebox{\linewidth}{!}{
\tabcolsep=3pt
\begin{tabular}{llclccrrlcc}
\hline
\hline
Source & ~~~~~~Short-Low RA/DEC & (PA) & ~~~~~~Long-Low RA/DEC & (PA) & 6.2$\mu m$ EQW($\sigma$) & s$_{9.7\mu m}$($\sigma$)~ & $F_{\nu}[30\mu m]/F_{\nu}[15\mu m]$($\sigma$) & Scale & Merger & MS\\
Name & ~~~~~~~~~~~~~~~~~~~[J2000] & [$^{\circ}$]~ & ~~~~~~~~~~~~~~~~~~~[J2000] & [$^{\circ}$]~ & [$\mu m$] & & & Factor & Stage & (HST)\\
\hline
IRAS05223+1908 & 05h25m16.7s +19d10m48.0s & (-10.1) & 05h25m16.7s +19d10m48.0s & (-93.8) & $<$0.01 (---) & 0.12 (0.02) &  2.00 (0.01) ~~~~~~~~~~~~~ &  0.99 & N & \\
MCG+08-11-002 & 05h40m43.7s +49d41m41.3s & (169.6) & 05h40m43.7s +49d41m41.3s & (85.9) & 0.56 (0.01) & -0.89 (0.03) &  9.47 (0.08) ~~~~~~~~~~~~~ &  1.22 & d & \\
NGC1961 & 05h42m04.7s +69d22m43.1s & (162.9) & 05h42m04.7s +69d22m43.2s & (79.2) & 0.24 (0.01) & 0.03 (0.09) &  4.18 (0.05) ~~~~~~~~~~~~~ &  1.93 & d & \\
UGC03351 & 05h45m48.2s +58d42m03.1s & (155.1) & 05h45m48.1s +58d42m03.2s & (71.4) & 0.53 (0.01) & -0.68 (0.04) &  6.88 (0.10) ~~~~~~~~~~~~~ &  1.07 & a & \\
IRAS05442+1732 & 05h47m11.2s +17d33m46.5s & (-9.3) & 05h47m11.2s +17d33m46.6s & (-93.0) & 0.66 (0.01) & -0.38 (0.02) &  7.52 (0.03) ~~~~~~~~~~~~~ &  1.16 & a & \\
IRASF06076-2139 & 06h09m45.8s -21d40m23.9s & (166.3) & 06h09m45.8s -21d40m23.8s & (110.0) & 0.33 (0.02) & -1.39 (0.09) &  7.98 (0.10) ~~~~~~~~~~~~~ & ~~--- & c & \\
UGC03410\_W & 06h13m58.8s +80d28m35.2s & (154.7) & 06h13m58.8s +80d28m35.2s & (70.8) & 0.62 (0.01) & 0.15 (0.04) &  3.38 (0.06) ~~~~~~~~~~~~~ &  1.60 & a & \\
UGC03410\_E & 06h14m30.5s +80d26m59.9s & (154.6) & 06h14m30.5s +80d26m59.9s & (71.0) & 0.63 (0.01) & 0.14 (0.05) &  5.19 (0.03) ~~~~~~~~~~~~~ &  1.79 & a & \\
NGC2146 & 06h18m37.5s +78d21m24.3s & (17.7) & 06h18m37.7s +78d21m25.8s & (-62.4) & 0.67 (0.01) & -0.43 (0.02) &  9.96 (0.01) ~~~~~~~~~~~~~ &  2.43 & d & \\
ESO255-IG007\_W & 06h27m21.7s -47d10m36.5s & (-128.8) & 06h27m21.7s -47d10m36.4s & (147.4) & 0.62 (0.01) & -0.26 (0.02) &  8.31 (0.03) ~~~~~~~~~~~~~ & ~~--- & b & 3\\
ESO255-IG007\_E & 06h27m22.5s -47d10m47.6s & (-128.8) & 06h27m22.5s -47d10m47.5s & (147.4) & 0.65 (0.02) & 0.26 (0.02) &  8.05 (0.04) ~~~~~~~~~~~~~ & ~~--- & b & 3\\
ESO255-IG007\_S & 06h27m23.1s -47d11m02.9s & (-128.8) & 06h27m23.1s -47d11m02.9s & (147.4) & 0.69 (0.02) & -0.10 (0.02) &  7.27 (0.04) ~~~~~~~~~~~~~ & ~~--- & b & 3\\
ESO557-G002\_S & 06h31m45.7s -17d38m44.8s & (-167.6) & 06h31m45.7s -17d38m44.7s & (108.7) & 0.70 (0.07) & 0.34 (0.06) &  6.42 (0.64) ~~~~~~~~~~~~~ &  1.09 & a & \\
ESO557-G002\_N & 06h31m47.2s -17d37m16.5s & (-167.6) & 06h31m47.2s -17d37m16.4s & (108.7) & 0.60 (0.01) & -0.73 (0.04) & 12.18 (0.09) ~~~~~~~~~~~~~ &  1.04 & a & \\
UGC03608 & 06h57m34.4s +46d24m10.7s & (174.4) & 06h57m34.4s +46d24m10.7s & (90.6) & 0.53 (0.01) & -0.32 (0.03) &  7.40 (0.07) ~~~~~~~~~~~~~ &  1.13 & b & \\
IRASF06592-6313 & 06h59m40.3s -63d17m52.5s & (-164.8) & 06h59m40.3s -63d17m52.5s & (111.5) & 0.48 (0.01) & -0.31 (0.04) &  6.98 (0.08) ~~~~~~~~~~~~~ &  1.10 & N & \\
AM0702-601\_N & 07h03m24.1s -60d15m21.8s & (-163.8) & 07h03m24.1s -60d15m21.8s & (112.4) & 0.04 (0.002)~~ & -0.10 (0.02) &  3.11 (0.01) ~~~~~~~~~~~~~ &  1.24 & a & 1\\
AM0702-601\_S & 07h03m28.5s -60d16m43.6s & (-163.8) & 07h03m28.5s -60d16m43.6s & (112.4) & 0.68 (0.01) & -0.05 (0.03) &  7.21 (0.72) ~~~~~~~~~~~~~ &  1.37 & a & 1\\
NGC2342 & 07h09m18.1s +20d38m09.5s & (2.6) & 07h09m18.1s +20d38m09.9s & (99.4) & 0.67 (0.01) & -0.06 (0.03) &  8.07 (0.11) ~~~~~~~~~~~~~ &  1.28 & a & \\
NGC2369 & 07h16m37.9s -62d20m36.4s & ~~(50.9)$*$ & 07h16m37.8s -62d20m34.2s & ~~(43.2)$*$ & 0.48 (0.01) & -0.60 (0.02) &  5.88 (0.02) ~~~~~~~~~~~~~ &  1.65 & N & \\
IRAS07251-0248 & 07h27m37.6s -02d54m54.2s & (-170.4) & 07h27m37.6s -02d54m54.2s & (105.9) & 0.09 (0.01) & -2.35 (0.14) & 13.38 (0.12) ~~~~~~~~~~~~~ &  1.10 & d & \\
NGC2388 & 07h28m53.4s +33d49m08.9s & & ~~~~~~~~~~~~~~~~~~~~~--- & & 0.53 (0.005)~~ & 0.07 (0.06) & ---~~~~~~~~~~~~~~~~~~~~ & ~~--- & a & \\
MCG+02-20-003\_S & 07h35m41.5s +11d36m42.0s & (-173.9) & 07h35m41.5s +11d36m42.0s & (102.3) & 0.58 (0.05) & 0.47 (0.18) &  4.62 (0.66) ~~~~~~~~~~~~~ &  1.74 & a & \\
MCG+02-20-003\_N & 07h35m43.5s +11d42m34.7s & (-173.9) & 07h35m43.4s +11d42m34.7s & (102.3) & 0.17 (0.003)~~ & -0.49 (0.04) & 10.08 (0.14) ~~~~~~~~~~~~~ &  1.26 & a & \\
IRAS08355-4944 & 08h37m01.8s -49d54m30.3s & (-155.8) & 08h37m01.8s -49d54m30.2s & (120.5) & 0.19 (0.003)~~ & -0.32 (0.02) &  6.59 (0.04) ~~~~~~~~~~~~~ &  1.16 & d & 3\\
IRASF08339+6517 & ~~~~~~~~~~~~~~~~~~~~~--- & & 08h38m23.2s +65d07m14.5s & (109.0) & --- & ---~~~~~~ &  7.37 (0.05) ~~~~~~~~~~~~~ & ~~--- & N & \\
NGC2623 & 08h38m24.1s +25d45m17.4s & (11.9) & 08h38m24.1s +25d45m17.2s & (-71.8) & 0.27 (0.01) & -1.12 (0.05) & 14.66 (0.13) ~~~~~~~~~~~~~ &  1.05 & d & 5\\
ESO432-IG006\_W & 08h44m27.2s -31d41m50.9s & (-155.6) & 08h44m27.2s -31d41m50.8s & (120.7) & 0.64 (0.01) & 0.02 (0.04) &  6.23 (0.11) ~~~~~~~~~~~~~ &  1.29 & b & \\
ESO432-IG006\_E & 08h44m28.9s -31d41m30.4s & (-155.6) & 08h44m28.9s -31d41m30.3s & (120.7) & 0.41 (0.01) & -0.01 (0.07) &  6.91 (0.14) ~~~~~~~~~~~~~ &  1.05 & b & \\
ESO60-IG016 & 08h52m31.8s -69d01m55.8s & (-84.5) & 08h52m31.8s -69d01m55.7s & (-168.2) & 0.11 (0.004)~~ & -1.76 (0.04) &  7.71 (0.04) ~~~~~~~~~~~~~ &  1.09 & b & 3\\
\hline
\end{tabular}}
\end{sidewaystable}
\begin{sidewaystable}
\resizebox{\linewidth}{!}{
\tabcolsep=3pt
\begin{tabular}{llclccrrlcc}
\hline
\hline
Source & ~~~~~~Short-Low RA/DEC & (PA) & ~~~~~~Long-Low RA/DEC & (PA) & 6.2$\mu m$ EQW($\sigma$) & s$_{9.7\mu m}$($\sigma$)~ & $F_{\nu}[30\mu m]/F_{\nu}[15\mu m]$($\sigma$) & Scale & Merger & MS\\
Name & ~~~~~~~~~~~~~~~~~~~[J2000] & [$^{\circ}$]~ & ~~~~~~~~~~~~~~~~~~~[J2000] & [$^{\circ}$]~ & [$\mu m$] & & & Factor & Stage & (HST)\\
\hline
IRASF08572+3915 & 09h00m25.4s +39d03m54.6s & (19.0) & 09h00m25.4s +39d03m55.1s & (-64.7) & $<$0.03 (---) & -3.58 (0.04) &  5.24 (0.01) ~~~~~~~~~~~~~ &  1.01 & d & 3\\
IRAS09022-3615 & 09h04m12.7s -36d27m01.6s & (23.2) & 09h04m12.7s -36d27m01.7s & (-60.5) & 0.14 (0.004)~~ & -0.88 (0.03) &  7.97 (0.05) ~~~~~~~~~~~~~ &  1.05 & d & \\
IRASF09111-1007\_W & 09h13m36.5s -10d19m30.0s & (-160.9) & 09h13m36.5s -10d19m29.9s & (115.4) & 0.47 (0.02) & -0.73 (0.10) & 12.54 (0.39) ~~~~~~~~~~~~~ &  1.13 & b & \\
IRASF09111-1007\_E & 09h13m38.9s -10d19m19.9s & (-160.9) & 09h13m38.9s -10d19m19.9s & (115.4) & 0.52 (0.03) & -0.15 (0.07) &  5.91 (0.26) ~~~~~~~~~~~~~ &  1.30 & b & \\
UGC04881\_W & 09h15m54.7s +44d19m50.7s & (-161.0) & 09h15m54.7s +44d19m50.7s & (115.3) & 0.61 (0.01) & -0.27 (0.04) &  9.17 (0.31) ~~~~~~~~~~~~~ &  1.10 & c & \\
UGC04881\_E & 09h15m55.5s +44d19m57.2s & (-161.0) & 09h15m55.5s +44d19m57.3s & (115.3) & 0.40 (0.01) & -0.81 (0.04) & 10.22 (0.17) ~~~~~~~~~~~~~ &  1.01 & c & \\
UGC05101 & 09h35m51.6s +61d21m11.7s & (47.4) & 09h35m51.7s +61d21m12.0s & (-36.4) & 0.13 (0.005)~~ & -0.78 (0.05) &  7.69 (0.05) ~~~~~~~~~~~~~ &  0.99 & d & \\
MCG+08-18-013 & ~~~~~~~~~~~~~~~~~~~~~--- & & 09h36m37.2s +48d28m27.9s & (119.2) & --- & ---~~~~~~ & 7.48 (0.05) ~~~~~~~~~~~~~ & ~~--- & a & \\
Arp303\_S & 09h46m20.3s +03d02m44.3s & (-163.3) & 09h46m20.3s +03d02m44.4s & (112.9) & 0.60 (0.02) & 0.06 (0.04) &  6.06 (0.11) ~~~~~~~~~~~~~ &  2.55 & a & \\
Arp303\_N & 09h46m21.1s +03d04m15.9s & (-163.3) & 09h46m21.1s +03d04m16.0s & (112.9) & 0.57 (0.06) & 0.08 (0.09) &  4.56 (0.10) ~~~~~~~~~~~~~ &  1.58 & a & \\
NGC3110 & 10h04m02.0s -06d28m29.5s & ~~(161.6)$*$ & 10h04m02.2s -06d28m31.9s & ~~(65.7)$*$ & 0.64 (0.01) & 0.19 (0.03) &  5.91 (0.04) ~~~~~~~~~~~~~ &  2.10 & a & \\
ESO374-IG032 & 10h06m04.7s -33d53m06.3s & (-128.3) & 10h06m04.7s -33d53m06.2s & (148.0) & 0.03 (0.002)~~ & -2.74 (0.04) &  9.36 (0.03) ~~~~~~~~~~~~~ &  1.13 & d & \\
IRASF10173+0828 & 10h20m00.2s +08d13m33.8s & (-162.9) & 10h20m00.2s +08d13m33.6s & (113.3) & 0.35 (0.04) & -1.20 (0.12) & 35.40 (1.38) ~~~~~~~~~~~~~ &  1.07 & a & \\
NGC3221 & 10h22m20.3s +21d34m22.1s & (-162.9) & 10h22m20.3s +21d34m22.1s & (113.3) & 0.75 (0.01) & -0.12 (0.04) &  5.94 (0.05) ~~~~~~~~~~~~~ &  1.44 & N & \\
NGC3256 & 10h27m51.2s -43d54m13.8s & (-5.5) & 10h27m51.3s -43d54m13.9s & (-57.4) & 0.61 (0.01) & -0.28 (0.02) &  8.16 (0.04) ~~~~~~~~~~~~~ &  1.33 & d & 5\\
ESO264-G036 & 10h43m07.5s -46d12m44.3s & (-111.2) & 10h43m07.5s -46d12m44.3s & (165.0) & 0.44 (0.01) & -0.04 (0.02) &  5.49 (0.03) ~~~~~~~~~~~~~ &  1.87 & N & \\
ESO264-G057 & 10h59m01.8s -43d26m25.8s & (-137.5) & 10h59m01.8s -43d26m25.8s & (138.8) & 0.60 (0.01) & -0.19 (0.04) &  7.46 (0.10) ~~~~~~~~~~~~~ &  1.34 & d & \\
IRASF10565+2448 & 10h59m18.1s +24d32m34.9s & (25.3) & 10h59m18.1s +24d32m35.4s & (-58.4) & 0.51 (0.01) & -0.75 (0.04) &  9.51 (0.07) ~~~~~~~~~~~~~ &  1.08 & d & 2\\
MCG+07-23-019 & 11h03m54.0s +40d51m00.7s & (-159.7) & 11h03m54.0s +40d51m00.7s & (116.6) & 0.64 (0.01) & -0.55 (0.03) & 12.40 (1.24) ~~~~~~~~~~~~~ &  1.14 & d & \\
CGCG011-076 & 11h21m12.2s -02d59m02.4s & ~~(164.4)$*$ & 11h21m12.3s -02d59m03.0s & ~~(157.9)$*$ & 0.32 (0.02) & -0.42 (0.03) &  4.35 (0.05) ~~~~~~~~~~~~~ &  1.30 & a & \\
IRASF11231+1456 & 11h25m45.1s +14d40m35.8s & (-166.8) & 11h25m45.1s +14d40m36.4s & (117.0) & 0.60 (0.01) & -0.22 (0.03) &  8.90 (0.20) ~~~~~~~~~~~~~ &  0.93 & a & 1\\
ESO319-G022 & 11h27m54.1s -41d36m52.4s & (-126.3) & 11h27m54.1s -41d36m52.4s & (150.0) & 0.42 (0.02) & -0.30 (0.05) & 10.81 (0.17) ~~~~~~~~~~~~~ &  1.11 & d & \\
NGC3690\_W & 11h28m31.1s +58d33m41.6s & ~~(108.8)$*$ & 11h28m31.1s +58d33m43.6s & ~~(44.3)$*$ & 0.12 (0.002)~~ & -0.77 (0.02) &  4.48 (0.00) ~~~~~~~~~~~~~ & ~~--- & c & 3\\
NGC3690\_E & 11h28m33.8s +58d33m47.0s & ~~(51.5)$*$ & 11h28m33.7s +58d33m48.3s & ~~(44.3)$*$ & 0.38 (0.003)~~ & -1.65 (0.02) & 12.12 (0.01) ~~~~~~~~~~~~~ &  1.44 & c & 3\\
ESO320-G030 & 11h53m11.6s -39d07m47.3s & ~~(166.3)$*$ & 11h53m11.6s -39d07m45.6s & ~~(158.2)$*$ & 0.58 (0.01) & -0.22 (0.02) & 13.24 (0.05) ~~~~~~~~~~~~~ &  1.44 & N & \\
ESO440-IG058\_N & 12h06m51.7s -31d56m46.4s & (-152.0) & 12h06m51.7s -31d56m46.3s & (124.3) & 0.56 (0.02) & 0.37 (0.04) &  9.80 (0.98) ~~~~~~~~~~~~~ & ~~--- & b & \\
ESO440-IG058\_S & 12h06m51.9s -31d56m59.2s & (-152.0) & 12h06m51.9s -31d56m59.1s & (124.3) & 0.66 (0.01) & -0.45 (0.03) &  7.56 (0.06) ~~~~~~~~~~~~~ &  1.21 & b & \\
IRASF12112+0305 & 12h13m46.0s +02d48m40.8s & (-161.5) & 12h13m46.0s +02d48m40.6s & (114.8) & 0.30 (0.03) & -1.24 (0.14) & 16.65 (0.34) ~~~~~~~~~~~~~ &  1.13 & d & 4\\
ESO267-G030\_W & 12h13m52.3s -47d16m25.9s & (-132.3) & 12h13m52.3s -47d16m25.8s & (144.0) & 0.68 (0.01) & -0.08 (0.04) &  7.39 (0.10) ~~~~~~~~~~~~~ &  1.15 & a & \\
NGC4194 & 12h14m09.6s +54d31m34.3s & (179.7) & 12h14m09.5s +54d31m33.8s & (96.0) & 0.55 (0.005)~~ & -0.29 (0.02) &  7.65 (0.05) ~~~~~~~~~~~~~ &  1.37 & d & \\
\hline
\end{tabular}}
\end{sidewaystable}
\begin{sidewaystable}
\resizebox{\linewidth}{!}{
\tabcolsep=3pt
\begin{tabular}{llclccrrlcc}
\hline
\hline
Source & ~~~~~~Short-Low RA/DEC & (PA) & ~~~~~~Long-Low RA/DEC & (PA) & 6.2$\mu m$ EQW($\sigma$) & s$_{9.7\mu m}$($\sigma$)~ & $F_{\nu}[30\mu m]/F_{\nu}[15\mu m]$($\sigma$) & Scale & Merger & MS\\
Name & ~~~~~~~~~~~~~~~~~~~[J2000] & [$^{\circ}$]~ & ~~~~~~~~~~~~~~~~~~~[J2000] & [$^{\circ}$]~ & [$\mu m$] & & & Factor & Stage & (HST)\\
\hline
ESO267-G030\_E & 12h14m12.8s -47d13m43.0s & (-132.4) & 12h14m12.8s -47d13m42.9s & (143.9) & 0.50 (0.01) & -0.01 (0.03) &  4.86 (0.06) ~~~~~~~~~~~~~ &  1.61 & a & \\
IRAS12116-5615 & 12h14m22.1s -56d32m32.9s & (-132.4) & 12h14m22.1s -56d32m32.8s & (143.9) & 0.36 (0.01) & -0.85 (0.03) &  7.93 (0.07) ~~~~~~~~~~~~~ &  1.11 & N & 0\\
IRASF12224-0624 & 12h25m03.9s -06d40m52.2s & (17.5) & 12h25m03.9s -06d40m52.1s & (-66.3) & 0.07 (0.02) & -2.12 (0.15) & 28.34 (1.46) ~~~~~~~~~~~~~ &  0.95 & N & \\
NGC4418 & 12h26m54.6s -00d52m40.0s & (18.9) & 12h26m54.6s -00d52m40.1s & (-64.8) & $<$0.02 (---) & -3.51 (0.09) &  7.83 (0.01) ~~~~~~~~~~~~~ &  1.04 & N & \\
UGC08058 & 12h56m14.3s +56d52m25.4s & (80.3) & 12h56m14.4s +56d52m24.7s & (-3.4) & 0.01 (0.001)~~ & -0.48 (0.02) &  3.67 (0.01) ~~~~~~~~~~~~~ &  1.02 & d & \\
NGC4922 & 13h01m25.3s +29d18m50.0s & & ~~~~~~~~~~~~~~~~~~~~~--- & & 0.16 (0.003)~~ & -0.60 (0.24) & ---~~~~~~~~~~~~~~~~~~~~ &  ~~--- & c & \\
CGCG043-099 & 13h01m50.3s +04d20m00.1s & (-164.9) & 13h01m50.3s +04d20m00.2s & (111.4) & 0.55 (0.01) & -0.81 (0.03) &  9.12 (0.17) ~~~~~~~~~~~~~ &  1.14 & d & \\
MCG-02-33-098\_W & 13h02m19.7s -15d46m04.0s & (-160.9) & 13h02m19.7s -15d46m03.9s & (115.3) & 0.55 (0.01) & -0.04 (0.03) &  5.77 (0.07) ~~~~~~~~~~~~~ &  1.15 & b & \\
MCG-02-33-098\_E & 13h02m20.4s -15d45m59.4s & (-160.9) & 13h02m20.4s -15d45m59.3s & (115.3) & 0.70 (0.02) & 0.04 (0.05) &  8.47 (0.21) ~~~~~~~~~~~~~ &  1.29 & b & \\
ESO507-G070 & 13h02m52.4s -23d55m18.3s & (-153.1) & 13h02m52.4s -23d55m18.2s & (123.2) & 0.56 (0.01) & -1.24 (0.10) & 13.58 (0.19) ~~~~~~~~~~~~~ &  1.10 & d & \\
IRAS13052-5711 & 13h08m18.7s -57d27m29.9s & (-144.6) & 13h08m18.7s -57d27m29.9s & (131.7) & 0.61 (0.01) & -0.51 (0.03) & 12.09 (0.21) ~~~~~~~~~~~~~ &  1.12 & a & \\
IC0860 & 13h15m03.5s +24d37m08.0s & (-177.8) & 13h15m03.5s +24d37m08.1s & (98.4) & 0.43 (0.01) & -0.98 (0.11) & 33.15 (0.29) ~~~~~~~~~~~~~ &  1.14 & N & \\
IRAS13120-5453 & 13h15m06.5s -55d09m23.6s & (-137.2) & 13h15m06.5s -55d09m23.8s & (139.1) & 0.45 (0.01) & -0.91 (0.03) & 12.54 (0.07) ~~~~~~~~~~~~~ &  1.26 & d & 5\\
VV250a\_W & 13h15m30.7s +62d07m45.7s & (-137.5) & 13h15m30.7s +62d07m45.7s & (138.8) & 0.76 (0.03) & -0.23 (0.11) &  8.09 (0.33) ~~~~~~~~~~~~~ &  1.35 & b & 2\\
VV250a\_E & 13h15m35.0s +62d07m29.1s & (-137.5) & 13h15m35.0s +62d07m29.1s & (138.8) & 0.63 (0.01) & -0.67 (0.03) &  7.65 (0.05) ~~~~~~~~~~~~~ &  1.03 & b & 2\\
UGC08387 & 13h20m35.4s +34d08m22.1s & (-169.5) & 13h20m35.4s +34d08m22.2s & (105.6) & 0.62 (0.01) & -1.01 (0.03) & 10.80 (0.10) ~~~~~~~~~~~~~ &  1.07 & d & 4\\
NGC5104 & 13h21m23.1s +00d20m32.6s & (-167.4) & 13h21m23.2s +00d20m33.2s & (113.7) & 0.51 (0.01) & -0.47 (0.04) &  6.67 (0.11) ~~~~~~~~~~~~~ &  1.21 & N & \\
MCG-03-34-064 & 13h22m24.5s -16d43m42.7s & (-159.5) & 13h22m24.5s -16d43m42.6s & (116.8) & $<$0.01 (---) & -0.18 (0.02) &  2.13 (0.00) ~~~~~~~~~~~~~ &  0.96 & a & \\
NGC5135 & 13h25m44.1s -29d50m00.7s & (15.0) & 13h25m44.2s -29d50m00.4s & (-68.7) & 0.49 (0.01) & -0.24 (0.02) &  5.78 (0.03) ~~~~~~~~~~~~~ &  1.53 & N & \\
ESO173-G015 & 13h27m23.8s -57d29m21.7s & (-148.7) & 13h27m23.8s -57d29m21.6s & (127.5) & 0.53 (0.01) & -1.31 (0.08) & 15.04 (0.07) ~~~~~~~~~~~~~ & ~~--- & N & \\
IC4280 & 13h32m53.4s -24d12m25.3s & (-157.6) & 13h32m53.4s -24d12m25.2s & (118.6) & 0.59 (0.01) & 0.13 (0.04) &  5.24 (0.09) ~~~~~~~~~~~~~ &  1.56 & N & \\
NGC5256 & 13h38m17.7s +48d16m32.9s & (-162.4) & 13h38m17.6s +48d16m33.9s & (113.9) & 0.44 (0.02) & -0.47 (0.03) &  6.41 (0.03) ~~~~~~~~~~~~~ & ~~--- & b & 3\\
Arp240\_W & 13h39m53.0s +00d50m25.3s & (19.1) & 13h39m53.0s +00d50m25.3s & (-64.6) & 0.63 (0.01) & -0.01 (0.03) &  7.18 (0.19) ~~~~~~~~~~~~~ &  1.66 & b & 2\\
Arp240\_E & 13h39m57.7s +00d49m51.0s & (19.1) & 13h39m57.7s +00d49m51.0s & (-64.6) & 0.54 (0.01) & 0.12 (0.03) &  6.31 (0.19) ~~~~~~~~~~~~~ &  1.69 & b & \\
UGC08696 & 13h44m42.2s +55d53m13.2s & (93.5) & 13h44m42.3s +55d53m12.6s & (9.8) & 0.12 (0.01) & -1.37 (0.05) & 12.36 (0.07) ~~~~~~~~~~~~~ &  1.06 & d & 4\\
UGC08739 & 13h49m13.9s +35d15m26.6s & (-169.7) & 13h49m13.9s +35d15m26.6s & (106.5) & 0.51 (0.01) & -0.59 (0.03) &  6.74 (0.12) ~~~~~~~~~~~~~ &  2.14 & N & \\
ESO221-IG010 & 13h50m56.9s -49d03m19.2s & (-153.3) & 13h50m56.9s -49d03m19.1s & (123.0) & 0.69 (0.01) & 0.03 (0.03) &  7.26 (0.08) ~~~~~~~~~~~~~ &  1.48 & N & \\
NGC5331\_S & 13h52m16.2s +02d06m05.5s & (-171.0) & 13h52m16.2s +02d06m05.5s & (105.2) & 0.61 (0.01) & -0.43 (0.03) &  7.06 (0.10) ~~~~~~~~~~~~~ &  1.19 & c & 3\\
NGC5331\_N & 13h52m16.5s +02d06m31.3s & (-171.0) & 13h52m16.4s +02d06m31.3s & (105.2) & 0.60 (0.01) & 0.13 (0.03) &  5.24 (0.24) ~~~~~~~~~~~~~ &  1.11 & c & 3\\
Arp84\_N & 13h58m33.6s +37d27m13.2s & (-178.0) & 13h58m33.6s +37d27m13.2s & (98.2) & 0.62 (0.01) & -0.08 (0.03) &  6.84 (0.03) ~~~~~~~~~~~~~ &  1.09 & c & \\
\hline
\end{tabular}}
\end{sidewaystable}
\begin{sidewaystable}
\resizebox{\linewidth}{!}{
\tabcolsep=3pt
\begin{tabular}{llclccrrlcc}
\hline
\hline
Source & ~~~~~~Short-Low RA/DEC & (PA) & ~~~~~~Long-Low RA/DEC & (PA) & 6.2$\mu m$ EQW($\sigma$) & s$_{9.7\mu m}$($\sigma$)~ & $F_{\nu}[30\mu m]/F_{\nu}[15\mu m]$($\sigma$) & Scale & Merger & MS\\
Name & ~~~~~~~~~~~~~~~~~~~[J2000] & [$^{\circ}$]~ & ~~~~~~~~~~~~~~~~~~~[J2000] & [$^{\circ}$]~ & [$\mu m$] & & & Factor & Stage & (HST)\\
\hline
Arp84\_S & 13h58m37.9s +37d25m28.2s & (-178.0) & 13h58m37.9s +37d25m28.2s & (98.2) & 0.17 (0.02) & 0.52 (0.07) &  3.82 (0.24) ~~~~~~~~~~~~~ &  2.10 & c & \\
CGCG247-020 & 14h19m43.3s +49d14m11.5s & (-152.2) & 14h19m43.3s +49d14m11.6s & (115.2) & 0.56 (0.01) & -0.20 (0.03) &  7.64 (0.10) ~~~~~~~~~~~~~ &  1.13 & N & \\
NGC5653 & 14h30m10.4s +31d12m56.1s & (172.7) & 14h30m10.4s +31d12m56.1s & (89.0) & 0.54 (0.01) & 0.25 (0.03) &  7.35 (0.04) ~~~~~~~~~~~~~ & ~~--- & N & \\
IRASF14348-1447 & 14h37m38.4s -15d00m24.0s & (-167.2) & 14h37m38.3s -15d00m24.6s & (109.1) & 0.25 (0.01) & -1.36 (0.16) & 16.76 (0.36) ~~~~~~~~~~~~~ &  1.14 & d & 4\\
IRASF14378-3651 & 14h40m59.0s -37d04m33.1s & (-160.3) & 14h40m58.9s -37d04m33.0s & (116.0) & 0.39 (0.03) & -1.14 (0.16) & 15.71 (0.30) ~~~~~~~~~~~~~ &  1.08 & d & 6\\
NGC5734\_N & 14h45m09.0s -20d52m13.2s & (-167.1) & 14h45m09.0s -20d52m13.2s & (109.1) & 0.47 (0.01) & 0.12 (0.03) &  5.01 (0.14) ~~~~~~~~~~~~~ &  1.66 & a & \\
NGC5734\_S & 14h45m11.0s -20d54m48.7s & (-167.1) & 14h45m11.0s -20d54m48.5s & (109.1) & 0.49 (0.01) & -0.04 (0.04) &  5.07 (0.08) ~~~~~~~~~~~~~ &  2.15 & a & \\
VV340a\_S & 14h57m00.3s +24d36m24.2s & (177.8) & 14h57m00.3s +24d36m24.3s & (94.1) & 0.58 (0.02) & 0.33 (0.04) &  4.82 (0.31) ~~~~~~~~~~~~~ &  1.37 & b & 1\\
VV340a\_N & 14h57m00.7s +24d37m05.4s & (177.8) & 14h57m00.7s +24d37m05.5s & (94.1) & 0.58 (0.01) & -0.63 (0.03) &  6.21 (0.10) ~~~~~~~~~~~~~ &  0.82 & b & 1\\
CGCG049-057 & 15h13m13.1s +07d13m33.1s & ~~(155.9)$*$ & 15h13m13.0s +07d13m35.2s & ~~(59.8)$*$ & 0.51 (0.04) & -0.83 (0.03) & 31.40 (0.39) ~~~~~~~~~~~~~ &  1.05 & N & \\
VV705 & 15h18m06.4s +42d44m36.6s & (-145.5) & ~~~~~~~~~~~~~~~~~~~~~--- & & 0.75 (0.06) & 0.32 (0.06) & ---~~~~~~~~~~~~~~~~~~~~ & ~~--- & b & \\
ESO099-G004 & 15h24m58.0s -63d07m29.2s & (-169.3) & 15h24m58.0s -63d07m29.1s & (106.9) & 0.53 (0.01) & -0.78 (0.04) &  7.74 (0.06) ~~~~~~~~~~~~~ &  1.04 & d & 3\\
IRASF15250+3608 & 15h26m59.4s +35d58m37.7s & (170.1) & 15h26m59.4s +35d58m37.2s & (86.3) & 0.03 (0.01) & -2.69 (0.07) & 10.52 (0.03) ~~~~~~~~~~~~~ &  0.97 & d & 5\\
NGC5936 & 15h30m00.8s +12d59m22.3s & (-172.2) & 15h30m00.8s +12d59m22.4s & (104.0) & 0.62 (0.01) & -0.13 (0.03) &  6.41 (0.07) ~~~~~~~~~~~~~ &  1.25 & N & \\
Arp220 & 15h34m57.3s +23d30m11.7s & (178.0) & 15h34m57.2s +23d30m11.1s & (94.2) & 0.17 (0.004)~~ & -2.26 (0.06) & 20.38 (1.67) ~~~~~~~~~~~~~ &  1.08 & d & 4\\
NGC5990 & 15h46m16.4s +02d24m55.7s & (-171.3) & 15h46m16.4s +02d24m55.8s & (104.9) & 0.15 (0.002)~~ & -0.11 (0.03) &  4.74 (0.02) ~~~~~~~~~~~~~ &  1.22 & a & \\
NGC6052 & 16h05m13.1s +20d32m35.6s & ~~(59.0)$*$ & 16h05m12.9s +20d32m35.5s & ~~(51.7)$*$ & 0.69 (0.03) & -0.06 (0.03) &  6.63 (0.04) ~~~~~~~~~~~~~ &  3.08 & c & \\
NGC6090 & 16h11m40.2s +52d27m24.9s & (26.8) & 16h11m40.3s +52d27m24.7s & (-57.0) & 0.73 (0.02) & -0.04 (0.03) &  7.54 (0.03) ~~~~~~~~~~~~~ & ~~--- & c & 4\\
IRASF16164-0746 & 16h19m11.8s -07d54m02.7s & (-177.8) & 16h19m11.8s -07d54m02.7s & (102.1) & 0.61 (0.01) & -1.18 (0.07) & 11.58 (0.20) ~~~~~~~~~~~~~ &  1.03 & d & 5\\
CGCG052-037 & 16h30m56.5s +04d04m58.5s & (16.1) & 16h30m56.5s +04d04m58.8s & (101.2) & 0.62 (0.01) & -0.21 (0.02) &  7.07 (0.09) ~~~~~~~~~~~~~ &  1.16 & N & \\
NGC6156 & 16h34m52.5s -60d37m07.5s & (-179.8) & 16h34m52.5s -60d37m07.5s & (96.4) & 0.37 (0.01) & 0.47 (0.02) &  5.89 (0.02) ~~~~~~~~~~~~~ &  1.23 & N & \\
ESO069-IG006 & 16h38m11.8s -68d26m08.0s & (178.6) & 16h38m11.8s -68d26m07.9s & (94.8) & 0.64 (0.01) & -0.40 (0.04) &  7.91 (0.10) ~~~~~~~~~~~~~ &  1.17 & b & 2\\
IRASF16399-0937 & 16h42m40.1s -09d43m13.5s & (-176.4) & 16h42m40.1s -09d43m13.4s & (99.9) & 0.43 (0.01) & -1.12 (0.04) &  9.09 (0.09) ~~~~~~~~~~~~~ &  1.14 & d & 3\\
ESO453-G005\_N & 16h47m29.4s -29d19m06.8s & (-175.6) & 16h47m29.4s -29d19m06.7s & (100.7) & 0.74 (0.07) & 0.47 (0.05) &  5.21 (0.14) ~~~~~~~~~~~~~ &  1.51 & N & \\
ESO453-G005\_S & 16h47m31.1s -29d21m21.6s & (-175.6) & 16h47m31.1s -29d21m21.5s & (100.6) & 0.42 (0.01) & -0.47 (0.09) & 27.11 (2.71) ~~~~~~~~~~~~~ &  1.22 & N & \\
NGC6240 & 16h52m58.9s +02d24m03.7s & (-177.4) & 16h52m58.9s +02d24m03.0s & (98.9) & 0.35 (0.01) & -0.92 (0.07) &  7.72 (0.04) ~~~~~~~~~~~~~ &  1.05 & d & 4\\
IRASF16516-0948 & 16h54m23.9s -09d53m20.6s & (-177.9) & 16h54m23.9s -09d53m20.5s & (98.4) & 0.69 (0.01) & 0.18 (0.03) &  7.02 (0.15) ~~~~~~~~~~~~~ &  1.67 & d & \\
NGC6286\_N & 16h58m24.0s +58d57m21.4s & (-117.4) & 16h58m24.0s +58d57m21.9s & (116.3) & 0.66 (0.02) & 0.08 (0.03) &  6.81 (0.07) ~~~~~~~~~~~~~ &  1.14 & b & \\
NGC6286\_S & 16h58m31.3s +58d56m10.7s & (27.8) & 16h58m31.7s +58d56m13.5s & (116.3) & 0.59 (0.01) & -0.40 (0.03) &  5.97 (0.04) ~~~~~~~~~~~~~ &  0.91 & b & \\
IRASF17132+5313 & ~~~~~~~~~~~~~~~~~~~~~--- & & 17h14m20.2s +53d10m30.4s & (-102.2) & --- & ---~~~~~~ &  9.67 (0.97) ~~~~~~~~~~~~~ & ~~--- & b & \\
\hline
\end{tabular}}
\end{sidewaystable}
\begin{sidewaystable}
\resizebox{\linewidth}{!}{
\tabcolsep=3pt
\begin{tabular}{llclccrrlcc}
\hline
\hline
Source & ~~~~~~Short-Low RA/DEC & (PA) & ~~~~~~Long-Low RA/DEC & (PA) & 6.2$\mu m$ EQW($\sigma$) & s$_{9.7\mu m}$($\sigma$)~ & $F_{\nu}[30\mu m]/F_{\nu}[15\mu m]$($\sigma$) & Scale & Merger & MS\\
Name & ~~~~~~~~~~~~~~~~~~~[J2000] & [$^{\circ}$]~ & ~~~~~~~~~~~~~~~~~~~[J2000] & [$^{\circ}$]~ & [$\mu m$] & & & Factor & Stage & (HST)\\
\hline
IRASF17138-1017 & 17h16m35.6s -10d20m37.9s & (173.8) & 17h16m36.0s -10d20m41.2s & ~~(1.2)$*$ & 0.68 (0.01) & -0.37 (0.04) &  7.17 (1.51) ~~~~~~~~~~~~~ &  4.12 & d & 6\\
IRASF17207-0014 & 17h23m22.0s -00d17m00.8s & (173.2) & 17h23m22.0s -00d17m01.0s & (89.5) & 0.31 (0.01) & -1.26 (0.07) & 26.39 (0.24) ~~~~~~~~~~~~~ &  1.22 & d & 5\\
ESO138-G027 & 17h26m43.4s -59d55m54.8s & (-177.1) & 17h26m43.3s -59d55m54.7s & (99.1) & 0.52 (0.01) & -0.35 (0.06) &  8.13 (0.03) ~~~~~~~~~~~~~ &  1.14 & N & \\
UGC11041 & 17h54m51.9s +34d46m34.0s & (-14.4) & 17h54m51.9s +34d46m34.0s & (-98.2) & 0.58 (0.01) & 0.07 (0.03) &  5.27 (0.04) ~~~~~~~~~~~~~ &  2.26 & N & \\
CGCG141-034 & 17h56m56.6s +24d01m01.8s & (-11.9) & 17h56m56.6s +24d01m01.8s & (-95.6) & 0.48 (0.01) & -0.64 (0.06) &  9.35 (0.16) ~~~~~~~~~~~~~ &  1.18 & N & \\
IRAS17578-0400\_W & 18h00m24.3s -04d01m04.2s & (171.6) & 18h00m24.3s -04d01m04.1s & (87.9) & 0.62 (0.02) & 0.16 (0.09) &  8.56 (0.25) ~~~~~~~~~~~~~ &  2.28 & b & \\
IRAS17578-0400\_N & 18h00m31.8s -04d00m53.8s & (171.6) & 18h00m31.8s -04d00m53.7s & (87.9) & 0.68 (0.01) & -0.67 (0.05) & 21.18 (0.24) ~~~~~~~~~~~~~ &  1.45 & b & \\
IRAS17578-0400\_S & 18h00m34.1s -04d01m44.3s & (171.6) & 18h00m34.1s -04d01m44.2s & (87.9) & 0.78 (0.01) & 0.15 (0.07) &  6.38 (0.21) ~~~~~~~~~~~~~ &  1.83 & a & \\
IRAS18090+0130\_W & 18h11m33.4s +01d31m42.3s & (169.3) & 18h11m33.4s +01d31m42.4s & (85.5) & 0.52 (0.03) & -0.72 (0.10) &  8.04 (0.26) ~~~~~~~~~~~~~ &  1.19 & b & \\
IRAS18090+0130\_E & 18h11m38.4s +01d31m40.2s & (169.3) & 18h11m38.4s +01d31m40.3s & (85.5) & 0.61 (0.01) & -0.28 (0.04) &  7.16 (0.06) ~~~~~~~~~~~~~ &  1.22 & b & 2\\
NGC6621 & 18h12m55.2s +68d21m48.4s & ~~(21.6)$*$ & 18h12m54.8s +68d21m48.7s & ~~(5.2)$*$ & 0.56 (0.01) & -0.16 (0.02) &  5.83 (0.01) ~~~~~~~~~~~~~ &  1.26 & b & \\
IC4687 & 18h13m39.7s -57d43m30.5s & ~~(13.6)$*$ & 18h13m40.2s -57d43m33.5s & ~~(2.1)$*$ & 0.73 (0.02) & 0.12 (0.02) &  6.50 (0.02) ~~~~~~~~~~~~~ &  1.56 & b & \\
CGCG142-034\_W & 18h16m33.8s +22d06m38.9s & (172.9) & 18h16m33.8s +22d06m39.0s & (89.1) & 0.48 (0.01) & -0.24 (0.03) &  5.62 (0.21) ~~~~~~~~~~~~~ &  1.03 & a & \\
CGCG142-034\_E & 18h16m40.7s +22d06m46.4s & (172.9) & 18h16m40.7s +22d06m46.5s & (89.1) & 0.51 (0.01) & -0.48 (0.03) &  7.34 (0.11) ~~~~~~~~~~~~~ &  1.95 & a & \\
IRASF18293-3413 & 18h32m41.1s -34d11m27.1s & (-8.8) & 18h32m41.1s -34d11m27.1s & (-92.5) & 0.63 (0.01) & -0.51 (0.02) &  7.66 (0.05) ~~~~~~~~~~~~~ &  1.38 & c & 1\\
NGC6670\_W & 18h33m34.3s +59d53m17.9s & (-7.3) & 18h33m34.3s +59d53m17.9s & (-91.0) & 0.63 (0.01) & -0.22 (0.03) &  8.82 (0.11) ~~~~~~~~~~~~~ &  1.85 & b & 2\\
NGC6670\_E & 18h33m37.8s +59d53m22.8s & (-7.3) & 18h33m37.8s +59d53m22.8s & (-91.0) & 0.61 (0.01) & -0.34 (0.03) &  8.78 (0.10) ~~~~~~~~~~~~~ &  1.26 & b & 2\\
IC4734 & 18h38m25.8s -57d29m25.4s & (167.4) & 18h38m25.8s -57d29m25.3s & (83.7) & 0.51 (0.01) & -0.60 (0.04) &  8.26 (0.09) ~~~~~~~~~~~~~ &  1.35 & N & \\
NGC6701 & 18h43m12.5s +60d39m11.9s & (-6.0) & 18h43m12.5s +60d39m11.9s & (-89.8) & 0.55 (0.01) & -0.15 (0.09) &  7.10 (0.07) ~~~~~~~~~~~~~ &  1.27 & N & \\
VV414\_W & 19h10m53.9s +73d24m36.1s & (0.6) & 19h10m53.9s +73d24m36.1s & (-83.2) & 0.64 (0.01) & 0.10 (0.03) &  7.24 (0.06) ~~~~~~~~~~~~~ &  1.41 & c & 2\\
VV414\_E & 19h11m04.3s +73d25m32.6s & (0.6) & 19h11m04.3s +73d25m32.6s & (-83.1) & 0.29 (0.01) & -0.08 (0.02) &  5.46 (0.02) ~~~~~~~~~~~~~ &  1.08 & c & 2\\
ESO593-IG008 & 19h14m31.2s -21d19m06.4s & (-13.2) & 19h14m31.2s -21d19m06.3s & (-96.9) & 0.56 (0.01) & -0.59 (0.03) &  8.92 (0.12) ~~~~~~~~~~~~~ &  1.38 & d & 4\\
IRASF19297-0406 & 19h32m22.3s -04d00m00.2s & (156.3) & 19h32m22.3s -04d00m00.2s & (72.6) & 0.30 (0.02) & -1.05 (0.13) & 15.07 (0.25) ~~~~~~~~~~~~~ &  1.16 & d & 4\\
IRAS19542+1110 & 19h56m35.8s +11d19m05.4s & (-6.4) & 19h56m35.8s +11d19m05.4s & (-90.1) & 0.29 (0.02) & -0.74 (0.08) & 14.53 (0.31) ~~~~~~~~~~~~~ &  1.07 & N & 0\\
ESO339-G011 & 19h57m37.6s -37d56m08.4s & (-22.0) & 19h57m37.6s -37d56m08.4s & (-105.8) & 0.27 (0.01) & -0.31 (0.03) &  4.64 (0.04) ~~~~~~~~~~~~~ &  1.22 & N & \\
NGC6907 & 20h25m06.5s -24d48m32.6s & (-22.2) & 20h25m06.5s -24d48m32.5s & (-105.9) & 0.57 (0.01) & 0.10 (0.03) &  6.67 (0.04) ~~~~~~~~~~~~~ &  1.62 & N & \\
MCG+04-48-002 & 20h28m35.1s +25d44m00.2s & (2.8) & 20h28m35.1s +25d44m00.2s & (-80.9) & 0.57 (0.01) & -0.47 (0.03) &  5.22 (0.05) ~~~~~~~~~~~~~ &  1.73 & a & \\
NGC6926 & 20h33m06.1s -02d01m38.8s & (-12.0) & 20h33m06.1s -02d01m38.7s & (-95.7) & 0.37 (0.01) & -0.49 (0.03) &  5.07 (0.05) ~~~~~~~~~~~~~ &  1.45 & d & \\
IRAS20351+2521 & 20h37m17.7s +25d31m39.1s & (3.8) & 20h37m17.7s +25d31m39.1s & (-79.9) & 0.57 (0.01) & -0.24 (0.04) &  7.50 (0.10) ~~~~~~~~~~~~~ &  1.68 & N & 0\\
CGCG448-020\_W & 20h57m24.0s +17d07m35.1s & (-11.9) & 20h57m24.0s +17d07m35.2s & (-95.6) & 0.48 (0.01) & -0.42 (0.02) &  9.14 (0.04) ~~~~~~~~~~~~~ & ~~--- & c & 3\\
\hline
\end{tabular}}
\end{sidewaystable}
\begin{sidewaystable}
\resizebox{\linewidth}{!}{
\tabcolsep=3pt
\begin{tabular}{llclccrrlcc}
\hline
\hline
Source & ~~~~~~Short-Low RA/DEC & (PA) & ~~~~~~Long-Low RA/DEC & (PA) & 6.2$\mu m$ EQW($\sigma$) & s$_{9.7\mu m}$($\sigma$)~ & $F_{\nu}[30\mu m]/F_{\nu}[15\mu m]$($\sigma$) & Scale & Merger & MS\\
Name & ~~~~~~~~~~~~~~~~~~~[J2000] & [$^{\circ}$]~ & ~~~~~~~~~~~~~~~~~~~[J2000] & [$^{\circ}$]~ & [$\mu m$] & & & Factor & Stage & (HST)\\
\hline
CGCG448-020\_E & 20h57m24.3s +17d07m39.1s & (-11.9) & 20h57m24.3s +17d07m39.2s & (-95.6) & 0.27 (0.01) & -0.92 (0.03) &  9.82 (0.04) ~~~~~~~~~~~~~ & ~~--- & c & 3\\
IRAS20551-4250 & 20h58m26.8s -42d39m01.7s & (162.1) & 20h58m26.7s -42d39m02.8s & (78.4) & 0.10 (0.01) & -2.52 (0.10) &  7.59 (0.02) ~~~~~~~~~~~~~ &  1.04 & d & 5\\
ESO286-G035 & 21h04m11.1s -43d35m34.4s & (-27.5) & 21h04m11.1s -43d35m34.4s & (-111.2) & 0.69 (0.01) & -0.24 (0.03) &  7.02 (0.09) ~~~~~~~~~~~~~ &  1.43 & a & \\
IRAS21101+5810 & 21h11m29.3s +58d23m07.6s & (10.5) & 21h11m29.3s +58d23m07.6s & (-73.2) & 0.55 (0.02) & -0.87 (0.07) & 11.06 (0.15) ~~~~~~~~~~~~~ &  1.04 & c & 2\\
ESO343-IG013\_S & 21h36m10.5s -38d32m42.4s & (-31.9) & 21h36m10.5s -38d32m42.4s & (-115.6) & 0.61 (0.01) & -0.09 (0.03) &  9.21 (0.18) ~~~~~~~~~~~~~ & ~~--- & c & \\
ESO343-IG013\_N & 21h36m10.9s -38d32m32.6s & (-1.9) & 21h36m10.9s -38d32m32.6s & (-115.6) & 0.47 (0.01) & -0.48 (0.02) &  6.71 (0.07) ~~~~~~~~~~~~~ & ~~--- & c & \\
NGC7130 & 21h48m19.6s -34d57m01.9s & (154.1) & 21h48m19.5s -34d57m02.1s & (70.4) & 0.30 (0.01) & -0.27 (0.03) &  5.67 (0.03) ~~~~~~~~~~~~~ &  1.42 & N & \\
ESO467-G027 & 22h14m39.9s -27d27m51.4s & (-36.7) & 22h14m39.9s -27d27m51.3s & (-120.4) & 0.62 (0.01) & 0.29 (0.02) &  5.48 (0.09) ~~~~~~~~~~~~~ &  2.24 & N & \\
IC5179 & 22h16m09.0s -36d50m38.1s & ~~(121.5)$*$ & 22h16m09.3s -36d50m33.8s & ~~(115.9)$*$ & 0.63 (0.01) & 0.07 (0.02) &  5.42 (0.02) ~~~~~~~~~~~~~ &  2.43 & N & \\
ESO602-G025 & 22h31m25.5s -19d02m04.0s & (-31.4) & 22h31m25.4s -19d02m04.0s & (-117.1) & 0.45 (0.01) & -0.66 (0.04) &  6.23 (0.08) ~~~~~~~~~~~~~ &  1.17 & N & \\
UGC12150 & 22h41m12.2s +34d14m56.9s & (-20.1) & 22h41m12.2s +34d14m56.8s & (-98.7) & 0.53 (0.01) & -0.42 (0.02) &  8.14 (0.09) ~~~~~~~~~~~~~ &  1.27 & N & \\
ESO239-IG002 & 22h49m39.8s -48d50m58.4s & (-48.3) & 22h49m39.8s -48d50m58.4s & (-132.0) & 0.45 (0.02) & -0.49 (0.04) & 10.35 (0.12) ~~~~~~~~~~~~~ &  1.08 & d & 5\\
IRASF22491-1808 & 22h51m49.4s -17d52m24.5s & (155.2) & 22h51m49.2s -17d52m24.8s & (71.5) & 0.48 (0.02) & -1.04 (0.13) & 17.63 (0.33) ~~~~~~~~~~~~~ &  1.14 & d & 4\\
NGC7469 & 23h03m15.6s +08d52m26.2s & (-29.7) & 23h03m15.6s +08d52m27.3s & (-113.4) & 0.23 (0.002)~~ & 0.06 (0.02) &  4.81 (0.03) ~~~~~~~~~~~~~ &  1.12 & a & 2\\
CGCG453-062 & 23h04m56.5s +19d33m07.6s & (-24.9) & 23h04m56.6s +19d33m07.8s & (-103.6) & 0.58 (0.02) & -0.46 (0.04) & 12.83 (0.10) ~~~~~~~~~~~~~ &  1.50 & N & \\
ESO148-IG002 & 23h15m47.0s -59d03m17.0s & (136.3) & 23h15m47.0s -59d03m18.4s & (52.6) & 0.31 (0.01) & -0.66 (0.03) &  6.49 (0.05) ~~~~~~~~~~~~~ &  1.03 & c & 4\\
IC5298 & 23h16m00.7s +25d33m24.4s & (-19.5) & 23h16m00.7s +25d33m24.5s & (-103.3) & 0.12 (0.004)~~ & -0.37 (0.02) &  7.41 (0.02) ~~~~~~~~~~~~~ &  1.01 & N & 0\\
NGC7552 & 23h16m10.6s -42d35m03.8s & ~~(112.9)$*$ & 23h16m10.6s -42d35m05.8s & ~~(107.8)$*$ & 0.56 (0.01) & -0.21 (0.02) &  6.62 (0.01) ~~~~~~~~~~~~~ &  1.72 & N & \\
NGC7591 & 23h18m16.3s +06d35m08.7s & (-27.4) & 23h18m16.2s +06d35m09.2s & (-109.7) & 0.48 (0.01) & -0.37 (0.05) &  7.96 (0.13) ~~~~~~~~~~~~~ &  1.21 & N & \\
NGC7592\_W & 23h18m21.7s -04d24m57.6s & ~~(119.1)$*$ & 23h18m21.5s -04d24m54.2s & ~~(112.9)$*$ & 0.30 (0.01) & -1.08 (0.03) &  5.13 (0.02) ~~~~~~~~~~~~~ &  1.10 & b & \\
NGC7592\_E & 23h18m22.7s -04d24m57.8s & ~~(119.1)$*$ & 23h18m22.8s -04d25m02.3s & ~~(112.9)$*$ & 0.69 (0.01) & 0.11 (0.02) &  8.38 (0.07) ~~~~~~~~~~~~~ &  1.20 & b & \\
ESO077-IG014\_W & 23h21m03.7s -69d13m00.9s & (-55.8) & 23h21m03.7s -69d13m00.9s & (-139.5) & 0.63 (0.01) & -0.59 (0.04) & 11.84 (0.17) ~~~~~~~~~~~~~ & ~~--- & b & 2\\
ESO077-IG014\_E & 23h21m05.4s -69d12m47.2s & (-55.8) & 23h21m05.4s -69d12m47.2s & (-139.5) & 0.49 (0.01) & -0.61 (0.03) & 12.54 (0.12) ~~~~~~~~~~~~~ & ~~--- & b & 2\\
NGC7674 & 23h27m56.7s +08d46m44.5s & ~~(116.5)$*$ & 23h27m56.6s +08d46m41.8s & ~~(20.5)$*$ & 0.02 (0.01) & -0.13 (0.02) &  2.52 (0.02) ~~~~~~~~~~~~~ &  1.13 & a & 2\\
NGC7679 & 23h28m46.6s +03d30m41.7s & (-27.4) & 23h28m46.6s +03d30m41.7s & (-111.1) & 0.64 (0.01) & 0.34 (0.02) &  5.78 (0.06) ~~~~~~~~~~~~~ &  1.41 & a & \\
IRASF23365+3604 & 23h39m01.3s +36d21m10.2s & (-35.5) & 23h39m01.4s +36d21m11.3s & (-119.2) & 0.41 (0.02) & -1.41 (0.17) & 10.04 (0.11) ~~~~~~~~~~~~~ &  1.89 & d & 5\\
MCG-01-60-022 & ~~~~~~~~~~~~~~~~~~~~~--- & & 23h42m00.7s -03d36m54.7s & (-114.0) & --- & ---~~~~~~ &  7.24 (0.13) ~~~~~~~~~~~~~ & ~~--- & a & \\
IRAS23436+5257 & 23h46m05.4s +53d14m01.3s & (-3.9) & 23h46m05.4s +53d14m01.3s & (-87.7) & 0.37 (0.01) & -0.22 (0.03) &  7.00 (0.06) ~~~~~~~~~~~~~ &  1.24 & c & 4\\
Arp86\_S & 23h46m58.5s +29d27m31.6s & (-19.8) & 23h46m58.5s +29d27m31.8s & (-103.5) & 0.72 (0.01) & 0.12 (0.02) &  6.14 (0.11) ~~~~~~~~~~~~~ &  1.61 & b & \\
Arp86\_N & 23h47m04.8s +29d28m59.6s & (-19.7) & 23h47m04.8s +29d28m59.8s & (-103.5) & 0.32 (0.01) & 0.14 (0.04) &  6.72 (0.18) ~~~~~~~~~~~~~ &  1.53 & b & \\
\hline
\end{tabular}}
\end{sidewaystable}
\begin{sidewaystable}
\resizebox{\linewidth}{!}{
\tabcolsep=3pt
\begin{tabular}{llclccrrlcc}
\hline
\hline
Source & ~~~~~~Short-Low RA/DEC & (PA) & ~~~~~~Long-Low RA/DEC & (PA) & 6.2$\mu m$ EQW($\sigma$) & s$_{9.7\mu m}$($\sigma$)~ & $F_{\nu}[30\mu m]/F_{\nu}[15\mu m]$($\sigma$) & Scale & Merger & MS\\
Name & ~~~~~~~~~~~~~~~~~~~[J2000] & [$^{\circ}$]~ & ~~~~~~~~~~~~~~~~~~~[J2000] & [$^{\circ}$]~ & [$\mu m$] & & & Factor & Stage & (HST)\\
\hline
NGC7771\_W & 23h51m04.0s +20d09m02.0s & (-19.8) & 23h51m04.0s +20d09m02.0s & (-103.6) & 0.39 (0.01) & 0.27 (0.06) &  6.26 (0.11) ~~~~~~~~~~~~~ &  1.34 & N & \\
NGC7771\_S & 23h51m22.5s +20d05m47.0s & (-19.8) & 23h51m22.5s +20d05m47.0s & (-103.6) & 0.36 (0.01) & -0.20 (0.04) &  4.42 (0.05) ~~~~~~~~~~~~~ &  2.11 & c & \\
NGC7771\_N & 23h51m24.9s +20d06m43.0s & (-19.8) & 23h51m24.9s +20d06m43.1s & (-103.6) & 0.52 (0.01) & -0.24 (0.03) &  8.43 (0.04) ~~~~~~~~~~~~~ &  1.95 & a & \\
MRK0331 & 23h51m26.7s +20d35m10.1s & (-19.6) & 23h51m26.7s +20d35m10.3s & (-103.3) & 0.63 (0.01) & -0.35 (0.03) &  9.40 (0.05) ~~~~~~~~~~~~~ &  1.19 & a & 1\\
\hline
\end{tabular}}
~~MIR Spectral Parameters of the GOALS Sample. Column (1): Source Name, Columns (2)-(5): the central right ascension, declination, and
position angle of the field of view
for the SL and LL observations (for staring mode data:
RA\_FOV, DEC\_FOV, \& PA\_FOV from the headers with a
pointing accuracy within 1\arcsec; for mapping mode data: calculated
from the four corners of the extraction aperture used in CUBISM), Column (6): the equivalent width of the
6.2\micron~PAH feature in \micron, Column (7): the apparent depth of the
9.7\micron~silicate absorption feature, Column (8): the MIR slope calculated
using F$_{\nu}$ at 15 and 30\micron, Column (9): the SL-to-LL scale factor, Column (10): the merger stage (N $=$ nonmerger, a $=$ pre-merger, b $=$ early stage merger, c $=$ mid-stage merger, and d $=$ late stage merger, see Section \ref{mergsec} for details), and Column (11): merger stage as derived from the high resolution HST data (0 $=$ nonmerger, 1 $=$ pre-merger, 2 $=$ ongoing merger with separable progenitor galaxies, 3 $=$ ongoing merger with progenitors sharing a common envelope, 4 $=$ ongoing merger with double nuclei plus tidal tail, 5 $=$ post-merger with single nucleus plus prominent tail, and 6 $=$ post-merger with single nucleus with disturbed morphology, as described in \cite{haanHST}).\\
~~$*$Data was taken in mapping mode, and so the PA was user-selected. 
\end{sidewaystable}
\end{document}